\documentclass[12pt]{elsarticle}
\makeatletter
\def\ps@pprintTitle{%
 \let\@oddhead\@empty
 \let\@evenhead\@empty
 \def\@oddfoot{\centerline{\thepage}}%
 \def\@evenfoot{\thepage}
 \let\@evenfoot\@oddfoot}
\makeatother

\usepackage[final]{changes} 
\definecolor{C0}{HTML}{1F77B4} 
\definecolor{C1}{HTML}{FF7F0E}
\definecolor{C2}{HTML}{008000}
\definechangesauthor[color=C0]{R1}
\definechangesauthor[color=C1]{R2}
\definechangesauthor[color=C2]{R1R2}

\usepackage{hyperref}

\usepackage[T1]{fontenc}
\usepackage{lmodern}
\usepackage{dirtytalk}
\usepackage{microtype}
\usepackage{comment}
\usepackage[margin=1in]{geometry}
\usepackage[]{algorithm2e}
\usepackage{amsmath,amssymb,amsthm,bm}
\usepackage{mathtools}
\usepackage{graphicx}
 \usepackage{algpseudocode} 
\usepackage{subcaption}
\biboptions{sort&compress}
\usepackage{epstopdf}
\usepackage{float}

\usepackage[utf8x]{inputenc}
\usepackage{xcolor}

\begin{document}

\title{Dynamic and adaptive mesh-based graph neural network framework for simulating displacement and crack fields in phase field models}
\author[auburn]{Roberto Perera}
\author[auburn]{Vinamra Agrawal\corref{cor1}}
\cortext[cor1]{Corresponding author: vinagr@auburn.edu}
\address[auburn]{Department of Aerospace Engineering, Auburn University, Auburn, AL, USA}

\begin{abstract}
    Fracture is one of the main causes of failure in engineering structures. 
    Phase field methods coupled with adaptive mesh refinement (AMR) techniques have been widely used to model crack propagation due to their ease of implementation and scalability.
    However, phase field methods can still be computationally demanding making them unfeasible for high-throughput design applications.
    Machine learning (ML) models such as Graph Neural Networks (GNNs) have shown their ability to emulate complex dynamic problems with speed-ups orders of magnitude faster compared to high-fidelity simulators.
    In this work, we present a dynamic mesh-based GNN framework for emulating phase field simulations of {single-edge} crack propagation with AMR for different crack configurations. 
    The developed framework - ADAPTive mesh-based graph neural network (\textit{ADAPT}-GNN) -  exploits the benefits of both ML methods and AMR by describing the graph representation at each time step as the refined mesh itself.
    Using \textit{ADAPT}-GNN, we predict the evolution of displacement fields and scalar damage field (or phase field) with {good} accuracy compared to the conventional phase-field fracture model.
    We also compute crack stress fields with {good} accuracy using the predicted displacements and phase field parameter. 
\end{abstract}


\maketitle

\section{Introduction}\label{Introduction}

    Since the 1970's, crack initiation and propagation in engineering materials has been a crucial area of research both experimentally and computationally.    
    Initiation, propagation, and interaction of cracks in engineering materials is one of the leading causes of catastrophic structural failure. 
    To this effort, developing computationally efficient and reliable modeling techniques is an ongoing effort in solid mechanics and material science.
    Computational modeling techniques for material failure can be described by two methodologies: (i) \replaced[id=R1,comment={Q2,Q17}]{sharp interface}{discreet} methods where cracks are treated as discontinuities in the displacement field, and (ii) \replaced[id=R1,comment=Q2]{diffused interface}{continuous} methods where the discontinuity is regularized over a length scale using a continuous surrogate field. 
    \replaced[id=R1R2,comment={Q2,Q1}]{An extensive review and comparative analysis of these methods can be found in \cite{Sedmak2018Review} and references therein. 
    One of the most widely used sharp interface methods is the extended finite element method (XFEM) \cite{SUTULA2018205,SUTULA2018225,SUTULA2018257,Belytschko1999XFEM,Belytschko2000RockXFEM,li2018review}.}
    {Some of the most widely used discreet methods include the extended finite element method (XFEM) with cohesive zone modeling.
    An extensive review and comparative analysis of these methods can be found in and references therein.}
    Among the various \replaced[id=R1,comment=Q2]{diffused interface}{continuous} methods, one of the most popular is the phase field (PF) technique \cite{FRANCFORT19981319, app9122436}.
    This approach formulates an energy functional by regularizing the crack over a length scale $\epsilon$ using a smooth scalar damage field, $\phi$.
    The evolution of $\phi$ is governed by minimizing the energy function.
    PF fracture methods have widespread use due to their ease of implementation and scalability. 
    Multiple works have recently used the PF method to successfully simulate higher-complexity crack paths in both brittle and ductile materials
    \cite{Ambati2014Review, Ambati2016Phase, ERNESTI2020112793,ZHANG2022114282,clayton2022stress}.

    \replaced[id=R1,comment=Q3]{However, simulating crack propagation using diffused interface approaches requires high mesh resolution near the crack due to the characteristic length scale of material damage being much smaller than the domain size. }
    {However, phase field fracture methods require high mesh resolution near the crack tip to appropriately resolve crack, stress, and displacement fields.}
    To circumvent the associated computational cost, PF methods are typically used in conjunction with Adaptive Mesh Refinement (AMR) approaches.
    AMR techniques reduce the number of mesh elements and cells by using different mesh resolutions; coarser mesh at regions where little to no change occurs in the problem's physics to finer mesh in regions where significant changes are present. 
    Some works where AMR has played a crucial role in speeding up phase field models include \cite{RUNNELS2021110065, AGRAWAL2021114011,Ribot_2019, Norton2001PYRAMID, MACNEICE2000330}.
    Moreover, significant efforts over the years to improve the computational efficiency and robustness of PF fracture methods have led to staggered solvers, monolithic solvers, and fast Fourier transform-based solvers, each with its pros and cons. 
    Despite this, PF models still rely on solving complex systems of equations where computational costs increase with problem complexity to achieve convergence and accurate solutions.
    \added[id=R1,comment=Q3]{For instance, PF methods require solving an additional (pseudo) time-dependent PDE to evolve the scalar damage field, thus, adding computational expenses.}
    This limits the use of PF fracture approaches in  \replaced[id=R1,comment=Q1]{large-scale}{ high-throughput design and testing} applications\added[id=R1,comment=Q1]{ such as simulating fracture due to multiple cracks in bridges, ice glaciers, and evolution of subsurface fracture networks}.
    
    
    An attractive solution to circumvent these challenges involves reduced-order modeling techniques such as Machine Learning (ML).
    An extensive body of literature exists where ML models are explored in predictions of fatigue life, non-local damage, composites and lattice structures design, material properties for single crystals, optimal mesh configurations, two-phase flow dynamics, stress and strain fields, and stress hotspots \cite{feng2020stochastic, capuano2019smart,GU201819,C8MH00653A,ZHANG2022115233,D1MH01792F,ZHANG2020112725,hanna2022residual,ren2022phycrnet,wang2022structural,mangal2018applied,mangal2019applied,he2021deep,yang2021deep,saha2021hierarchical,IM2021114030}.
    Specifically to crack propagation problems, in 2019 \cite{HUNTER201987} developed a graph-theory-inspired artificial neural network to predict connecting cracks and the estimated time-to-failure of brittle materials with multiple microcracks.
    Additionally, in \cite{HSU2020197,lew2021deep} a ML model was introduced to simulate single-edge crack growth in graphene. 
    Other recent works have also used ML techniques in fracture mechanics to predict small fatigue crack driving forces, crack growth in graphene, and stress fields in brittle materials \cite{rovinelli2018using,elapolu2022novel,wang2021stressnet}. 
    ML methods have also found success in PF modeling predictions \cite{Zhang2020High, Zhu2021Linear,SAMANIEGO2020112790,TEICHERT2019666,FENG2021113885,karniadakis2022learning}.
    For instance, in \cite{montes2021accelerating}, authors integrated Long Short-Term Memory (LSTM) and Recurrent Neural Networks (RNNs) along with PF models of spinodal decomposition to simulate the evolution of two-phase mixtures.
    The developed ML model provided orders of magnitude speedups over the high-fidelity phase field model. 
    
    More recently, Graph Neural Networks (GNNs) have shown significantly faster performance capabilities to emulate complex dynamic problems \cite{sanchez2020learning}.
    In the GNN methodology, graph theory from mathematics is integrated with neural networks allowing the representation of physics problems using nodes and communicating (or connecting) edges. 
    This intuitive method has made GNNs the prime candidates for representing various physics problems using node- and edge-based configurations.
    Various non-dynamic materials science and chemistry works have used GNNs to predict material properties, extract Perovskite synthesizability, and for inverse design of glass structures \cite{frankel2022mesh, dai2021graph, cryst12020280, choudhary2021atomistic, Stylianos2022workflow, fung2021benchmarking, rosen2022high,HEIDER2020112875, Gu2022Perovskite, wang2021inverse}.
    Additionally, GNNs have also shown success in dynamic simulation problems due to their significantly accelerated performance \cite{BLACK2022115120, park2021accurate,VLASSIS2020113299,mayr2021boundary,perera2022graph}. 
    In a recent work, \cite{pfaff2020learning} authors developed a mesh-based GNN model where the mesh points and edges from FEM simulations were directly used to develop the graph architecture.
    The developed model showed high accuracy against FEM for simulating flag dynamics, plate bending, and flow over rigid bodies while providing speeds of one to two orders of magnitude.
    While mesh-based GNNs have shown promising results, adapting this approach for PF simulations with AMR has not been attempted in the past.
    An AMR mesh-based GNN framework would require the dynamic inclusion of new nodes and edges (representative of the refined mesh) in the graph at each time-step, making this problem a unique challenge.
    
    In this work, we present a dynamic and adaptive mesh-based GNN (\textit{ADAPT}-GNN) capable of emulating PF models of {single-edge} crack propagation with simulation speed-up of up to 36x.
    We leverage the second-order PF model with AMR in \cite{GOSWAMI2020112808} to develop the training, validation, and test datasets by varying the initial cracks' lengths, positions, and orientations.
    The developed \textit{ADAPT}-GNN is able to generate new graph architectures at each time-step by dynamically adding/removing nodes and edges using the AMR approach.
    This methodology guarantees us to capture the advantages of working with smaller mesh sizes, (i.e., lower number of cells, nodes, and edges) as well as the computational speed improvement from GNNs.
    Using this technique, we predict displacements, $({u}, \nu)$, and crack field, $\phi$, for each point in the adaptive mesh at each time-step. 
    We then use the predicted displacement fields and scalar phase field parameter to compute the stress evolution in the domain. 
    We believe this framework will pave the way for obtaining significantly faster mesh-based simulations across various materials science and mechanics problems in the future.
    While this work focuses on PF fracture problems, this approach is extensible to {other} materials and mechanics problem with a {similar} PF formulation.
    
    This paper is organized as follows. 
    In Section \ref{sec:Methods}, we describe the graph network representation\replaced[id=R1,comment=Q9]{, }{ and }message-passing process\added[id=R1,comment=Q9]{, and data generation approach}.
    We \replaced[id=R1,comment=Q9]{provide}{describe the data generation in Section \ref{sect:Setup} followed by} the description of individual prediction stages in Section \ref{sect:Framework} and cross-validation in Section \ref{sec:Cross_Validation}.
    Finally, we present our results in Section \ref{sec:Results} and conclusions in Section \ref{sec:Conclusion}.
    
    \begin{figure} 
        \centering
        \includegraphics[width=0.98\linewidth]{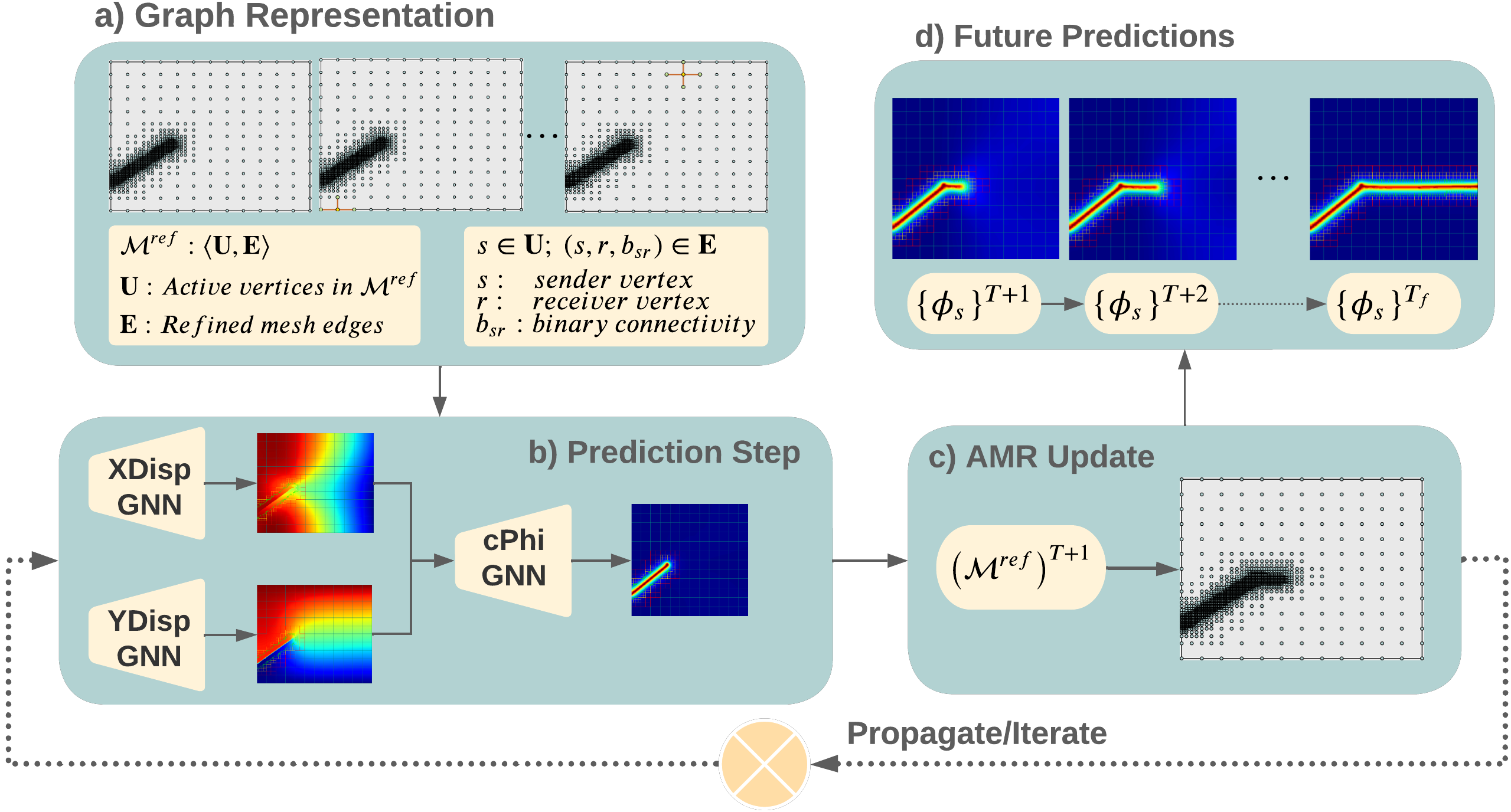}
        \caption{Flowchart of the phase-field AMR-based GNN framework's structure. a) Graph representation of nodes and edges for the refined mesh. b) Prediction step architecture including XDisp-GNN and YDisp-GNN for predicting displacements at $t+1$, and cPhi-GNN which uses the predicted displacements as input to then predict the scalar damage field at $t+1$. c) AMR update step depicting the adaptive mesh refinement step for $t+1$. d) Future predictions illustration for $T+1, T+2, ..., T_{f}$}
        \label{fig:GNN_Flowchart}
    \end{figure}

\section{Methods}\label{sec:Methods}
    
    \subsection{AMR phase field fracture model}\label{subsec:PhaseField_Model}
        We use the open source AMR PF fracture model presented in \cite{GOSWAMI2020112808} for simulating various cases of fracture mechanics in brittle materials involving single-edge notched cracks under tension (Figure \ref{fig:a_geometry}). 
        A second-order PF fracture model uses an energy functional $\mathcal{F}$ expressed as 
        {\begin{flalign}
        \mathcal{F} = \int_{\Omega} \left[ \mathcal{W} \left( \mathbf{\varepsilon} (\mathbf{u}),\phi \right) +\frac{\mathcal{G}_{c}}{2 \epsilon} \left( \phi^{2} + \epsilon^{2} | \nabla \phi |^{2} \right) \right] d\Omega
        \label{eq:energy_functional}
        \end{flalign}}
        where $\Omega$ is the domain, $\mathcal{W}$ is the strain energy density which depends on strain $\mathbf{\varepsilon}$,\added[id=R1,comment=17]{ $\mathbf{u}$ is the displacement field, $\phi$ is the scalar damage field,} and $\mathcal{G}_{c}$ is the fracture energy.
        The PF modeling framework \cite{GOSWAMI2020112808} is written in MATLAB using the isogeometric analysis (IGA) numerical tool. 
        The framework uses polynomial splines over hierarchical T-meshes (PHT-splines) local refinement method. 
        An adaptive \textit{h}-refinement scheme is then coupled with PHT-splines to dynamically rearrange the mesh resolution in locations of high gradients and singularities. 
        The framework employs the hybrid-staggered algorithm to refine mesh resolution until a convergence threshold is met. 
        This approach ensures that changes in local zones are captured at each time-step while decreasing computational costs. 
        We use the second-order phase field fracture model to generate a large dataset of two-dimensional fracture simulations with single-edge notched cracks.  
        
    \subsection{Graph network representation}\label{subsec:Graph_Representation}
\begin{figure} 
            \begin{subfigure}[c]{0.32\textwidth}
                \centering
                \begin{subfigure}[t]{1\textwidth}    
                   \centering \includegraphics[width=0.98\linewidth]{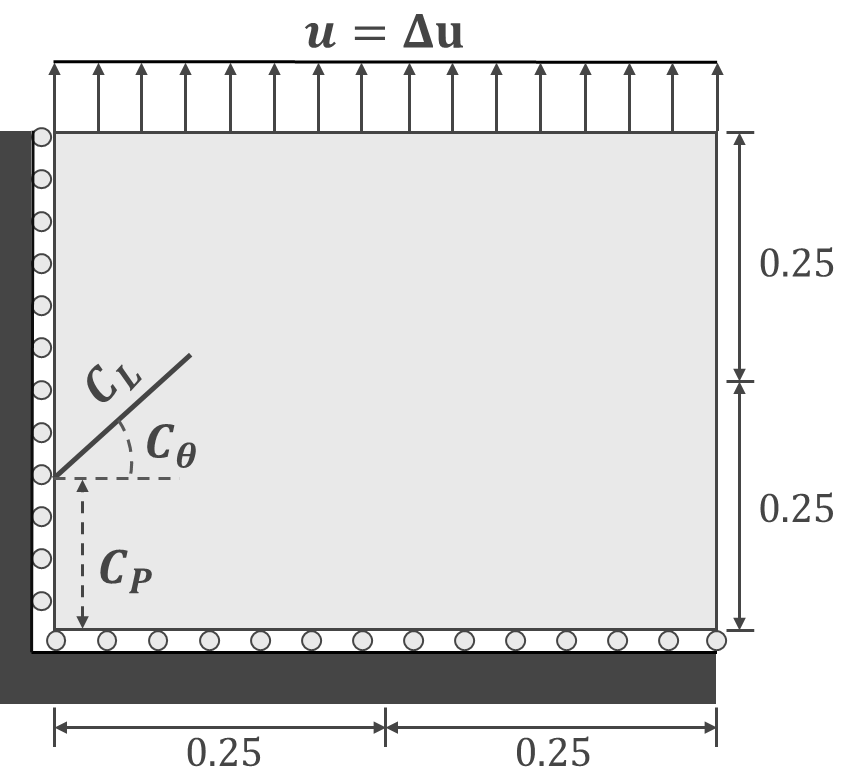}
                    \caption{Geometrical configuration}
                    \label{fig:a_geometry}
                \end{subfigure}
            \end{subfigure}
            \begin{subfigure}[c]{0.32\textwidth}
                \centering
                \begin{subfigure}[t]{1\textwidth}    
                   \centering \includegraphics[width=0.98\linewidth]{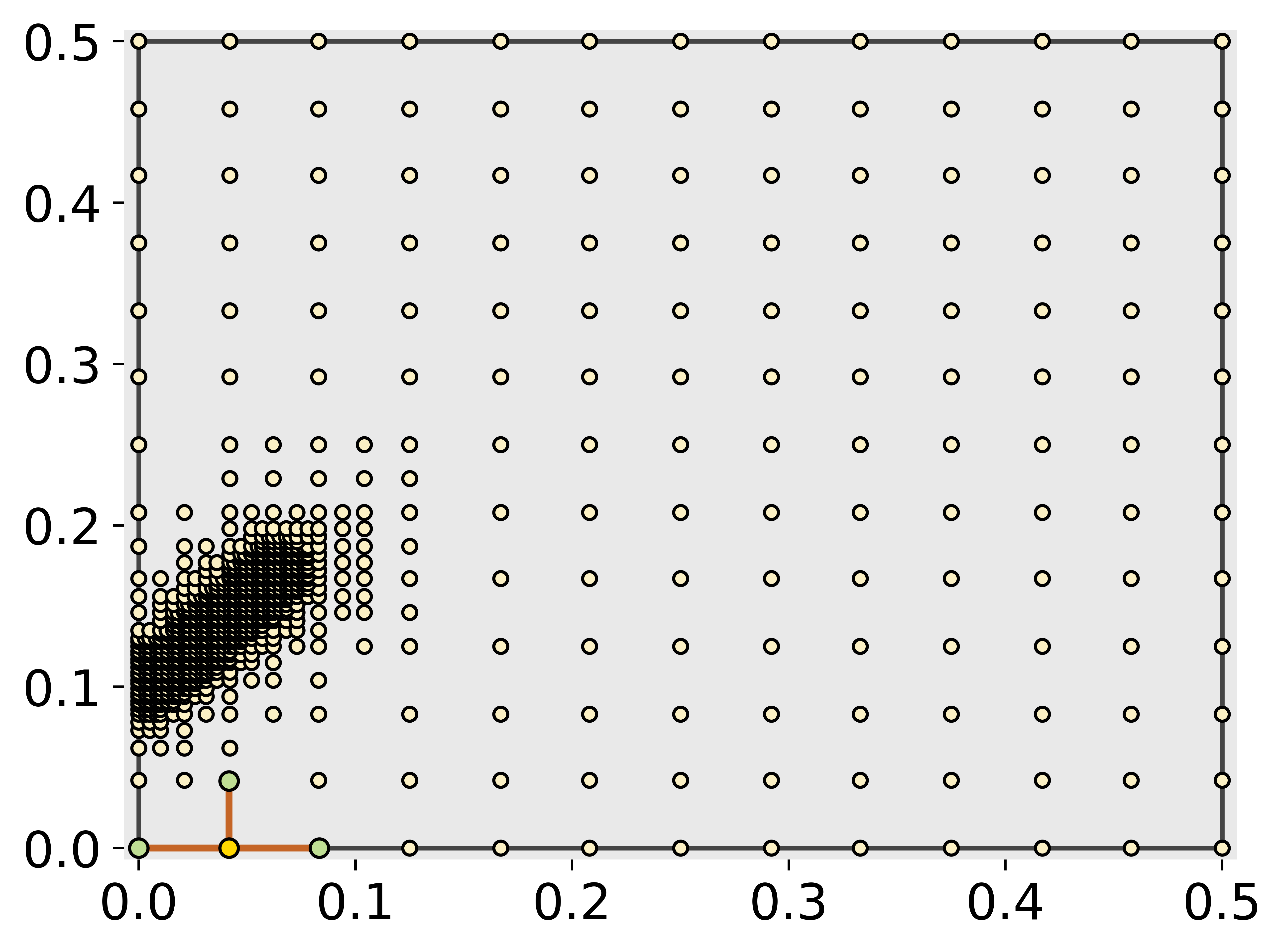}
                    \caption{Vertex No.16 Neighbors}
                    \label{fig:a_vertex}
                \end{subfigure}
            \end{subfigure}
            \begin{subfigure}[c]{0.32\textwidth}
                \centering
                \begin{subfigure}[t]{1\textwidth}    
                  \centering
                   \includegraphics[width=0.98\linewidth]{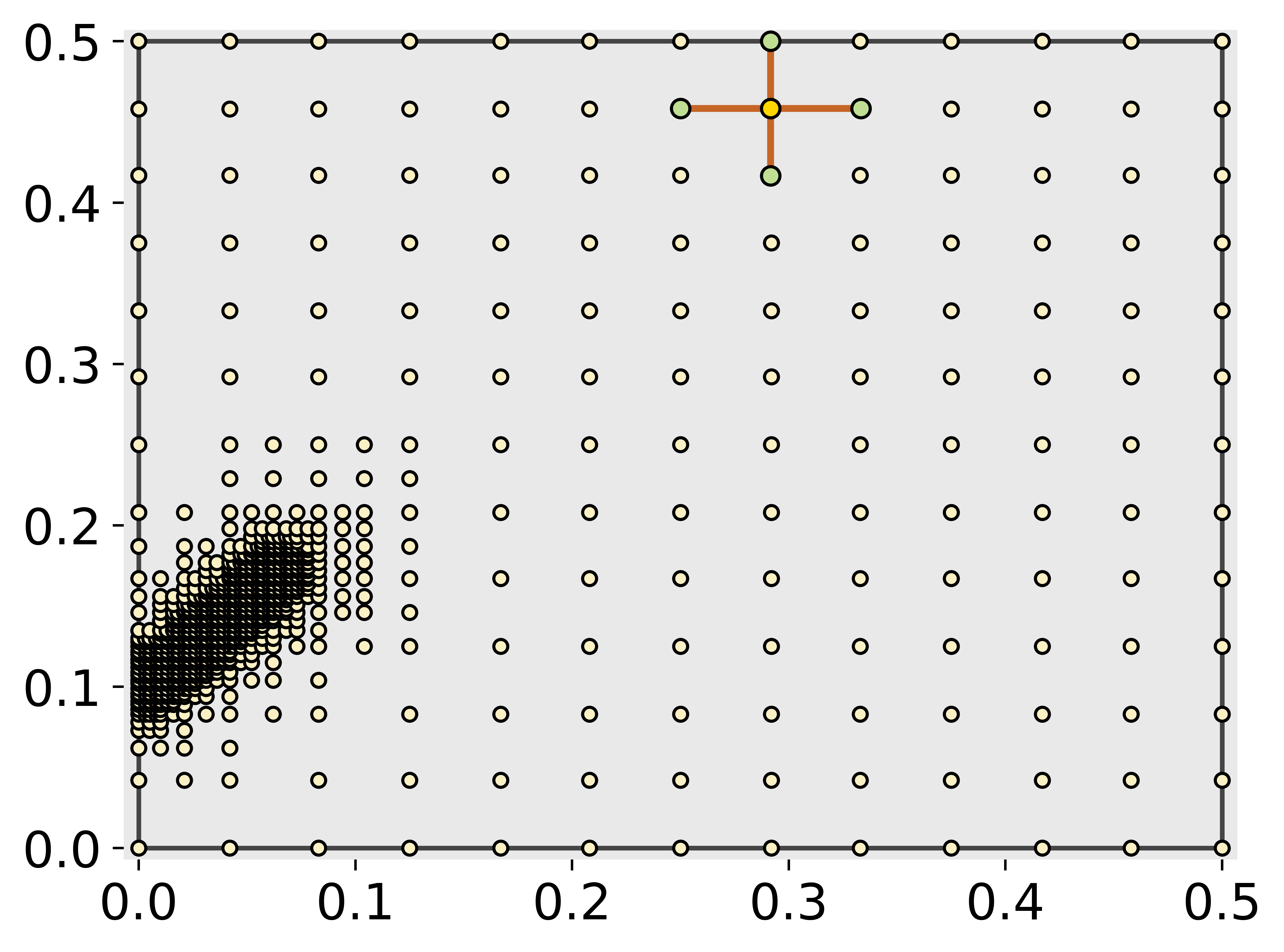}
                    \caption{Vertex No.34,080 Neighbors}
                    \label{fig:b_vertex}
                \end{subfigure}
            \end{subfigure}
            \caption{Representation of a) problem geometry and set-up of input parameters $C_{L},C_{P},C_{\theta}$, and b-c) active nodes and active edges connecting to their neighboring mesh vertices.}
            \label{fig:neighbors_vertex_edge}
        \end{figure}

        A key feature of the developed GNN, is its ability to dynamically reconfigure the graph's architecture at each instance in time according to the resulting refined mesh. 
        This approach grants the developed GNN with the computational efficiency of the AMR approach.
        In \cite{GOSWAMI2020112808}, the \textit{h}-refinement approach implemented  defines the two-dimensional geometrical representation of the mesh, $\mathcal{M}$,  as $\mathcal{U}^{i} = \{ \xi_{1}^{i}, \xi_{2}^{i}, \dots, \xi_{n_{i}+1}^{i} \}$, 
        where $\mathcal{U}^{i}$ contains the set of active vertices, $\xi$, and $n_{i}$ the number of active elements in each parametric direction (i.e., $i\in \{ 1,2\}$ for a two-dimensional case).
        We adopt and simplify this notation for the design of the GNN and describe the instantaneous refined graphs as {$\mathcal{M}^{ref} : \langle {\mathcal{\mathbf{U}}},\mathbf{E} \rangle$}, where $\mathcal{\mathbf{U}}$ indicates the active vertices in $\mathcal{M}^{ref}$, and $\mathbf{E}$ describes the resulting edges in $\mathcal{M}^{ref}$.
        Additionally, $\mathbf{E}$, includes edges connecting each active vertex in the refined mesh, $\xi_{s} \in \mathcal{\mathbf{U}}$\deleted[id=R1,comment=Q17]{,  (i.e., for any positive integer $s: \left\{1,2,\ldots, N \right\}$, where $N$ corresponds to the total number active vertices in $\mathcal{M}^{ref}$)} to the adjacent active vertices (as shown in Figures \ref{fig:a_vertex} - \ref{fig:b_vertex}).
        \added[id=R1,comment=Q17]{We note that for any positive integer $s: \left\{1,2,\ldots, N \right\}$, where $N$ corresponds to the total number active vertices in $\mathcal{M}^{ref}$.}

        The vertices at each instance of time are described by their spatial positions $\hat{\mathcal{P}}_{s}$, their adjacent active mesh nodes (neighboring vertices) $\hat{\mathcal{A}}_{s}$, their displacement values $\hat{\mathcal{D}}_{s}$\added[id=R1,comment=17]{ (x- and y-displacement fields, $u_{s}$ and $\nu_{s}$)}, and their energy- and physics-informed parameters $\hat{\Pi}_{s}$ \replaced[id=R1,comment=Q17]{. 
        In essence, $\hat{\Pi}_{s}$ includes}{(i.e.,} scalar damage field variable values, ${\phi}_{s}$, von Mises stress values, ${\sigma}_{s}$, active/inactive binary values, ${\mathcal{I}}_{s} \in \{0,1\}$ - $0$s are inactive mesh nodes and $1$s are active mesh nodes - the laplacian of the scalar damage field, $\Delta \phi_{s}$, and the applied displacement loading ${u}_{0_{s}}$\deleted[id=R1,comment=Q17]{)}.
        {\begin{flalign}
        && \hat{\mathcal{P}}_{s} = \{ \left(x_{s}, y_{s} \right)\} && \{s \in \mathcal{\mathbf{U}}\}; \ \{\mathcal{\mathbf{U}} \in \mathcal{{M}}^{ref}\}, \nonumber\\
        && \hat{\mathcal{A}}_{s} = \{ \mathcal{A}_{s} \} && \{s \in \mathcal{\mathbf{U}}\}; \ \{\mathcal{\mathbf{U}} \in \mathcal{{M}}^{ref}\}, \nonumber\\
        && \hat{\mathcal{D}}_{s} = \{ \left(u_{s}, \nu_{s} \right) \} && \{s \in \mathcal{\mathbf{U}}\}; \ \{\mathcal{\mathbf{U}} \in \mathcal{{M}}^{ref}\},  \nonumber\\
        && \hat{\Pi}_{s} = \{ \left( {\phi}_{s}, {\sigma}_{s},{\mathcal{I}}_{s}, \Delta \phi_{s}, {u}_{0_{s}}  \right)\} && \{s \in \mathcal{\mathbf{U}}\}; \ 
        \{\mathcal{\mathbf{U}} \in \mathcal{{M}}^{ref}\},  \nonumber\\
        && \{\xi_{s}\} = \{(\hat{\mathcal{P}}_{s}, \hat{\mathcal{A}}_{s}, \hat{\mathcal{D}}_{s}, \hat{\Pi}_{s} ) \} && \{s \in \mathcal{\mathbf{U}}\}; \ \{\mathcal{\mathbf{U}} \in \mathcal{{M}}^{ref}\}. \label{eq:vertex_states}
        \end{flalign}}
        Additionally, the instantaneous edges in $\mathcal{M}^{ref}$ are indexed using a binary value 
        specifying whether the current (or ``sender'') vertex $\xi_{s}$, and all other active (or ``receiver'') vertices $\xi_{r}$ in $\mathcal{\mathbf{E}}$, form part of the same neighboring array $\hat{\mathcal{A}}_{s}$. 
        This binary value is defined as $(s, r, b_{sr}) \in \mathbf{E}$, where $s$ and $r$ denote the ``sender'' and ``receiver'' vertices, respectively (i.e., for any positive integer $s: \{1, 2, \dots, \mathcal{\mathbf{U}} \}$ and $r: \{1, 2, \dots, \mathcal{\mathbf{U}} \}$), and $b \in \{0,1\}$. 
        In the case where $s=r$, we set $b_{sr} = 1$.
        We show an example of the graph representation in Figure \ref{fig:neighbors_vertex_edge}, where the active vertices (green node) and their edges (orange lines) are shown in Figures \ref{fig:a_vertex} - \ref{fig:b_vertex} for vertex no. \replaced[id=R1,comment=Q17]{16}{0} and vertex no. 34,080, respectively.   
        Using this representation, we can define the indices of a series of neighbors pertaining to each active mesh node as shown in equation (\ref{eq:discreet_neighbors}). 
        \begin{flalign}
        && \hat{\beta}_{sr}  = \{ \left(\xi_{s}, \xi_{r}, b_{sr}\right)\} && \{(s,r,b_{sr}) \in \mathbf{E}\}; \ \{\mathcal{\mathbf{E}} \in \mathcal{{M}}^{ref}\}. \label{eq:discreet_neighbors}
        \end{flalign}
        For the case where $b_{sr} = 1$, we then define five initial edge features using the distances in the $x$- and $y$-directions ($\delta \mathcal{{X}}_{sr}=\left(x_{r} - x_{s} \right)$, and $\delta \mathcal{{Y}}_{sr}=\left(y_{r} - y_{s} \right)$), the magnitude of the distance (${\mathcal{L}}_{sr}=\left( \sqrt{\delta \mathcal{X}_{sr}^{2} + \delta \mathcal{Y}_{sr}^{2} } \right)$), the difference in the scalar damage field ($\delta {\phi}_{sr}=\left(\phi_{r} - \phi_{s} \right)$), and the difference in the stress field ($\delta {\sigma}_{sr}=\left(\sigma_{r} - \sigma_{s} \right)$), between the sender and receiver nodes as shown in equation (\ref{eq:edges_states_1}).
        {\begin{flalign}
        && \delta \mathcal{\hat{P}}_{sr} = \Big\{ \left( \delta \mathcal{{X}}_{sr}, \delta \mathcal{{Y}}_{sr}, \mathcal{{L}}_{sr} \right) \Big\} && \{(s,r,b_{sr}) \in \mathbf{E}\}; \ \{\mathcal{\mathbf{E}} \in \mathcal{{M}}^{ref}\}, \nonumber\\
        && \delta {\hat{\Pi}}_{sr} = \Big\{ \left( \delta {{\phi}}_{sr}, \delta {{\sigma}}_{sr} \right) \Big\} && \{(s,r,b_{sr}) \in \mathbf{E}\}; \ \{\mathcal{\mathbf{E}} \in \mathcal{{M}}^{ref}\}.
        \label{eq:edges_states_1}
        \end{flalign}}
        
        To incorporate physics based information into our GNN framework, we leverage the energy functional shown in equation (\ref{eq:energy_functional}) which involves gradients of $\phi$, and a term for the strain energy density which depends on displacements and stresses.
        Therefore, we define three additional gradient edge features for the the damage field, $\nabla  \hat{\phi}_{sr}$, the $x$- and $y$-displacements, $\nabla \hat{\mathcal{D}}_{sr}$, and the stress $\nabla \hat{\sigma}_{sr}$.
        Lastly, the remaining and resultant edges features of the graph become
        {\begin{flalign}
        && \nabla  \hat{\phi}_{sr} = \Bigg\{ \left(\frac{\delta \phi_{sr}}{\delta \mathcal{X}} + \frac{\delta \phi_{sr}}{\delta \mathcal{Y}} \right)\Bigg\} && \{(s,r,b_{sr}) \in \mathbf{E}\}; \{\mathcal{\mathbf{E}} \in \mathcal{{M}}^{ref}\} \nonumber\\ 
        && \nabla  \hat{\mathcal{D}}_{sr} = \Bigg\{ \left( \frac{\delta u_{sr}}{\delta \mathcal{X}} + \frac{\delta \nu_{sr}}{\delta \mathcal{X}} \right) ,\left(  \frac{\delta u_{sr}}{\delta \mathcal{Y}} + \frac{\delta \nu_{sr}}{\delta \mathcal{Y}}  \right) \Bigg\} && \{(s,r,b_{sr}) \in \mathbf{E}\}; \{\mathcal{\mathbf{E}} \in \mathcal{{M}}^{ref}\} \nonumber\\
        && \nabla  \hat{\sigma}_{sr} = \Bigg\{ \left(\frac{\delta \sigma_{sr}}{\delta \mathcal{X}} + \frac{\delta \sigma_{sr}}{\delta \mathcal{Y}} \right) \Bigg\} && \{(s,r,b_{sr}) \in \mathbf{E}\}; \{\mathcal{\mathbf{E}} \in \mathcal{{M}}^{ref}\} \nonumber\\
        && \{e_{sr}\} = \Big\{  \left(  \hat{\beta_{sr}}, \delta \mathcal{\hat{P}}_{sr}, \delta {\hat{\Pi}}_{sr}, \nabla {\hat{\phi}}_{sr}, \nabla \mathcal{\hat{D}}_{sr}, \nabla {\hat{\sigma}}_{sr}  \right)  \Big\} && \{(s,r,b_{sr}) \in \mathbf{E}\}; \{\mathcal{\mathbf{E}} \in \mathcal{{M}}^{ref}\}.
        \label{eq:edges_states_2}
        \end{flalign}}
        
    \subsection{Spatial Message-Passing Process}\label{subsect:message_passing} 
        In GNN formulation the spatial message-passing process plays a crucial role in order to learn the graphs' relationships (vertices, edges, and neighbors) in the latent space.
        For this purpose, the developed GNN framework first involves the Graph Isomorphism Network with Edge Features (GINE) \cite{Hu2019GINEConv} used as the message-passing network for each GNN.
        \added[id=R1,comment={Q8,Q11}]{The GINE message-passing network involves node and edge embedding operation steps to map the input nodes and edge features to arrays, which are then concatenated and input to an MLP with ReLU activation function.
        The input to GINE involves the current time-step's node features, $\xi_{s}$, and the current time-step's edge features, $e_{sr}$, defined in equations (2) and (5), respectively.
        In essence, the input to GINE involves nodal information and edge information at the current time-step $t$.
        The nodal information includes the active/inactive nodes, displacement fields, scalar damage field, loading and von Mises tress. 
        Additionally, the edge information includes the active edges and their lengths, change in scalar damage field and von Mises stress, and gradients of the scalar damage field, von Mises stress, and displacement fields.}
        The output from the message-passing network is defined as $\{\xi_{s}^{'},e_{sr}^{'}\}$, where $\xi_{s}^{'}$ describes the vertices and their attributes' information in the latent space, and $e_{sr}^{'}$ describes the edges and their attributes in the latent space for time-step $t$.
        We emphasize that for each GNN model in the framework \added[id=R1,comment=Q11]{(\textit{XDisp}-GNN, \textit{YDisp}-GNN, and \textit{cPhi}-GNN)}, the message-passing network may be tuned to achieve high accuracy and prevent loss of information of the graphs' relations \cite{klicpera2020directional,zhang2020dynamic,gilmer2017neural}. 
        To this end, we tuned each GNN's message-passing network with respect to the number of message-passing steps (i.e., number of iterations the vertices and edges pass through the encoder networks), and the number of nodes in the hidden layers.
        The procedures for tuning the \replaced[id=R1,comment=Q8]{GINE}{MLP} encoders and their results is described in Section \ref{sec:Cross_Validation}.

    
   \subsection{Training-set and Validation-set}\label{subsect:dataset}
        As mentioned in Section \ref{subsec:PhaseField_Model}, we used the second-order phase field fracture model from \cite{GOSWAMI2020112808} to generate the training set, validation set, and the test set.
        \added[id=R2,comment=Q2]{We note that \cite{GOSWAMI2020112808} includes the fourth-order phase field model, which would provide more accurate results. However, we chose the second-order method for computationally efficiency and proof of concept.}
        The problem set-up involved a domain of $0.5$ m by $0.5$ m with a maximum number of mesh nodes of $193$ by $193$ involving a single edge crack under tensile displacement loading. 
        The material properties were modeled using isotropic and homogeneous conditions with Young's Modulus, $E = 210$ N/mm$^{2}$, \deleted[id=R1,comment=Q9]{and }Poisson's ratio, $\nu = 0.3$\added[id=R1,comment=Q9]{, critical energy release rate, $G_{1c}=2.7$, and length scale parameter, $l_{0} = 0.0125$ m}.
        Further, our analysis did not account for dynamic effects such as crack-tip bifurcation. 
        Similarly to the example discussed in \cite{GOSWAMI2020112808}, we fixed the bottom edge of the domain and applied constant displacement increments in the positive y-direction (tensile load perpendicular to top edge) of $\Delta u = 1 \times 10^{-4}$ mm in the initial 45 displacement steps, and $\Delta u = 1 \times 10^{-6}$ mm in the remaining steps to avoid dynamic effects. 
        Using this, we generated a dataset of $1245$ unique simulations by varying the initial crack length, edge position, and crack angle.
        The crack lengths, edge positions, and crack angles were varied using $C_{L} : \{ 0.05, 0.10, \dots, 0.45 \}$ m, $C_{P} : \{ 0.1, 0.15, \dots, 0.4 \}$ m, and $C_{\theta} : \{ -65^{o}, -60^{o}, \dots, 65^{o} \}$, respectively.
        \added[id=R1,comment=17]{Figure \ref{fig:a_geometry} depicts the problem set-up and configurations of $C_{L}$, $C_{P}$, and $C_{\theta}$.}
        Additionally, we removed cases resulting in crack-tip locations beyond the domain's bounds.
        We note that each simulation contains 100 to 450 time-steps, and each input to the GNN framework involved a single time-frame, thus, resulting in a dataset size of $124,500$ to $560,250$. 
        
        Next, to perform a systematic error analysis of the test set, we \replaced[id=R1,comment=Q10]{randomly selected}{first picked} 30 simulations for the test set with even number of large \added[id=R1,comment=17]{($C_{L}\geq0.25$ m)} versus small \added[id=R1,comment=17]{($C_{L}<0.25$ m)} initial crack lengths, top \added[id=R1,comment=17]{($C_{P}\geq0.5$ m)} versus bottom \added[id=R1,comment=17]{($C_{P}<0.5$ m)} initial edge positions, and positive \added[id=R1,comment=17]{($C_{\theta}\geq0^{o}$)} versus negative \added[id=R1,comment=17]{($C_{\theta}<0^{o}$)} crack angles.
        We then split the remaining simulations as $1100$ for the training-set, and $115$ for the validation-set.
        {The} training-set {was separated} into shuffled batches of size $32$ and kept the validation set in sequential order for batch size of $1$.
        {Lastly, each model (\textit{XDisp}-GNN, \textit{YDisp}-GNN, and \textit{cPhi}-GNN) was trained for a total of 20 epochs.}

\section{Adaptive Mesh-based Graph Neural Network}\label{sect:Framework}

    As shown in Figure \ref{fig:GNN_Flowchart}, the \textit{ADAPT}-GNN framework
    involves three initial GNNs: (i) \textit{XDisp}-GNN, (ii) \textit{YDisp}-GNN, and (ii) \textit{cPhi}-GNN.
    At each time-step, the framework first predicts the displacement fields \added[id=R1,comment=Q11]{at time $t+1$ given the node and edge features (defined in equations (2) and (5)) at the time $t$}\deleted[id=R1,comment=Q11]{, followed by prediction of $\phi$}.
    \added[id=R1,comment=Q11]{The framework then predicts the scalar damage field $\phi$ at time $t+1$, given the node and edge features (defined in equations (2) and (5)) from time $t$, along with the predicted displacement fields at time $t+1$.}
    One of the key features of \textit{ADAPT}-GNN is the integration of AMR.
    The framework is able to increase the size of the graph dynamically at each time-step by using the refined mesh as the instantaneous graph representation itself.
    This approach leverages both the order reduction and GPU-usage from ML methods, and the computational efficiency from the AMR approach, thus, reducing computational requirements while increases simulation speed.
    We present the implementation of the ``Prediction Step'' (\textit{XDisp}-GNN, \textit{YDisp}-GNN, and \textit{cPhi}-GNN), and the ``AMR Update'' shown in Figure \ref{fig:GNN_Flowchart} in detail in the following sections.
 
    \subsection{\textit{XDisp}-GNN and \textit{YDisp}-GNN}\label{subsect:XDisp-GNN_YDisp-GNN}
        
        We implemented the {\textit{XDisp}-GNN} and \textit{YDisp}-GNN for predicting the displacement fields in the x and y directions, respectively, for each active node in the refined mesh.
        For this step, we first generated the input graph representation following the procedures described in \replaced[id=R1,comment=Q11]{Section}{Sections} \ref{subsec:Graph_Representation}\deleted[id=R1,comment=Q11]{and \ref{subsect:message_passing}}.
        \added[id=R1,comment=Q11]{We note that both \textit{XDisp}-GNN and \textit{YDisp}-GNN first involve a GINE message-passing model to transfer the node and edge information to the latent space (node and edge embeddings), as described in Section \ref{subsect:message_passing}.}
        To propagate the displacements to the future time-steps, we used the outputs from the message-passing networks of \textit{XDisp}-GNN and \textit{YDisp}-GNN as the input to two Attention Temporal Graph Convolutional Networks (ATGCN) \cite{Zhu2020A3TGCN}.
        \added[id=R1,comment={Q8,Q11}]{In essence, the ATGCN model is designed to capture both local and global spatiotemporal variation trends in states. 
        The ATGCN model first involves a Temporal Graph Convolutional Network (T-GCN) \cite{Zhao2020TGCN} comprised of Graph Convolutional Networks (GCNs) and Gated Recurrent Units (GRUs) in sequence to capture the local variation trends. 
        The ATGCN model then introduces a tanh activation function operation (attention model) to re-weight the influence of historical states in order to capture the global variation trends.
        }
        We chose the ATGCN model due to its integration of gated recurrent units into graph convolutional networks for learning time changes while maintaining the graphs' spatial relations\added[id=R1,comment={Q8,Q11}]{, as well as for its integration of the attention model to capture both local and global spatiotemporal variations}. 
        The resulting input graph for a given time-step, $t$, and the resulting predicted displacements at a future time-step, $t+1$, from the ATGCNs are described as
        \begin{flalign}
            &&\left(\hat{u}_{s},\hat{\nu}_{s}\right)^{t+1}  \longleftarrow ATGCN\left[\{\xi_{s}^{'},e_{sr}^{'}\}^{t}\right] &&  \{s \in \mathbf{V}\}; \{(s,r,b_{sr}) \in \mathbf{E}\}; \{\mathcal{\mathbf{E}} \in \mathcal{{M}}^{ref}\}.
            \label{eq:XDisp-GNN_YDisp-GNN}
        \end{flalign}
        \begin{figure}
            \begin{subfigure}[t]{1\textwidth}
                \centering
                \begin{subfigure}[t]{0.49\textwidth}
                    \centering
                    \includegraphics[width=\linewidth]{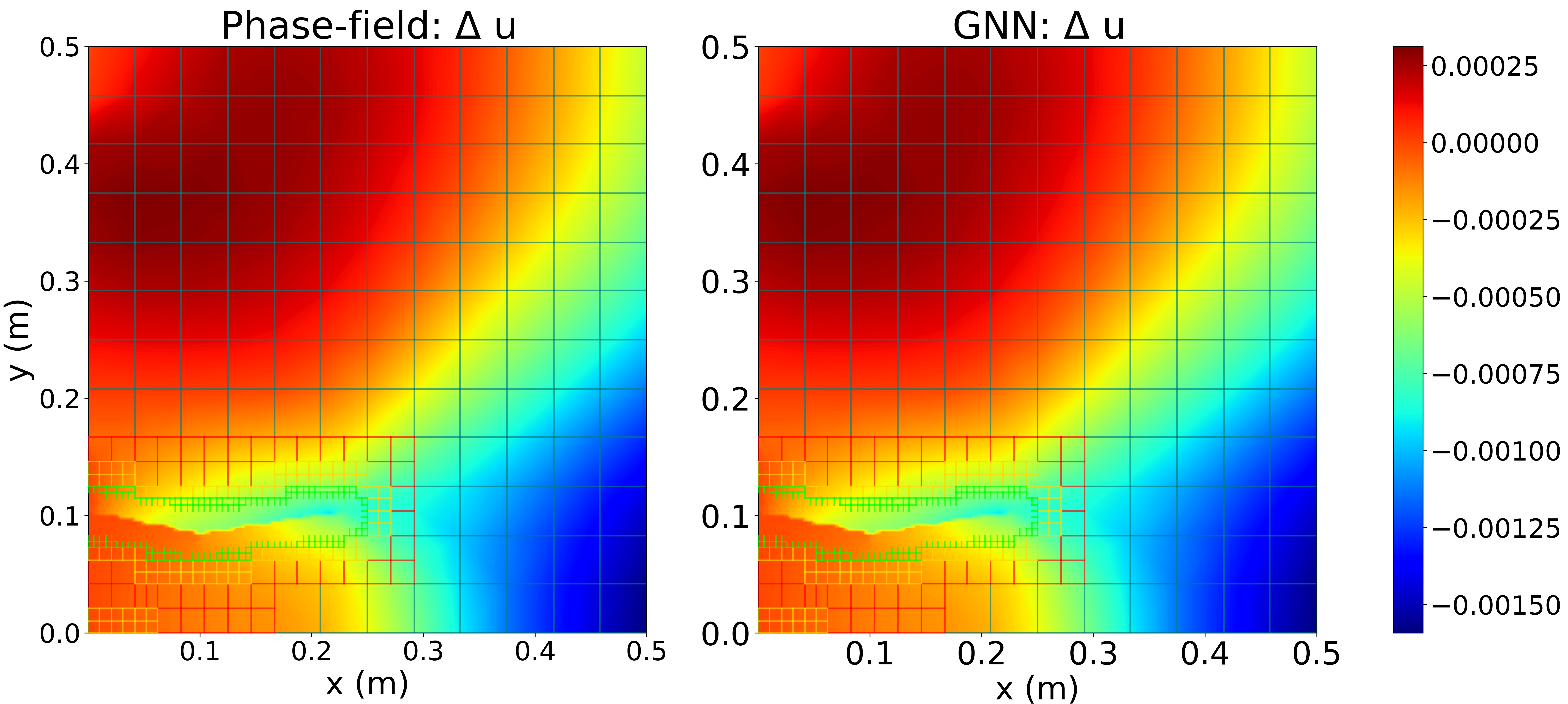}
                    \label{subfig:XDisp_GNN_time_165}
                \end{subfigure}
                \begin{subfigure}[t]{0.49\textwidth}
                    \centering
                    \includegraphics[width=\linewidth]{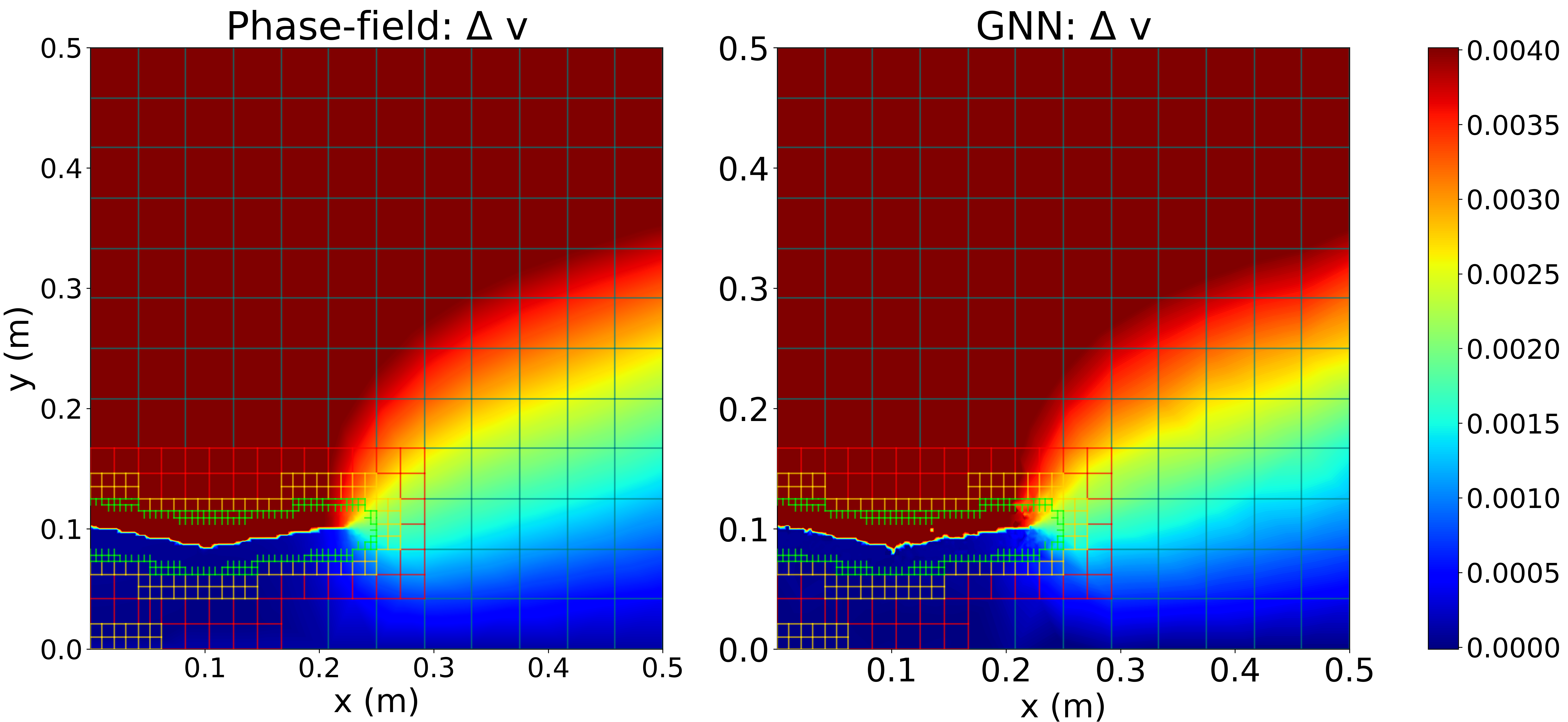}
                    \label{subfig:YDisp_GNN_time_165}
                \end{subfigure}
            \end{subfigure}
            \begin{subfigure}[b]{1\textwidth}
                \centering
                \begin{subfigure}[b]{0.49\textwidth}
                    \centering
                    \includegraphics[width=\linewidth]{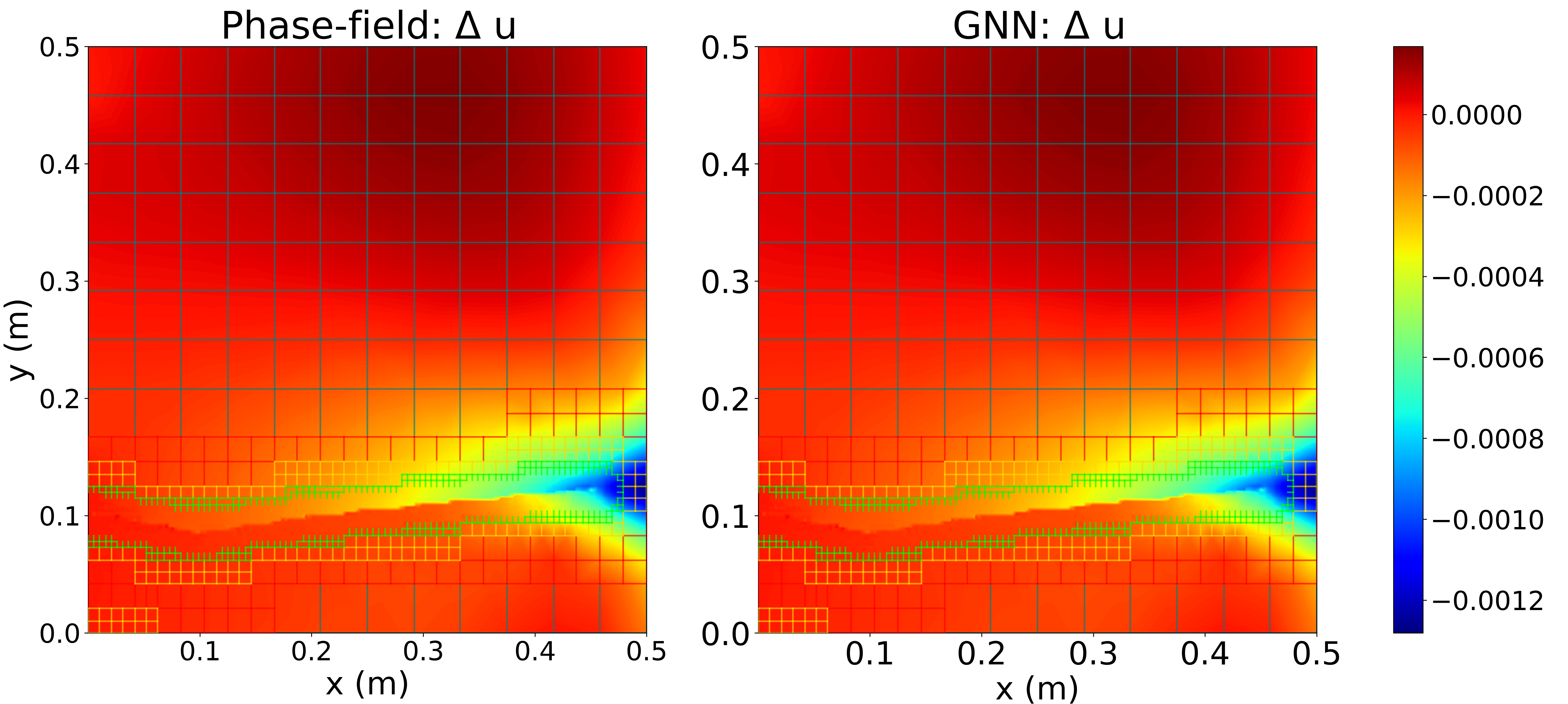}
                    \caption{PF versus \textit{XDisp}-GNN}
                    \label{subfig:XDisp_GNN_time_320}
                \end{subfigure}
                \begin{subfigure}[b]{0.49\textwidth}
                    \centering
                    \includegraphics[width=\linewidth]{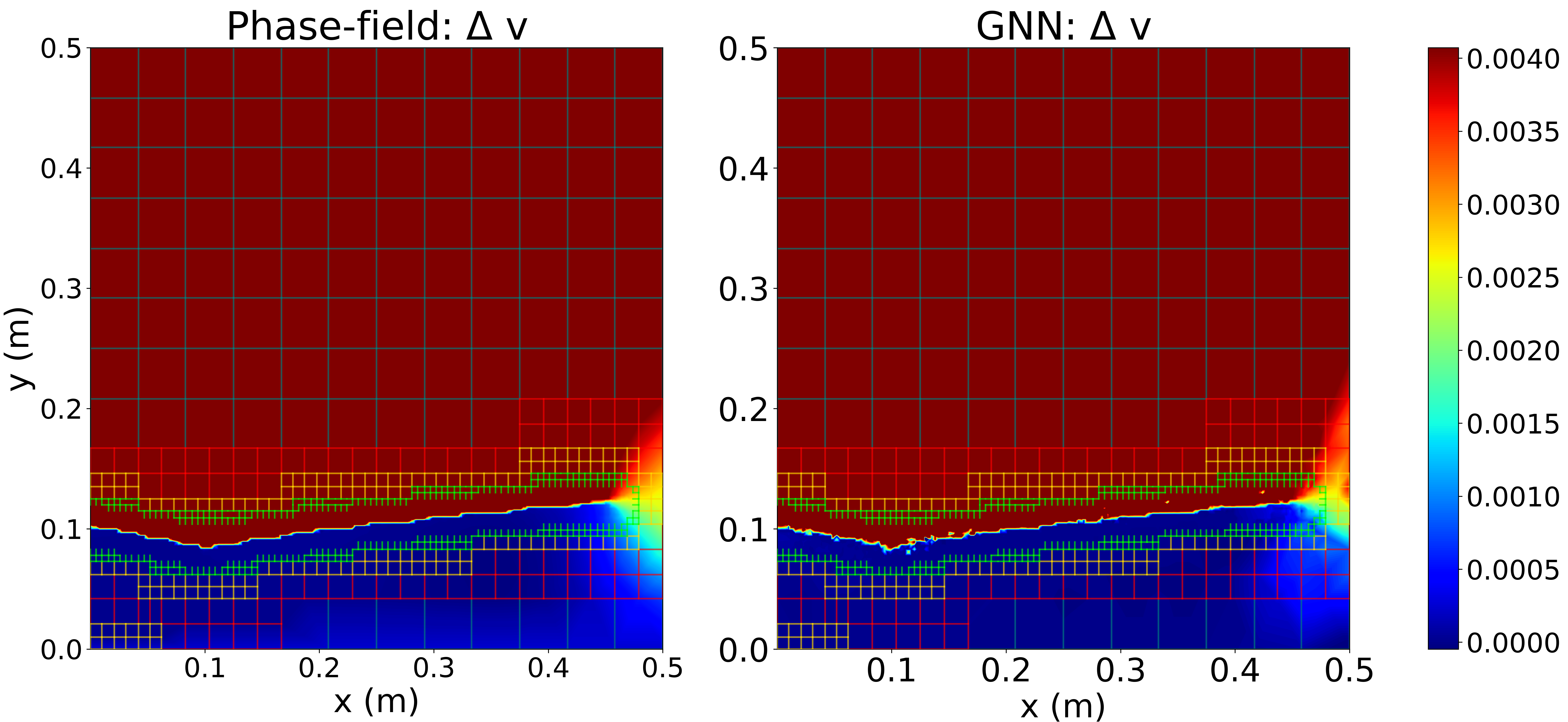}
                    \caption{PF vs. \textit{YDisp}-GNN}
                    \label{subfig:YDisp_GNN_time_320}
                \end{subfigure}
            \end{subfigure}
            \caption{PF fracture model versus a) \textit{XDisp}-GNN prediction and b) \textit{YDisp}-GNN prediction for a simulation from the test set involving a small crack ($C_{L}=0.1$ m) with negative angle and located at $C_{P}=0.1$ m.}
            \label{fig:PhaseField_GNN_XDisp_YDisp}
        \end{figure}

        \added[id=R1,comment=Q11]{We note that equation (6) depicts a single ATGCN for both \textit{XDisp}-GNN and \textit{YDisp}-GNN for simplicty (each GNN involves a GINE message passing model followed by an ATGCN model).
        Additionally, we used a SmoothL1Loss function to train \textit{XDisp}-GNN since $\Delta u$ values vary from negative to positive, and an MSELoss function to train \textit{YDisp}-GNN since $\Delta v$ values varied $\geq 0$.
        We used the Adam optimizer \cite{kingma2017adam} for both GNNs.}
        In essence, the ATGCNs take the vertices' features and the edges' features information in the latent space at the current time-step, $t$, as input to predict the real-space x- and y-displacements at the next time-step, $t+1$. 
        Figure \ref{fig:PhaseField_GNN_XDisp_YDisp} compares the PF fracture model versus predicted x- and y-displacements.

    \subsection{\textit{cPhi}-GNN}\label{subsect:cPhi-GNN}
    
        Next, we concatenated the predicted x-displacements and y-displacements and used them as input to the {\textit{cPhi}-GNN} (Figure \ref{fig:GNN_Flowchart}).
        As mentioned in previous sections, the purpose of \textit{cPhi}-GNN is to predict the scalar damage field (or crack field), $\phi_{s}$, at the future time-steps.  
        Similarly to the \textit{XDisp}-GNN and \textit{YDisp}-GNN models, {\textit{cPhi}-GNN} uses the vertices and edges features in the latent space outputted from the message-passing network as the first part of its input.
        However, we then concatenate the predicted x- and y-displacements at the future time-steps as the second part of its input.
        Therefore, the GNN model used for predicting $\phi$ is a modified ATGCN designed to include two additional vertex and edge features for the previously predicted displacements \added[id=R1,comment=Q11]{(XDisp- and YDisp-GNN concatenated output from equation (\ref{eq:XDisp-GNN_YDisp-GNN}))
        Similar to \textit{XDisp}-GNN and \textit{YDisp}-GNN, \textit{cPhi}-GNN was trained using the Adam optimizer \cite{kingma2017adam}.}.
        The \textit{cPhi}-GNN is then described as
        \begin{flalign}
            &&\left(\hat{\phi}_{s}\right)^{t+1}  \longleftarrow ATGCN\left[\{\xi_{s}^{'},e_{sr}^{'}\}^{t}, \left(\hat{u}_{s},\hat{\nu}_{s}\right)^{t+1} \right] &&  \{s \in \mathbf{V}\}; \{(s,r,b_{sr}) \in \mathbf{E}\}; \{\mathcal{\mathbf{E}} \in \mathcal{{M}}^{ref}\}.
            \label{eq:cPhi-GNN}
        \end{flalign}
        \begin{figure}
            \begin{subfigure}[t]{1\textwidth}
                \centering
                \begin{subfigure}[t]{0.49\textwidth}
                    \centering
                    \includegraphics[width=\linewidth]{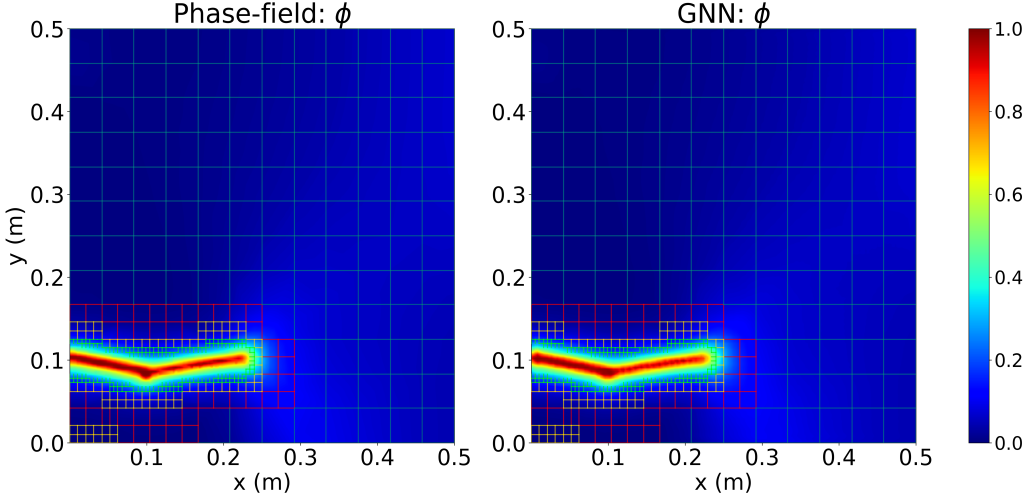}
                    \label{subfig:cPhi_GNN_time_165}
                \end{subfigure}
                \begin{subfigure}[t]{0.49\textwidth}
                    \centering
                    \includegraphics[width=\linewidth]{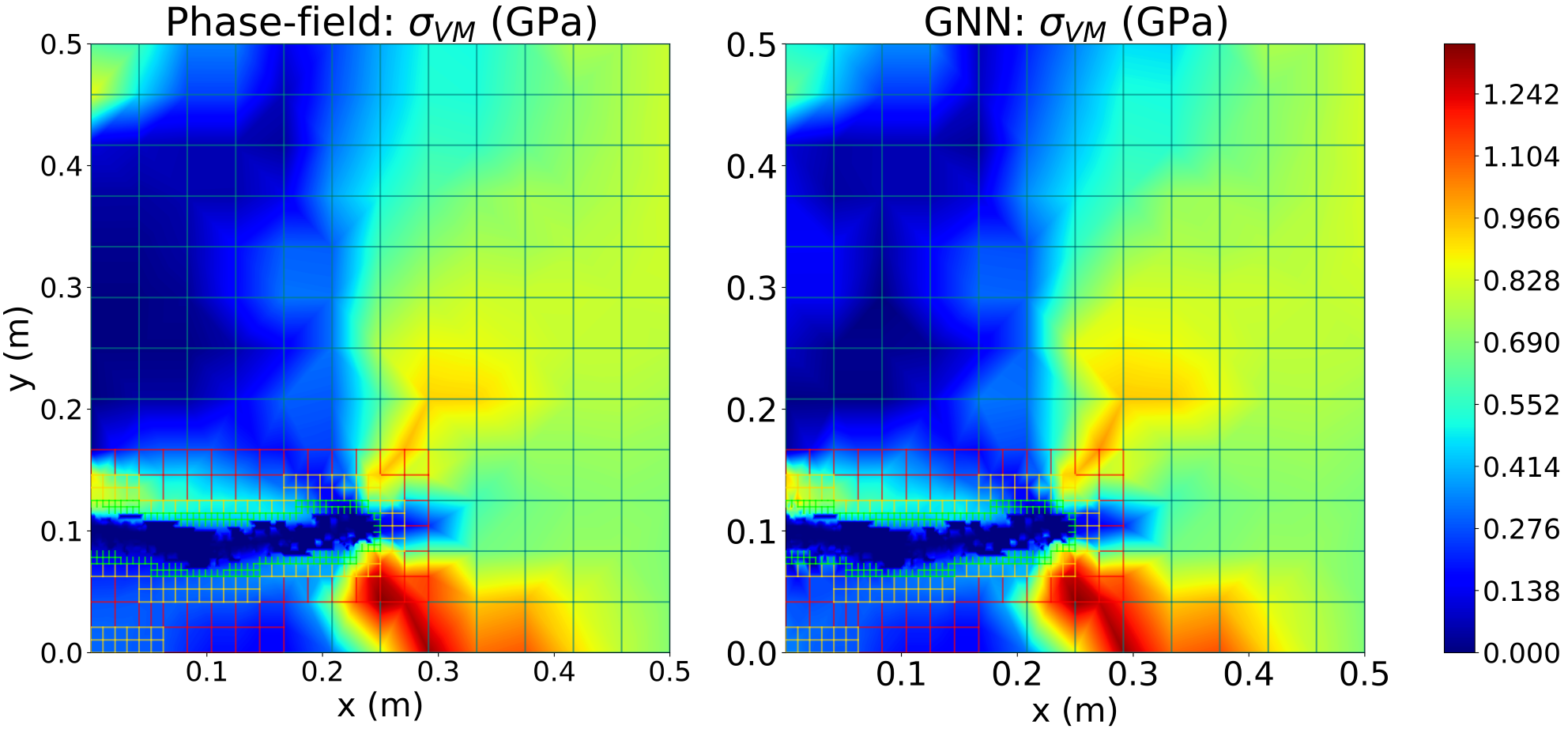}
                    \label{subfig:SVM_GNN_time_165}
                \end{subfigure}
            \end{subfigure}
            \begin{subfigure}[b]{1\textwidth}
                \centering
                \begin{subfigure}[b]{0.49\textwidth}
                    \centering
                    \includegraphics[width=\linewidth]{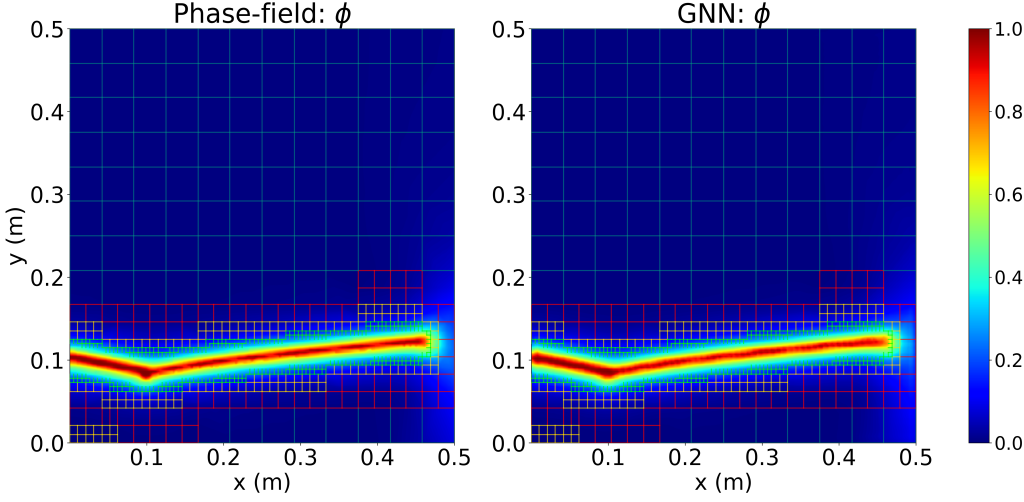}
                    \caption{PF vs. \textit{cPhi}-GNN}
                    \label{subfig:cPhi_GNN_time_320}
                \end{subfigure}
                \begin{subfigure}[b]{0.49\textwidth}
                    \centering
                    \includegraphics[width=\linewidth]{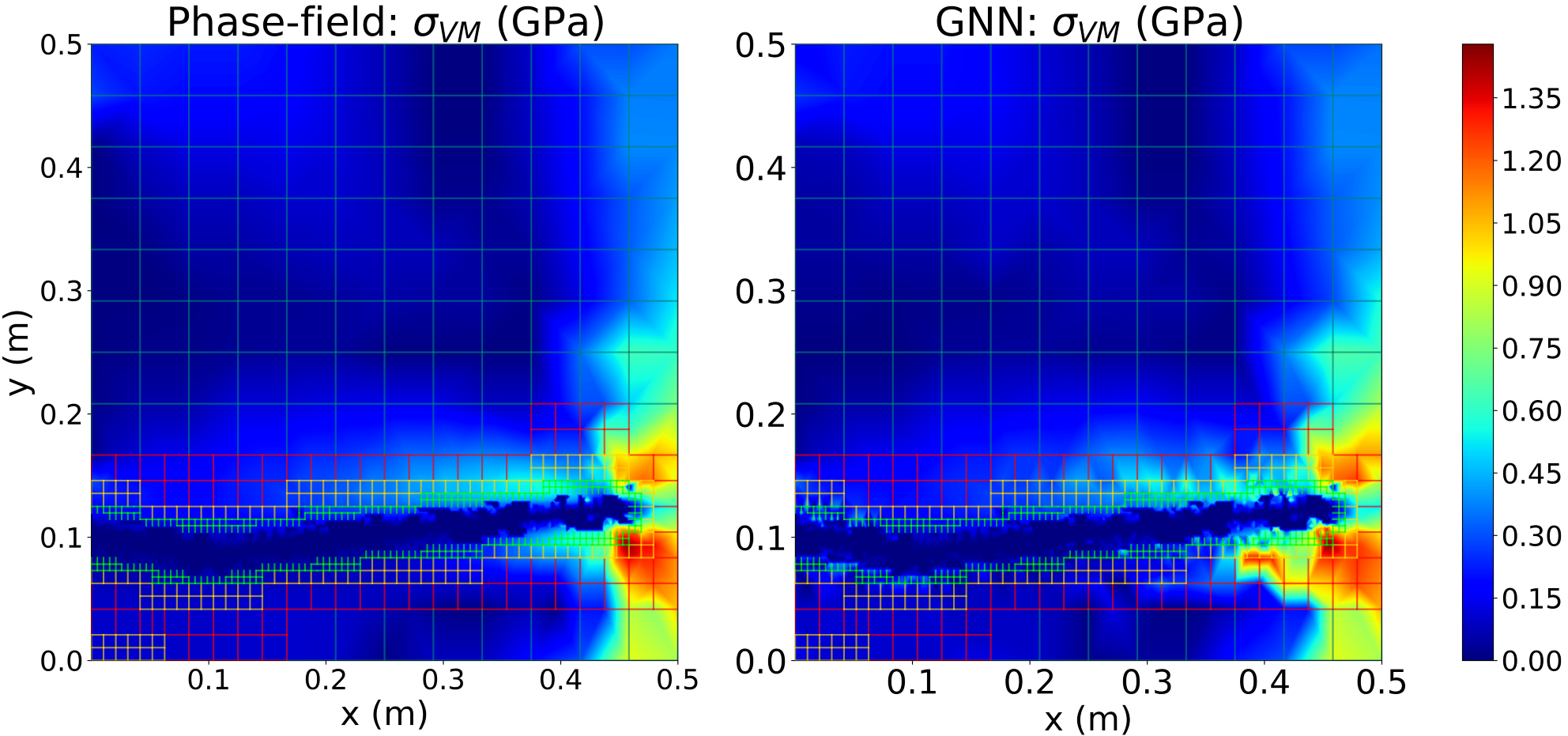}
                    \caption{PF vs. {$\sigma$}-GNN}
                    \label{subfig:SVM_GNN_time_320}
                \end{subfigure}
            \end{subfigure}
            \caption{PF fracture model versus a) \textit{cPhi}-GNN prediction and b) $\sigma_{VM}$ prediction for the same test case scenario shown in Figure \ref{fig:PhaseField_GNN_XDisp_YDisp} involving a small crack ($C_{L}=0.1$ m) with negative angle and located at $C_{P}=0.1$ m.}
            \label{fig:PhaseField_GNN_cPhi_SVM}
        \end{figure} 
        A key attribute of the developed framework, \textit{ADAPT}-GNN, is its capability to predict the stress evolution, $\sigma$, using the predicted x- and y-displacements, and $\phi$.
        From the second order PF fracture model \cite{GOSWAMI2020112808} the stress tensor is defined as $\underline{\underline{\mathbf{\sigma}}} = (1-\phi)^{2}\left[\lambda tr(\underline{\underline{\mathbf{\varepsilon}}}) \mathbf{I} + 2 \mu \underline{\underline{\mathbf{\varepsilon}}} \right]$, where $\lambda$ and $\mu$ are the Lam\'e constants, and $\mathlarger{\underline{\underline{\varepsilon}}}$ is the strain tensor.
        Using this definition along with \textit{ADAPT}-GNN predictions, we compute the evolution of stress. 
        Figure \ref{fig:PhaseField_GNN_cPhi_SVM} shows a comparison of PF fracture model versus predicted $\phi$, and predicted von Mises stress $\sigma_{VM}$.

    \subsection{AMR Update} \label{subsect:AMR_Update}
        The final and key component of the developed \textit{ADAPT}-GNN framework is the ``AMR Update'' step.
        \added[id=R1,comment={Q4,Q11}]{We note that AMR is necessary to improve the performance of ADAPT-GNN.
        Using a static fine mesh as the graph representation would have required a large number of edge connections (edges) because the solution at a point depends on faraway points. 
        However, this results in a graph with large number of edges which significantly increasing computational costs. 
        A possible solution is to increase the number of message passing steps according to the required hop distance. 
        In \cite{Hamilton2020Graph}, the authors show that in order to transfer information from two nodes which are “x” hops away, the GNN must include “x” message passing blocks. 
        However, the required number of hops for the finer mesh resolution in this problem would be very large, resulting in many message-passing steps.}
        
        To leverage the contribution of AMR along with GNNs, once the Prediction Step shown in Figure \ref{fig:GNN_Flowchart} is complete (\textit{XDisp}-GNN, \textit{YDisp}-GNN, and \textit{cPhi}-GNN), we refined the mesh by adding new nodes in regions where $\phi$ is greater than a threshold value (chosen as 0.5 from \cite{GOSWAMI2020112808}).
        Additionally, we formulated a new graph representation for the future time-step by introducing new vertices and edges. 
        We note that during training, we used a mask Boolean array to train for the active nodes explicitly and ignore the inactive nodes.
        This approach ensures a dynamic graph where edges are generated only between the adjacent active nodes (as depicted in Figure \ref{fig:neighbors_vertex_edge}), and training computations are only performed at the active nodes. 
        Lastly, once the new refined graph representation is generated for the following time-step, we repeated the procedures described in Sections \ref{subsect:XDisp-GNN_YDisp-GNN}, \ref{subsect:cPhi-GNN}, and \ref{subsect:AMR_Update} until failure has occurred throughout the entire domain.

\section{Cross-validation}\label{sec:Cross_Validation}
    For additional optimization of the framework, we performed cross-validation to \textit{XDisp}-GNN, \textit{YDisp}-GNN, and \textit{cPhi}-GNN using the 10-fold (k-fold) cross-validation approach \cite{Fushiki2011K-Fold}.
    The training parameters investigated were the learning rates, number of message-passing steps, and number of nodes in the hidden layers of the MLP network.
    The first step of the 10-fold cross-validation procedure was to shuffle the original training dataset (of 1100 simulations) into 10 unique groups.
    Next, we choose one group and set it aside as our `new validation dataset' and perform the training on the remaining 9 groups as our `new training dataset'.
    We perform the training for 5 epochs for this combination before choosing another combination of new validation and training groups.
    We repeat this process for each GNN model and for each of the training parameters investigated.
    The performance was computed using the averaged maximum percent errors in the predicted x-displacements, y-displacements, and $\phi$ field.
    \added[id=R1,comment=Q15]{We emphasize that cross-validation was only applied to the message passing GINE models pertaining to each GNN (\textit{XDisp}-GNN, \textit{YDisp}-GNN, and \textit{cPhi}-GNN).
    For the ATGCN models of each GNN, the only training parameter available for tuning was the filter size (or number of nodes).
    Therefore, we did not implement cross-validation in this work for the ATGCN models.
    We chose the filter size of each ATGCN to match the optimal number of hidden layer nodes obtained from cross-validation of the GINE message passing models.}
    
    \subsection{Cross-validation for \textit{XDisp}-GNN}
    
\begin{figure} 
            \begin{subfigure}[c]{0.329\textwidth}
                \centering
                \begin{subfigure}[t]{1\textwidth}    
                  \centering \includegraphics[width=0.98\linewidth]{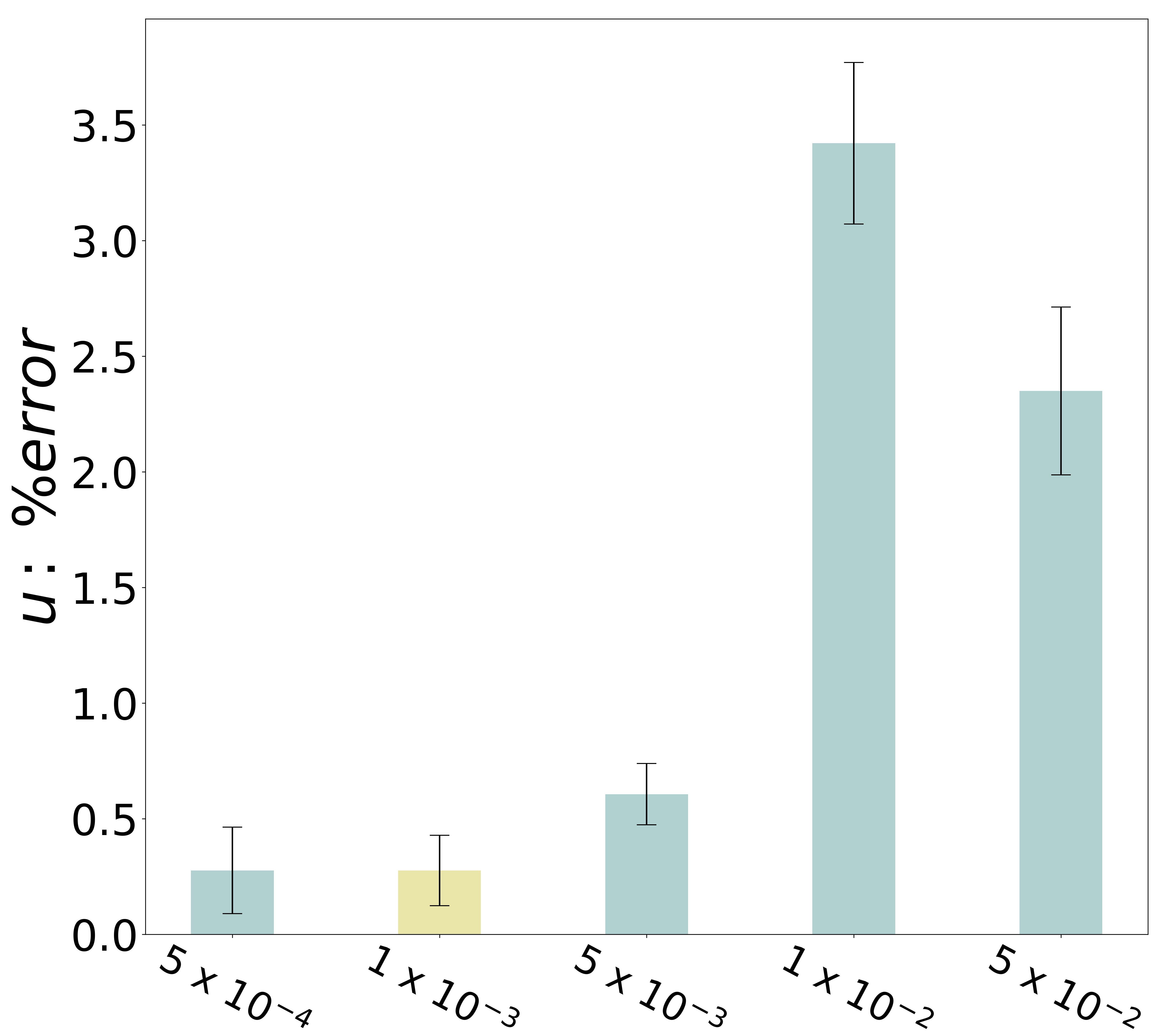}
                    \caption{Learning rates: $u$}
                    \label{subfig:XDisp_Cross_Validation_LR}
                \end{subfigure}
            \end{subfigure}
            \begin{subfigure}[c]{0.329\textwidth}
                \centering
                \begin{subfigure}[t]{1\textwidth}    
                  \centering
                  \includegraphics[width=0.98\linewidth]{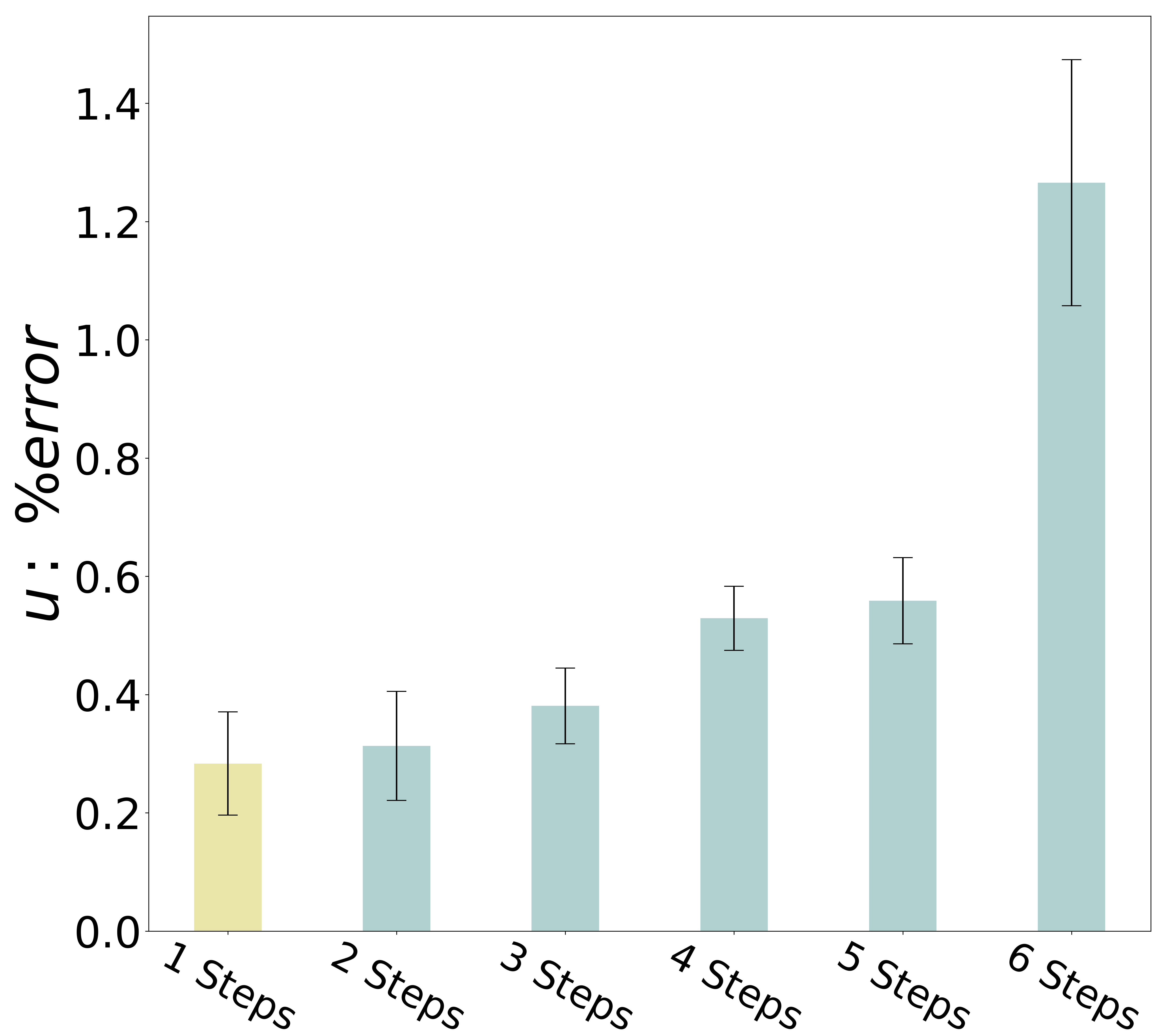}
                    \caption{Message-passing steps: $u$}
                    \label{subfig:XDisp_Cross_Validation_MSteps}
                \end{subfigure}
            \end{subfigure}
            \begin{subfigure}[c]{0.329\textwidth}
                \centering
                \begin{subfigure}[t]{1\textwidth}    
                  \centering
                  \includegraphics[width=0.98\linewidth]{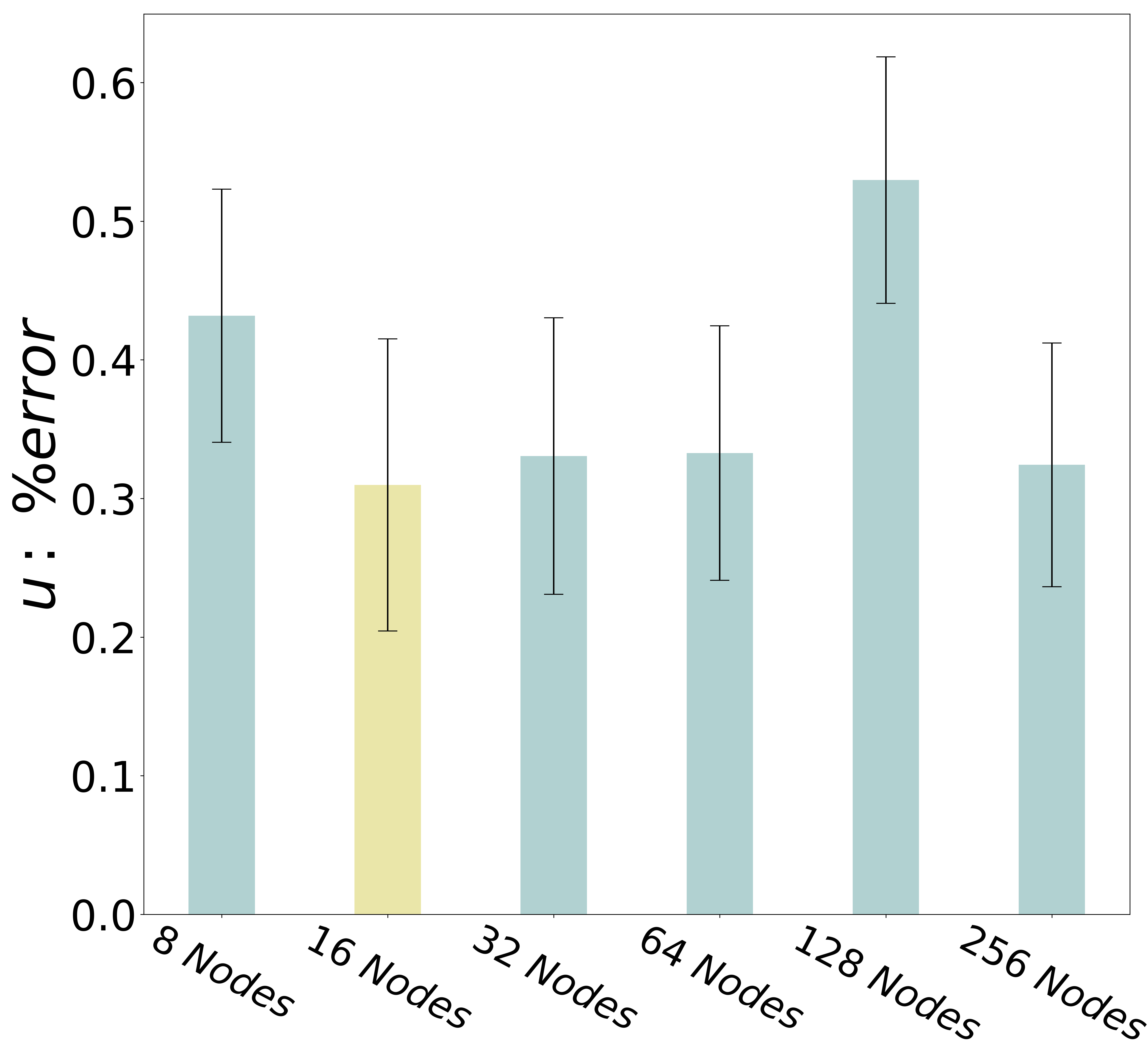}
                    \caption{Hidden layer nodes: $u$}
                    \label{subfig:XDisp_Cross_Validation_vertex_edge_filter}
                \end{subfigure}
            \end{subfigure}
            \caption{Cross-validation results for \textit{XDisp}-GNN: (a) learning rates 5 $\times 10^{-4}$, 5 $\times 10^{-3}$, 1 $\times 10^{-2}$, 5 $\times 10^{-2}$ shown in light blue, and our model’s learning rate 1 $\times 10^{-3}$ shown in yellow, (b) message-passing steps of 2, 3, 4, 5, and 6 shown in light blue, and our model’s message-passing steps of 1 shown in yellow, and (c) number of hidden layer nodes 8, 32, 64, 128, and 256 shown in light blue, and our model's hidden layer nodes of 16 shown in yellow.}
            \label{fig:XDisp_Cross_Validation}
        \end{figure}

        The resulting averaged percent errors for the x-displacement predictions are shown in Figure \ref{fig:XDisp_Cross_Validation}.
        From Figure \ref{subfig:XDisp_Cross_Validation_LR}, the learning rates of {$5\times 10^{-4}$, $5\times 10^{-3}$,  $1 \times 10^{-2}$ and $5 \times 10^{-2}$} (shown in light blue) depict higher errors for \textit{XDisp}-GNN compared to learning rate of $1\times 10^{-3}$ (shown in yellow).
        The highest error in x-displacement is seen for learning rate of {$1\times 10^{-2}$} at $3.42 \pm 0.35\%$, compared to {the smaller learning rate of $1\times 10^{-3}$ with} error of $0.28 \pm 0.15 \%$.
        {Therefore, we chose the optimal learning rate of $1\times 10^{-3}$ for the \textit{XDisp}-GNN model.}
        Figure \ref{subfig:XDisp_Cross_Validation_MSteps} shows the resultant averaged percent errors for message-passing steps of 1, {2,} 3, 4, 5, and 6.
        The model with the lowest percent error was for message-passing steps of $M = 1$ at $0.28 \pm 0.09 \%$, compared to {the highest number of message-passing steps of $M = 6$ with} error of $1.27 \pm 0.21 \%$.
        Similar to the cross-validation results for the learning rates, the optimal message-passing steps parameter of 1 {was} used in {this} work to further optimize the \textit{XDisp}-GNN model.
        We note that a lower number of message-passing steps requires less computational time, thus, decreasing training and simulation times for the \textit{XDisp}-GNN. 
        Lastly, we tested the number of nodes at the hidden layers of the MLP network as shown in Figure \ref{subfig:XDisp_Cross_Validation_vertex_edge_filter}.  
        We observe that when using 16 nodes the \textit{XDisp}-GNN achieved the least error at $0.31 \pm 0.11 \%$, compared to {128 nodes with} the highest error of $0.53 \pm 0.09 \%$.
        We also note that the higher the number of nodes, the more computational requirements are needed.
        \added[id=R1,comment=Q15]{Lastly, we chose the filter size for the ATGCN model in \textit{XDisp}-GNN  as 16 (the optimal number of hidden layer nodes for \textit{XDisp}-GNN's message passing GINE model).}
    
    \subsection{Cross-validation for \textit{YDisp}-GNN}
    
\begin{figure} 
            \begin{subfigure}[c]{0.329\textwidth}
                \centering
                \begin{subfigure}[t]{1\textwidth}    
                  \centering \includegraphics[width=0.98\linewidth]{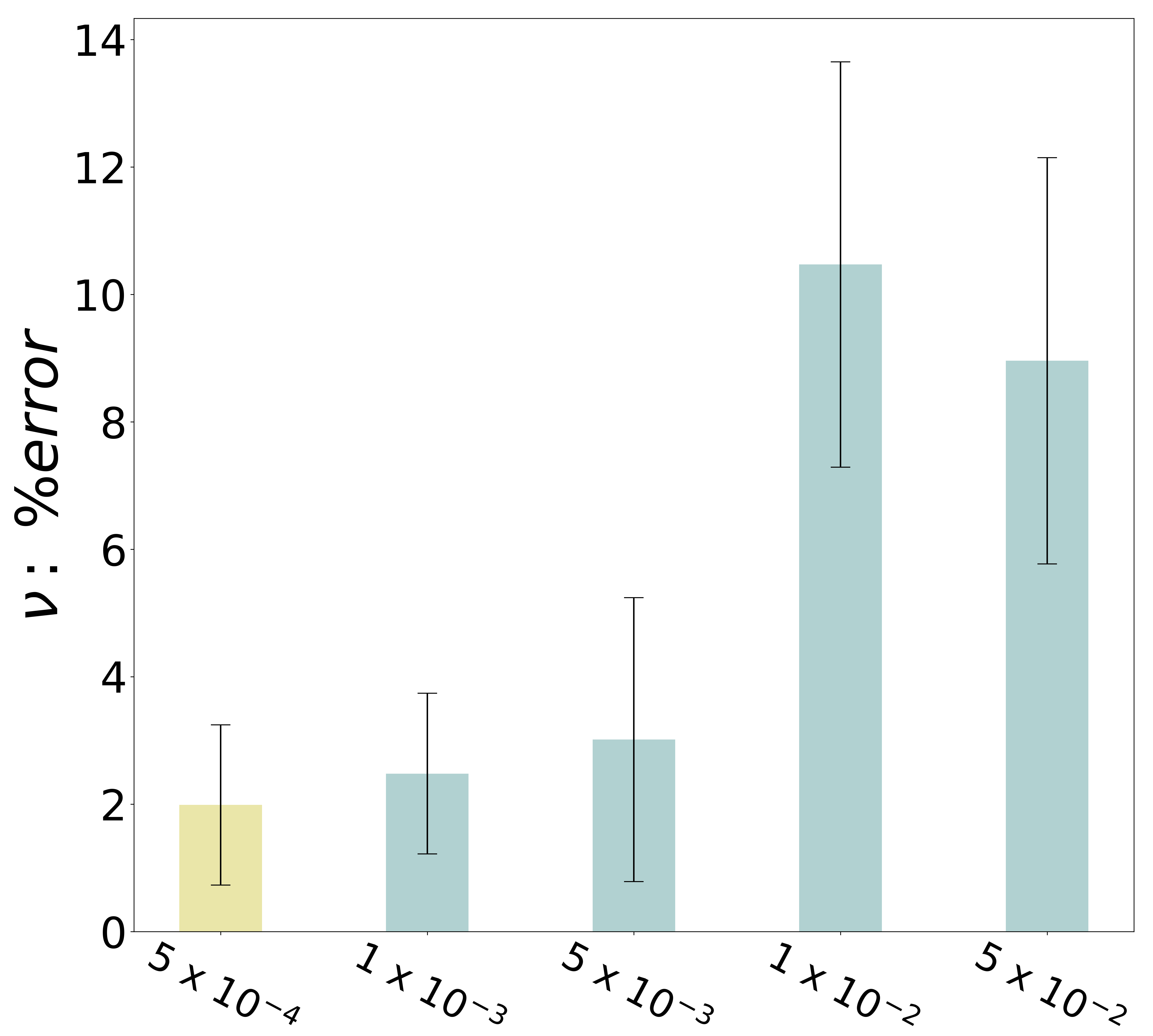}
                    \caption{Learning rates: $\nu$}
                    \label{subfig:YDisp_Cross_Validation_LR}
                \end{subfigure}
            \end{subfigure}
            \begin{subfigure}[c]{0.329\textwidth}
                \centering
                \begin{subfigure}[t]{1\textwidth}    
                  \centering
                  \includegraphics[width=0.98\linewidth]{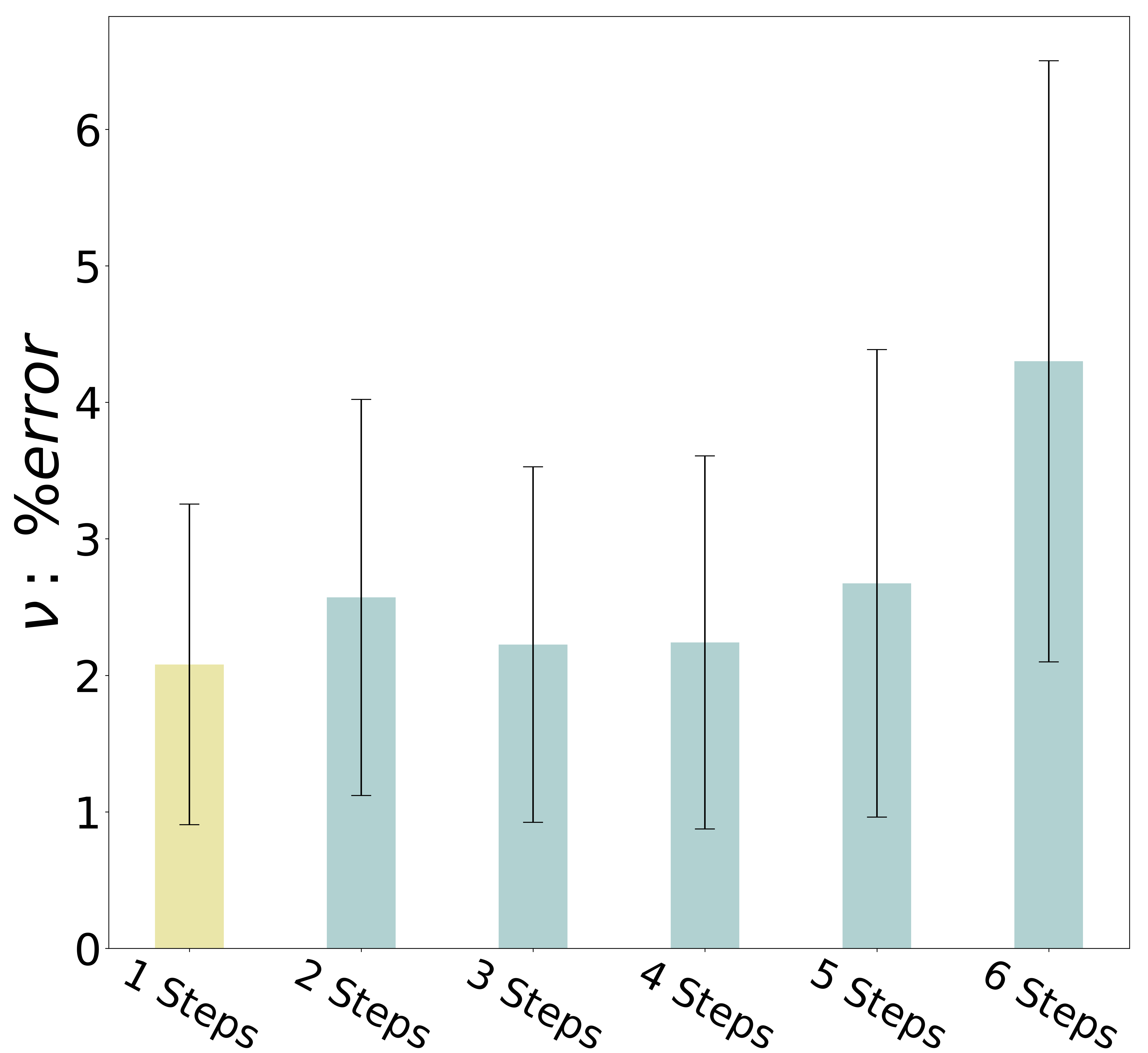}
                    \caption{Message-passing steps: $\nu$}
                    \label{subfig:YDisp_Cross_Validation_MSteps}
                \end{subfigure}
            \end{subfigure}
            \begin{subfigure}[c]{0.329\textwidth}
                \centering
                \begin{subfigure}[t]{1\textwidth}    
                  \centering
                  \includegraphics[width=0.98\linewidth]{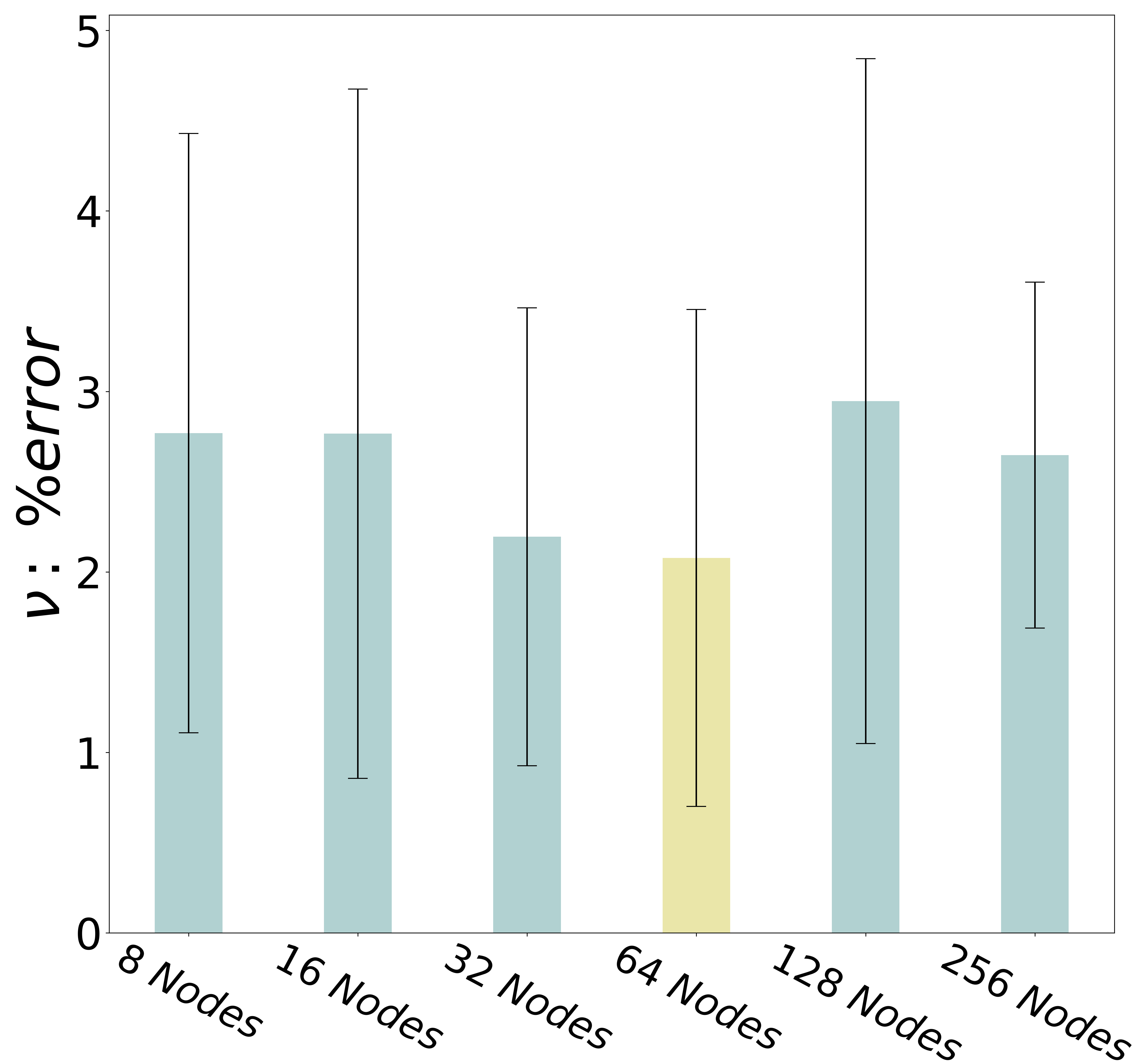}
                    \caption{Hidden layer nodes: $\nu$}
                    \label{subfig:YDisp_Cross_Validation_vertex_edge_filter}
                \end{subfigure}
            \end{subfigure}
            \caption{Cross-validation results for \textit{YDisp}-GNN: (a) learning rates 1 $\times 10^{-4}$, 5 $\times 10^{-3}$, 1 $\times 10^{-2}$, 5 $\times 10^{-2}$ shown in light blue, and our model’s learning rate 5 $\times 10^{-4}$ shown in yellow, (b) message-passing steps of 2, 3, 4, 5, and 6 shown in light blue, and our model’s message-passing steps of 1 shown in yellow, and (c) number of hidden layer nodes 8, 16, 32, 128, and 256 shown in light blue, and our model's hidden layer nodes of 64 shown in yellow.}
            \label{fig:YDisp_Cross_Validation}
        \end{figure}

        Similar to the cross-validation procedure followed for \textit{XDisp}-GNN, the resulting averaged percent errors for the y-displacement predictions were computed for various learning rates, message-passing steps, and number of nodes in the MLP networks as shown in Figure \ref{fig:YDisp_Cross_Validation}.
        From Figure \ref{subfig:YDisp_Cross_Validation_LR}, the learning rates of {$1\times 10^{-3}$, $5\times 10^{-3}$,  $1 \times 10^{-2}$ and $5 \times 10^{-2}$} (shown in light blue) depict higher errors compared to learning rate of $5\times 10^{-4}$ (shown in yellow).
        For the learning rate of {$1\times 10^{-2}$} the percent error is $10.47 \pm 3.18\%$, compared to {the smallest error for learning rate of $5\times 10^{-4}$ with} at $1.99 \pm 1.26 \%$.
        {Therefore, we chose learning rate of $5\times 10^{-4}$ for the \textit{YDisp}-GNN model.}
        Furthermore, from Figure \ref{subfig:YDisp_Cross_Validation_MSteps} the model with the lowest percent error with respect to the number of message-passing steps can be seen at $M = 1$ with error of $2.08 \pm 1.17 \%$, while {the highest number of message-passing steps of $M = 6$ achieved the highest} error of $4.30 \pm 2.20 \%$.
        Lastly, the optimal number of nodes in the MLP network for \textit{YDisp}-GNN are shown in Figure \ref{subfig:YDisp_Cross_Validation_vertex_edge_filter}.  
        When using 64 nodes \textit{YDisp}-GNN achieved the smallest error of $2.08 \pm 1.38 \%$, compared to {128 nodes showing} the highest error at $2.95 \pm 1.90 \%$.
        A key observation to make is that choosing the learning rate played a critical role to achieve higher accuracy in both \textit{XDisp}-GNN and \textit{YDisp}-GNN compared to the number of hidden layer nodes and message-passing steps.   
        \added[id=R1,comment=Q15]{We chose the filter size for the ATGCN model in \textit{YDisp}-GNN as 64 (the optimal number of hidden layer nodes for \textit{YDisp}-GNN's message passing GINE model.)}

    \subsection{Cross-validation for \textit{cPhi}-GNN}
    
\begin{figure} 
            \begin{subfigure}[c]{0.329\textwidth}
                \centering
                \begin{subfigure}[t]{1\textwidth}    
                  \centering \includegraphics[width=0.98\linewidth]{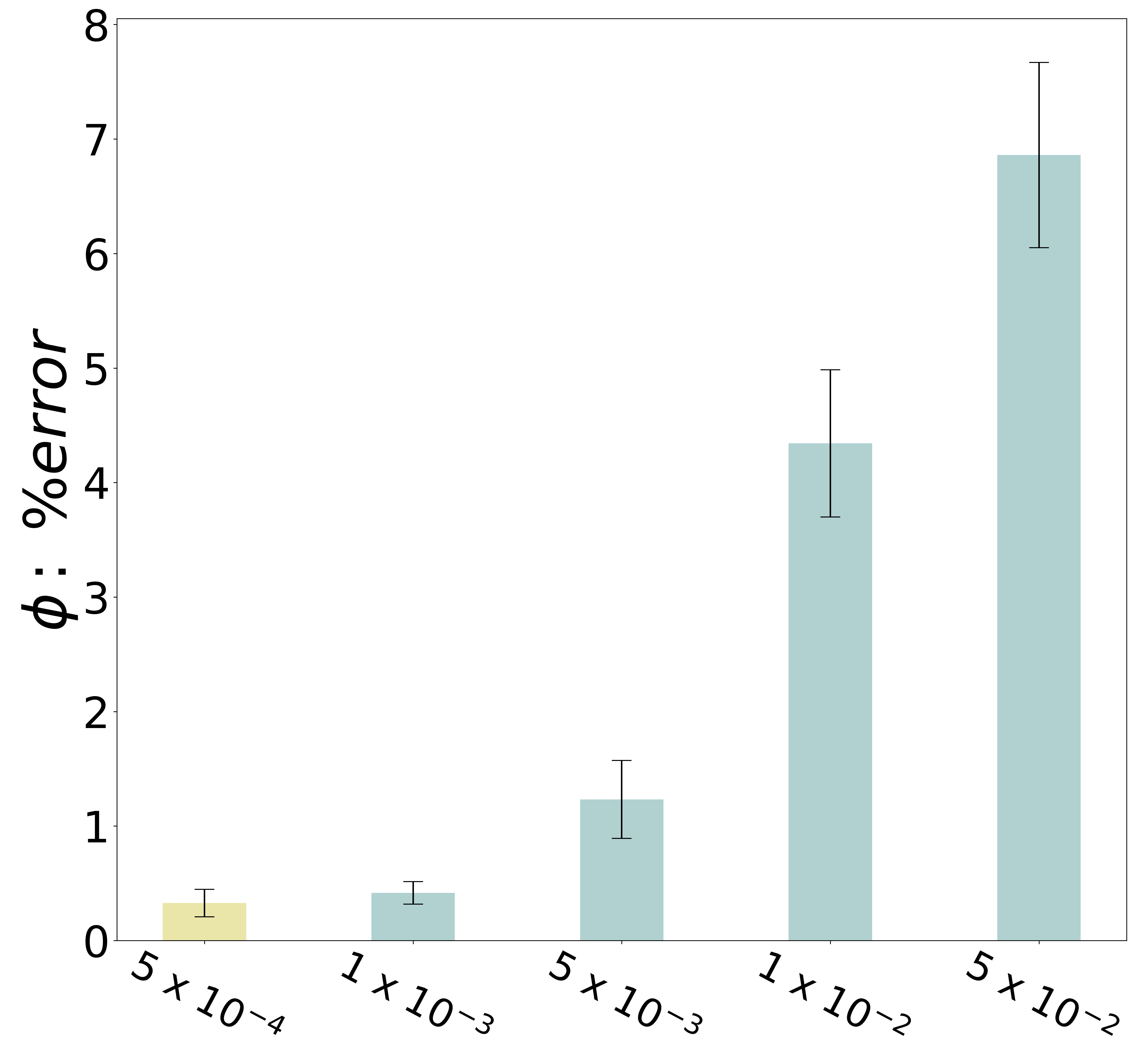}
                    \caption{Learning rates: $\phi$}
                    \label{subfig:cPhi_Cross_Validation_LR}
                \end{subfigure}
            \end{subfigure}
            \begin{subfigure}[c]{0.329\textwidth}
                \centering
                \begin{subfigure}[t]{1\textwidth}    
                  \centering
                  \includegraphics[width=0.98\linewidth]{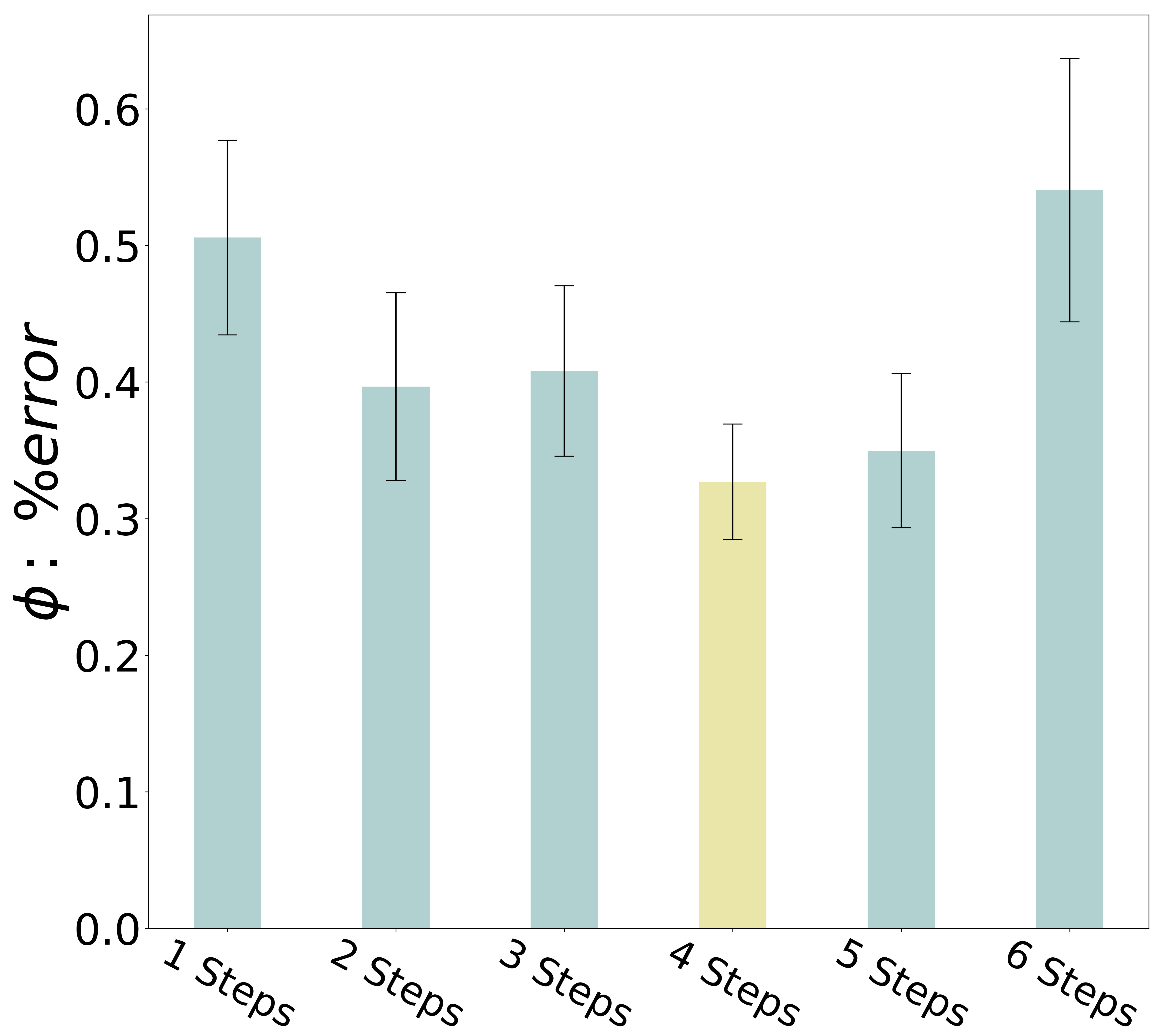}
                    \caption{Message-passing steps: $\phi$}
                    \label{subfig:cPhi_Cross_Validation_MSteps}
                \end{subfigure}
            \end{subfigure}
            \begin{subfigure}[c]{0.329\textwidth}
                \centering
                \begin{subfigure}[t]{1\textwidth}    
                  \centering
                  \includegraphics[width=0.98\linewidth]{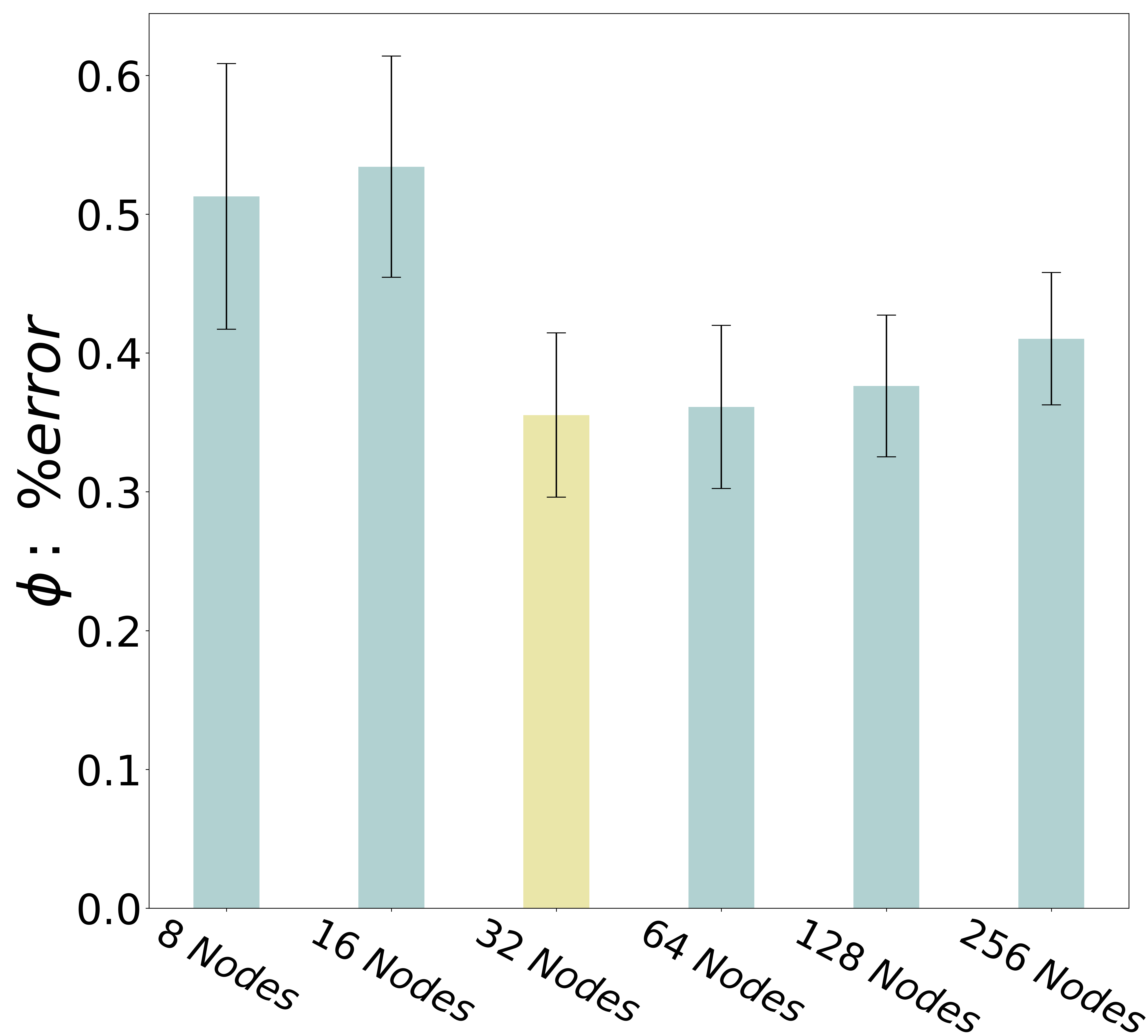}
                    \caption{Hidden layer nodes: $\phi$}
                    \label{subfig:cPhi_Cross_Validation_vertex_edge_filter}
                \end{subfigure}
            \end{subfigure}
            \caption{Cross-validation results for \textit{cPhi}-GNN: (a) learning rates 1 $\times 10^{-4}$, 5 $t\times 10^{-3}$, 1 $\times 10^{-2}$, 5 $\times 10^{-2}$ shown in light blue, and our model’s learning rate 5 $\times 10^{-4}$ shown in yellow, (b) message-passing steps of 1, 2, 3, 5, and 6 shown in light blue, and our model’s message-passing steps of 4 shown in yellow, and (c) number of hidden layer nodes 8, 16, 64, 128, and 256 shown in light blue, and our model's hidden layer nodes of 32 shown in yellow.}
            \label{fig:cPhi_Cross_Validation}
        \end{figure}

        The final GNN model of the cross-validation process was \textit{cPhi}-GNN as shown in Figure \ref{fig:cPhi_Cross_Validation}.
        Similarly to \textit{YDisp}-GNN, the optimal learning rate found was $5\times 10^{-4}$ (shown in yellow) with error of $0.33 \pm 0.12\%$, while the highest error was observed for learning rate $5\times 10^{-2}$ at $6.86 \pm 0.81\%$ (Figure \ref{subfig:cPhi_Cross_Validation_LR}). 
        From Figure \ref{subfig:cPhi_Cross_Validation_MSteps}, the number of message-passing steps resulting in the highest error was {$M = 6$} at $0.54 \pm 0.10\%$, compared to {the smallest error for $M = 4$ at} $0.33 \pm 0.04\%$.
        Additionally, the number of hidden layer nodes found with the lowest percent error of $0.36 \pm 0.06 \%$ was for the case of 32 nodes, and the highest percent error of $0.53 \pm 0.08\%$ for the case of 16 nodes (Figure \ref{subfig:cPhi_Cross_Validation_vertex_edge_filter}).
        \added[id=R1,comment=Q15]{Therefore, similar to \textit{XDisp}- and \textit{YDisp}-GNN, we chose the filter size for the ATGCN model in \textit{cPhi}-GNN as 32 (the optimal number of hidden layer nodes for \textit{cPhi}-GNN's message passing GINE model).}

\section{Results} \label{sec:Results}
    
    \subsection{ADAPT-GNN prediction of displacements, crack field and stresses}
\begin{figure} 
            \centering
            \includegraphics[width=1.\linewidth]{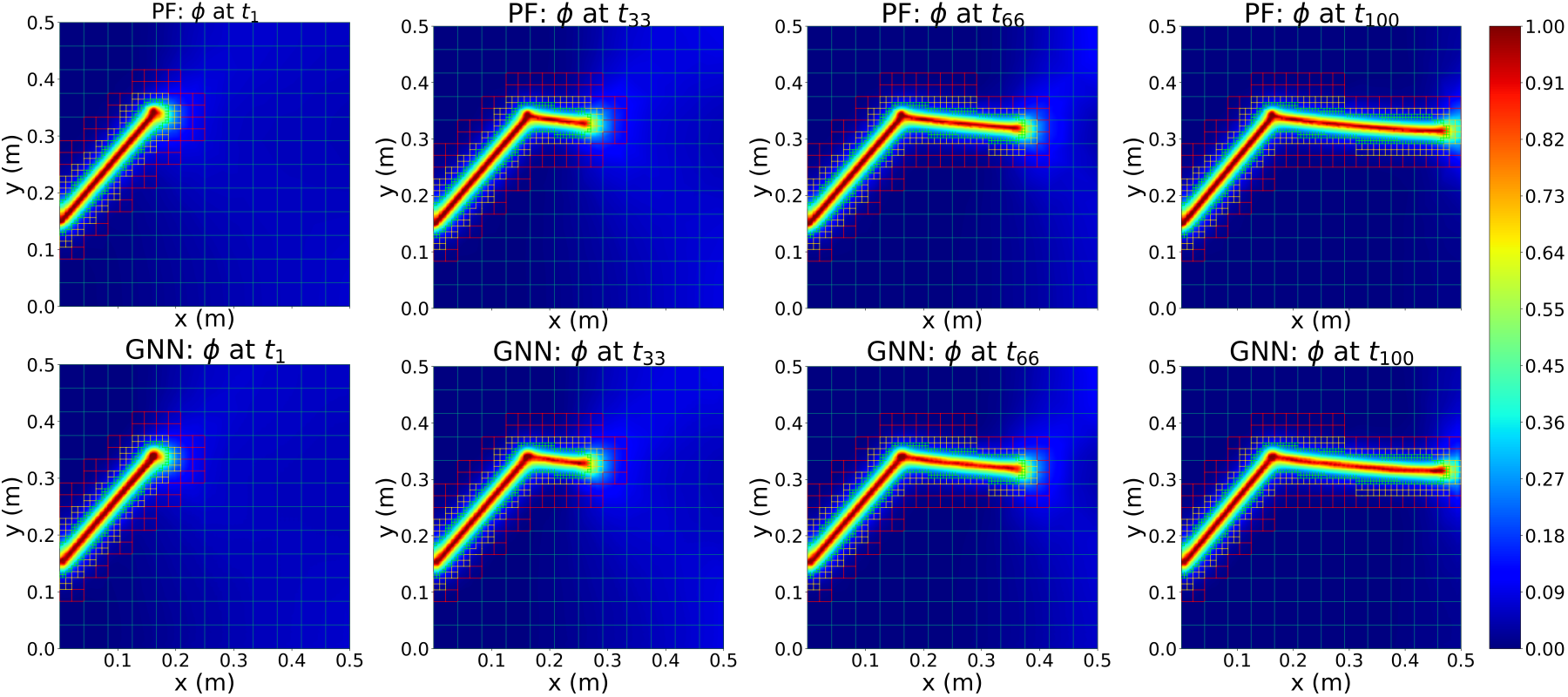}
            \caption{PF versus \textit{ADAPT}-GNN on the evolution of the scalar damage field, $\phi$ for a crack configuration from the test dataset involving a positive crack angle with large crack size ($C_{L}=0.25$ m) and bottom edge position ($C_{P}=0.15$ m).}
            \label{fig:Phi_Evolution}
        \end{figure}

\begin{figure} 
            \centering
            \includegraphics[width=1.\linewidth]{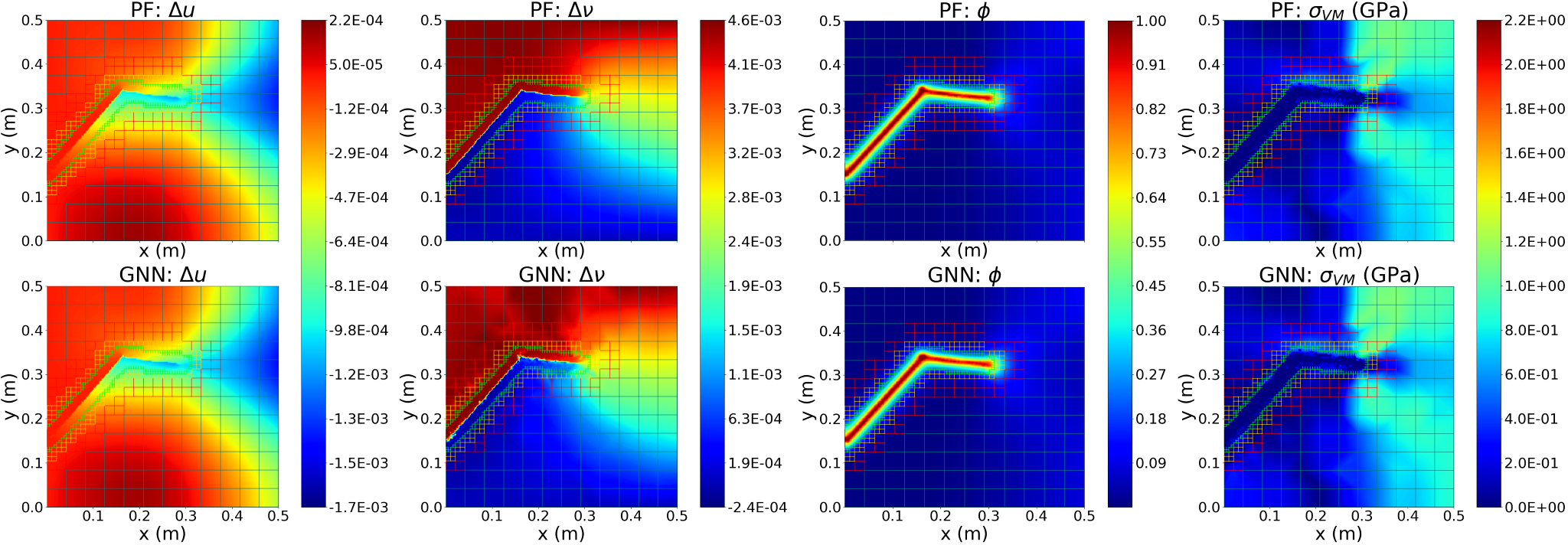}
            \caption{PF versus \textit{ADAPT}-GNN for predictions of x-displacements, $\Delta u$, y-displacements, $\Delta \nu$, scalar damage field, $\phi$, and computed von Mises stress, $\sigma_{VM}$ for the same test case scenario shown in Figure \ref{fig:Phi_Evolution} involving a positive crack angle with large crack size ($C_{L}=0.25$ m) and bottom edge position ($C_{P}=0.15$ m).}
            \label{fig:PF_versus_ADAPT-GNN_TestCase}
        \end{figure}

        Here we demonstrate the framework's capability to predict the evolution of the scalar damage field $\phi$, x-displacements $\Delta u$, y-displacements $\Delta \nu$, and von Mises stress $\sigma_{VM}$ for a crack configuration from the test dataset involving a positive crack angle with large crack size and bottom edge position.  
        Figure \ref{fig:Phi_Evolution} shows a qualitative comparison of the PF fracture model versus \textit{ADAPT}-GNN framework on the evolution of the scalar damage field.
        \added[id=R1,comment=Q16]{We emphasize that the results presented in Figures \ref{fig:Phi_Evolution} - \ref{fig:XDisp_YDisp_cPhi_Angles_Position_BarChart} for both the evolution and the computed errors in scalar damage field and displacement fields were obtained by propagating ADAPT-GNN from $t_{0}$ to $T_{f}$
        In other words, the predictions from the previous time-steps are used as input to the next time-step.}
        We\added[id=R1,comment=Q16]{ also} note that kinking of the predicted crack path during $t_{1}$ to $t_{33}$ is not as sharp as the path from PF.
        {\replaced[id=R1,comment=Q16]{Additionally, w}{W}e note that the crack field in the PF model contains numerous oscillations.
        These oscillations are associated with the second order model and the errors within the PF model implementation, which would then transfer to \textit{ADAPT}-GNN's prediction.}
        However, the results show nearly identical crack path prediction overall compared to the PF fracture model throughout the simulation.  
        These qualitative results show the developed GNN's capability to predict the evolution of scalar damage field with good accuracy.    
         
        Figure \ref{fig:PF_versus_ADAPT-GNN_TestCase} depicts a qualitative comparison of PF versus \textit{ADAPT}-GNN for x- and y-displacements, and von Mises stress at $t_{50}$ of the same test case scenario shown in Figure \ref{fig:Phi_Evolution}.   
        For x-displacements, it can be seen that the predicted field is virtually indistinguishable to the PF fracture model.  
        For y-displacements, there is a noticeable prediction error originating from inside the crack region's sharp interface of positive to negative y-displacements. 
        {For PF fracture models, the $y$ displacement exhibits this sharp jump within the crack - from negative to positive.
        We emphasize that errors inside the crack region do not play a significant role in PF fracture model.}  
        {Additionally, we} note that these plots were generated using the "\textit{tricontourf}" function along with the predicted values at the active mesh points. 
        Because the "\textit{tricontourf}" function performs interpolation between the active mesh points using the y-displacement values at the active mesh points, the highest y-displacement errors originating from inside the crack are interpolated to regions outside the crack. 
        Therefore, Figure \ref{fig:PF_versus_ADAPT-GNN_TestCase} shows the developed framework is able to predict the overall y-displacements with good accuracy outside of the crack region.

        Lastly, Figure \ref{subsec:Errors} shows a key feature of the developed GNN framework for generating the stress evolution in the domain.
        The von Mises stress can be computed using the predicted x- and y-displacements, and the scalar damage field.
        We note a good qualitative agreement between von Mises stress calculated from \textit{ADAPT}-GNN prediction and PF fracture model.
        Therefore, Figures \ref{fig:Phi_Evolution} and \ref{fig:PF_versus_ADAPT-GNN_TestCase} illustrate the framework's ability to predict the evolution of displacements, scalar damage field, and von Mises stress with good accuracy for a given crack configuration from the test dataset. 
        We have included animations of seven test cases as supplementary material.

    \subsection{Prediction errors}\label{subsec:Errors}
    
\begin{figure} 
            \begin{subfigure}[c]{0.32\textwidth}
                \centering
                \begin{subfigure}[t]{1\textwidth}    
                  \centering \includegraphics[width=0.98\linewidth]{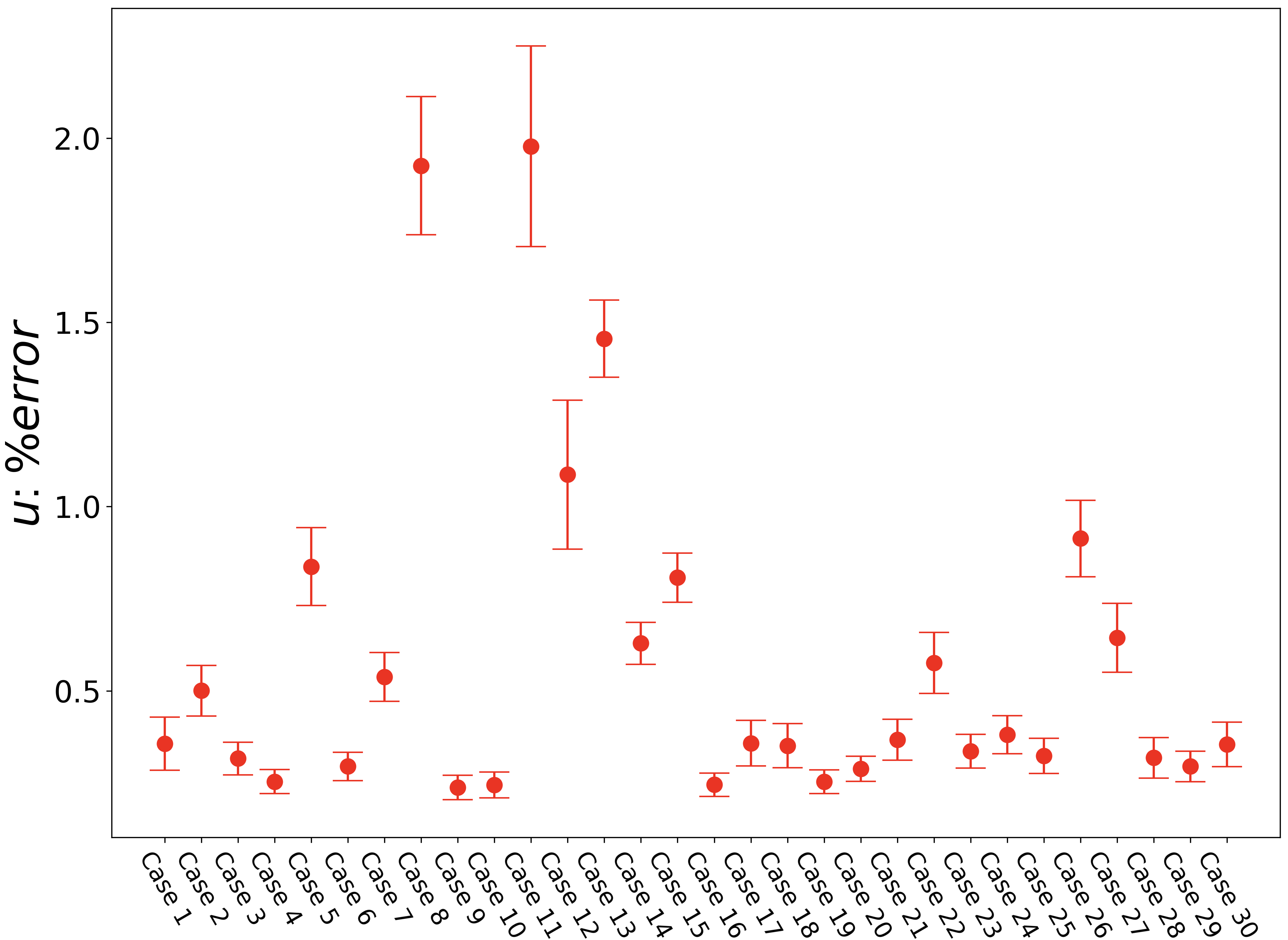}
                    \caption{Percent Errors in $u$}
                    \label{subfig:XDisp_All_Errors}
                \end{subfigure}
            \end{subfigure}
            \begin{subfigure}[c]{0.32\textwidth}
                \centering
                \begin{subfigure}[t]{1\textwidth}    
                  \centering
                  \includegraphics[width=0.98\linewidth]{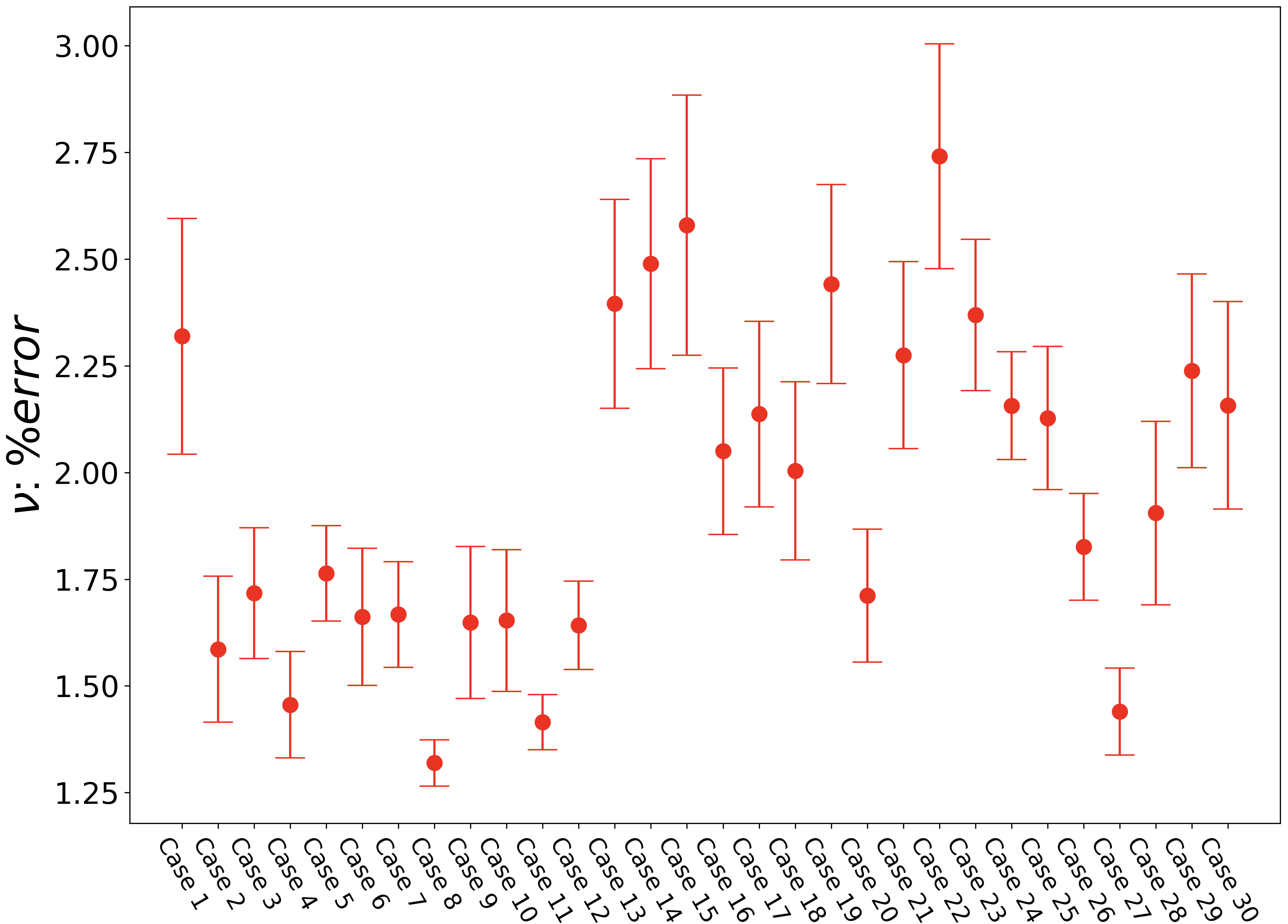}
                    \caption{Percent Errors in $\nu$}
                    \label{subfig:YDisp_All_Errors}
                \end{subfigure}
            \end{subfigure}
            \begin{subfigure}[c]{0.32\textwidth}
                \centering
                \begin{subfigure}[t]{1\textwidth}    
                  \centering \includegraphics[width=0.98\linewidth]{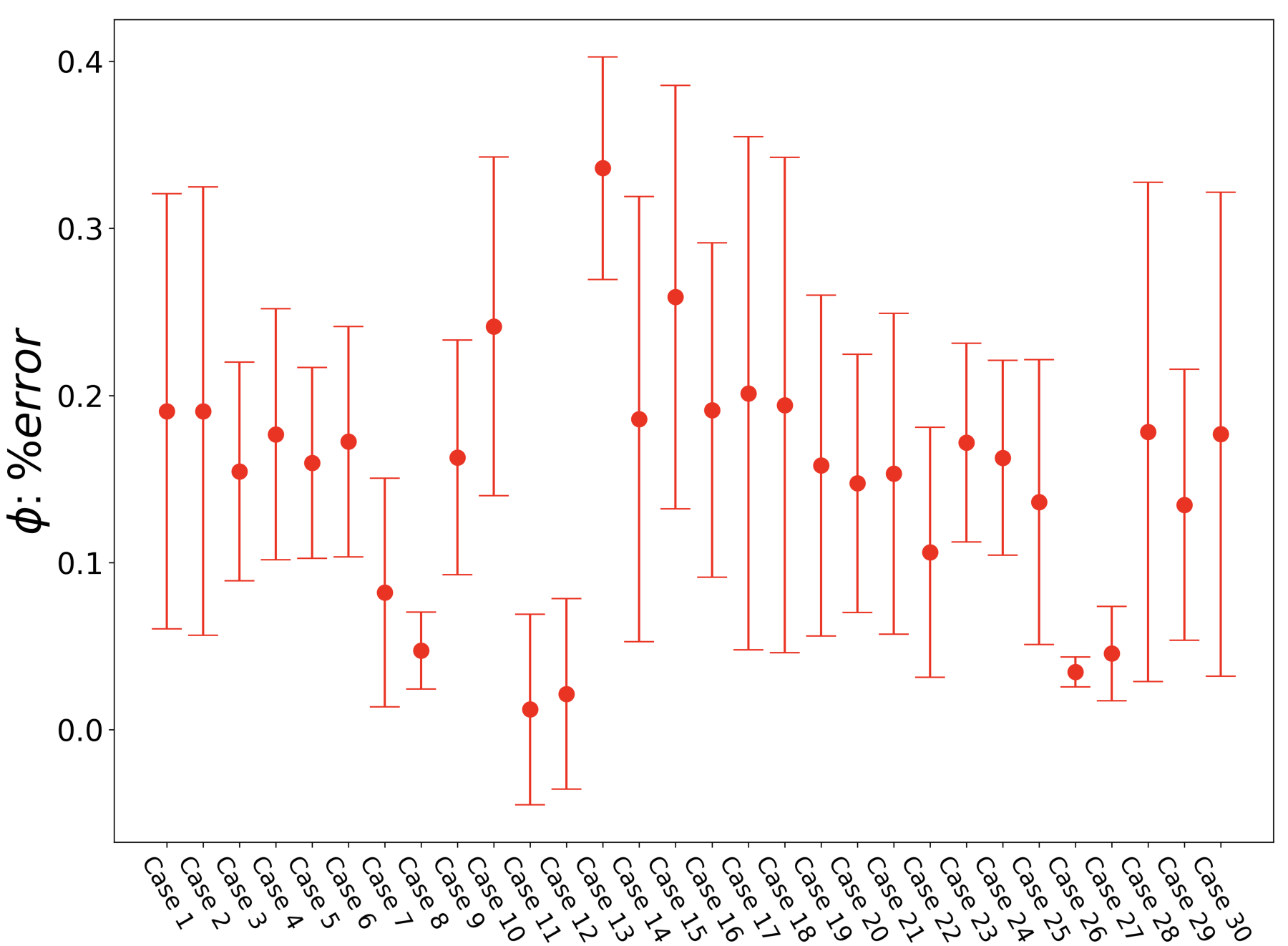}
                    \caption{Percent Errors in $\phi$}
                    \label{subfig:cPhi_All_Errors}
                \end{subfigure}
            \end{subfigure}
            \caption{Maximum percent error in (a)  $u$ predictions, (b) $\nu$ predictions, and (c) $\phi$ predictions across time for each simulation in the test set (Case 1 - Case 30)}
            \label{fig:All_Errors}
        \end{figure}

        To evaluate the errors generated by \textit{XDisp}-GNN, \textit{YDisp}-GNN, and \textit{cPhi}-GNN the \added[id=R1,comment=Q14]{maximum $\%$ errors were computed as}\deleted[id=R1,comment=Q14]{maximum $\%$ errors across time for each simulation in the test dataset are shown in Figures \ref{subfig:XDisp_All_Errors}, \ref{subfig:YDisp_All_Errors}, and \ref{subfig:cPhi_All_Errors}, respectively.}
        \begin{flalign}
            &&\% error = max\left[\Sigma_{i=1}^{\mathcal{M}}\frac{1}{\mathcal{M}}\left(\frac{|\phi_{pred}(t,i) - \phi_{true}(t,i)|}{\phi_{true}(t,i)}\right)\times 100\right] && \dots \{t \in {T_{f}}\},
            \label{eq:percent_error}
        \end{flalign}
        \added[id=R1,comment=Q14]{where $\phi_{pred}$ and $\phi_{true}$ are the predicted and true scalar damage fields, respectively, and $T_{f}$ is the final time once fracture is complete.
        We note that equation (\ref{eq:percent_error}) is shown for errors in the scalar damage field.
        Equation (\ref{eq:percent_error}) is also used for computing errors in x and y displacement fields.
        The resulting maximum $\%$ errors across time for each simulation in the test dataset are shown in Figures \ref{subfig:XDisp_All_Errors}, \ref{subfig:YDisp_All_Errors}, and \ref{subfig:cPhi_All_Errors}, respectively.
        We also emphasize that the results presented in Figure \ref{fig:All_Errors} depict the maximum percent error from the accumulated error versus iteration where the predictions from the previous time-steps are used as input to the next time-step.}
        {As mentioned in Section \ref{sec:Results} the PF model used in this work involves instability errors due to oscillations in the scalar damage field inside the crack's region (shown in Figure \ref{fig:Phi_Evolution}).
        Because these errors are localized at the refined mesh, $\mathcal{M}^{ref}$, error computations at these nodes may be inconsistent with the remaining nodes.}
        {Additionally,} \replaced[id=R1,comment=17]{b}{B}ecause \textit{ADAPT}-GNN makes predictions for all mesh points {in $\mathcal{M}$}, we first compute average error across all mesh points for each time-step and then choose the maximum $\%$ error across all time-steps\added[id=R1,comment=Q14]{ as shown in equation (\ref{eq:percent_error})}.
        {This error analysis ensures that errors in \textit{ADAPT}-GNN's are captured throughout all nodes in $\mathcal{M}$.}
        Following this approach, the test case with highest $\%$ error in the predicted x-displacement field, $u$, (Figure \ref{subfig:XDisp_All_Errors}) is shown for Case 11 at $1.98 \pm 0.27 \%$, while the lowest $\%$ error is shown for Case 9 at $0.24 \pm 0.33 \%$. 
        From Figure \ref{subfig:YDisp_All_Errors} the simulation with highest $\%$ error in y-displacement field predictions, $\nu$, is Case 22 with error of $2.74 \pm 0.26 \%$, while the lowest error is seen in Case 8 at $1.32 \pm  0.05 \%$.  
        Similarly, the highest prediction error in the scalar damage field, $\phi$, was observed for Case 13 with $0.19 \pm 0.13 \%$, and the lowest error for Case 3 at $0.01 \pm 0.06 \%$.
        These results demonstrate the ability of \textit{ADAPT}-GNN to predict both displacements and crack propagation with high accuracy.
        While \textit{YDisp}-GNN shows the highest obtained error compared to the remaining two implemented GNNs (\textit{XDisp}-GNN and \textit{cPhi}-GNN), a maximum $\%$ error of $2.74 \pm 0.26 \%$ is considerably low (micrometers) in a $0.5m \times 0.5m$ domain.

    \subsection{Parametric error analysis of crack angles, crack length, and edge position}\label{subsection:Parametric}
    
        We performed a systematic error analysis to study the effects of two possible combinations of initial configurations: (i) crack angle and crack length, and (ii) crack angle and edge position. 
        We parameterized each of these initial configurations by positive versus negative crack angles, and large versus small crack lengths. 
        Following this convention, we will discuss the maximum percent errors of each configuration for x-displacements, y-displacements, and $\phi$ in the following Sections.
        
        \subsubsection{Crack angle and crack length}\label{subsubsection:Parametric_angle_length}
        
\begin{figure} 
                \begin{subfigure}[c]{0.32\textwidth}
                    \centering
                    \begin{subfigure}[t]{1\textwidth}    
                      \centering \includegraphics[width=0.98\linewidth]{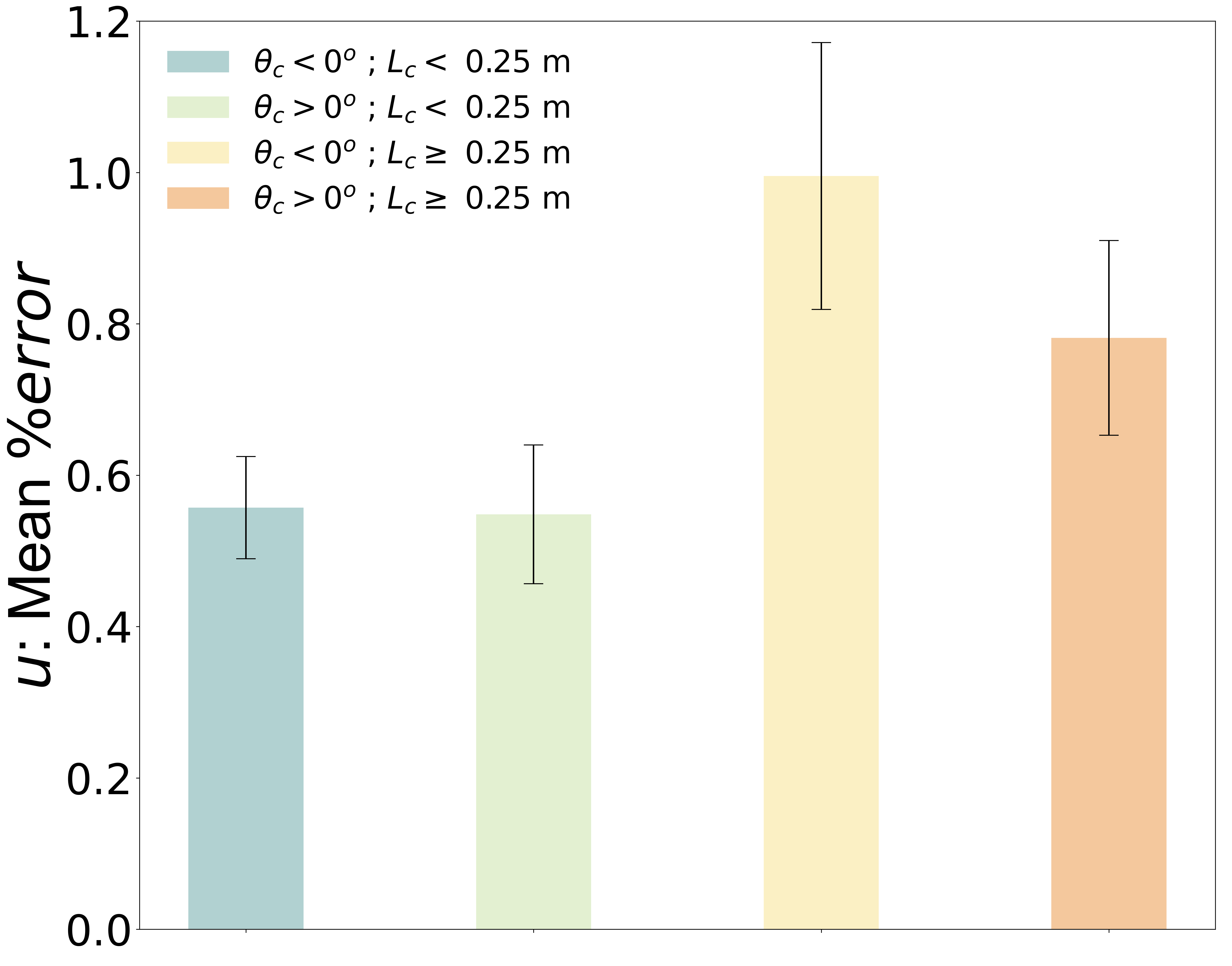}
                        \caption{Percent Errors in $u$}
                        \label{subfig:XDisp_Angles_Length_BarChart}
                    \end{subfigure}
                \end{subfigure}
                \begin{subfigure}[c]{0.32\textwidth}
                    \centering
                    \begin{subfigure}[t]{1\textwidth}    
                      \centering \includegraphics[width=0.98\linewidth]{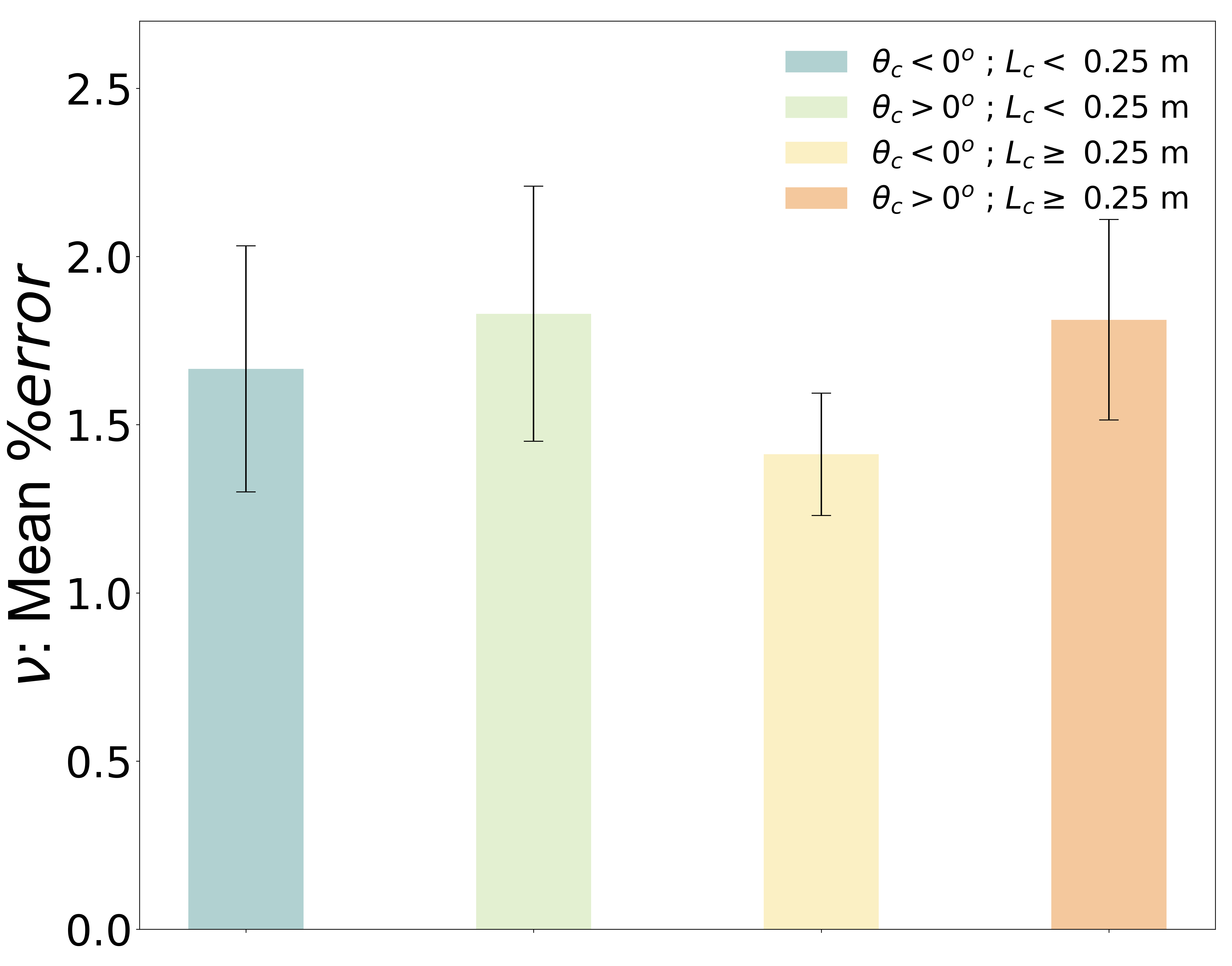}
                        \caption{Percent Errors in $\nu$}
                        \label{subfig:YDisp_Angles_Length_BarChart}
                    \end{subfigure}
                \end{subfigure}
                \begin{subfigure}[c]{0.32\textwidth}
                    \centering
                    \begin{subfigure}[t]{1\textwidth}    
                      \centering
                        \includegraphics[width=0.98\linewidth]{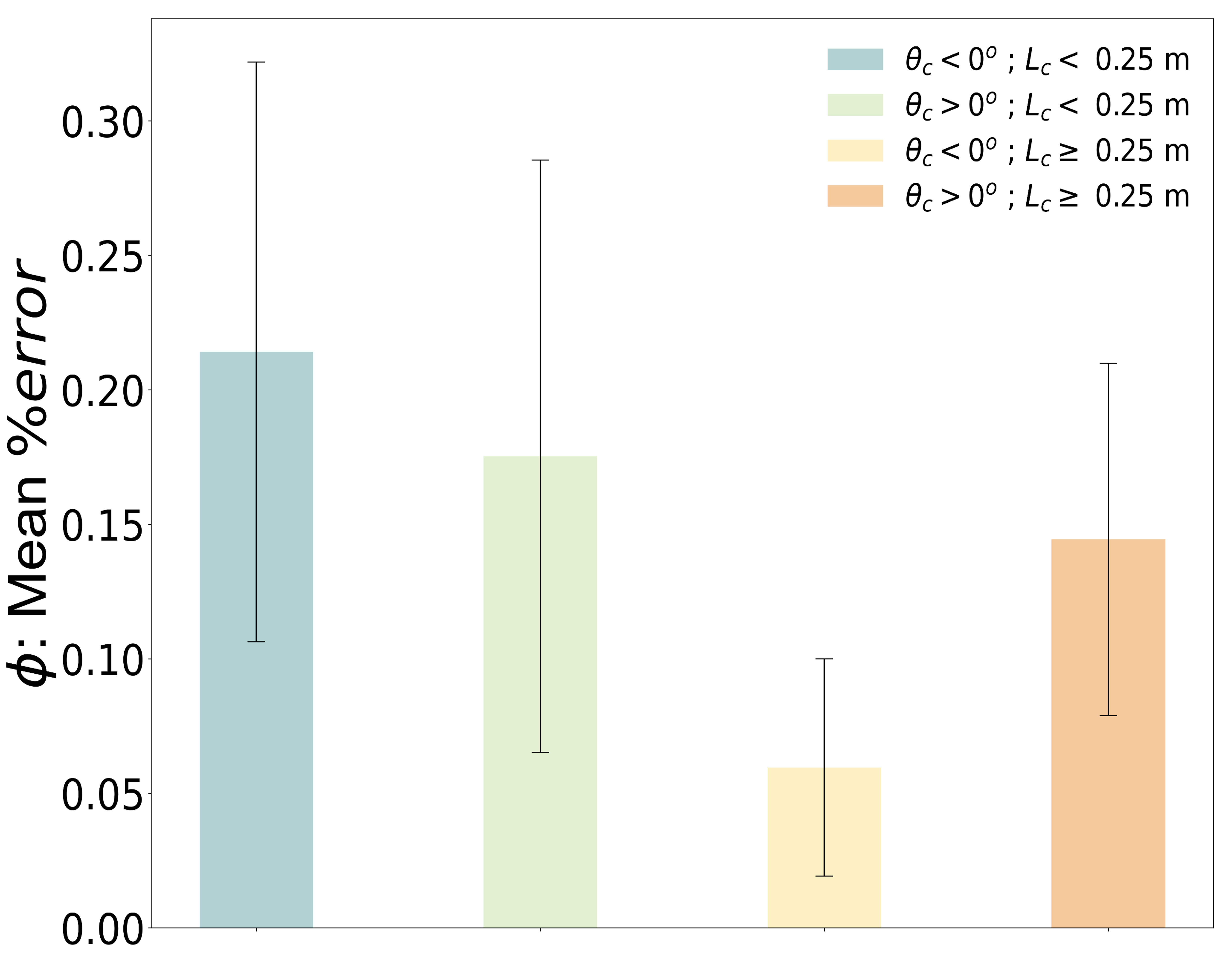}
                        \caption{Percent Errors in $\phi$}
                        \label{subfig:cPhi_Angles_Length_BarChart}
                    \end{subfigure}
                \end{subfigure}
                \caption{Parametric error analysis of the contribution of initial crack angles and crack lengths on (a) $u$ predictions, (b) $\nu$ predictions, and (c) $\phi$ predictions.}
                \label{fig:XDisp_YDisp_cPhi_Angles_Length_BarChart}
            \end{figure}

            To study the effects of varying initial crack orientations along with crack lengths on the resulting prediction errors, we split the test dataset into four groups: (i) negative crack angle + small crack length defined as $\theta_{c} < 0^{o}$; $L_{c} < 0.25$ m, (ii) positive crack angle + small crack length defined as $\theta_{c} > 0^{o}$; $L_{c} < 0.25$ m, (iii) negative crack angle + large crack length defined as $\theta_{c} < 0^{o}$; $L_{c} \geq 0.25$ m, and (iv) positive crack angle + large crack length defined as $\theta_{c} > 0^{o}$; $L_{c} \geq 0.25$ m.
            We then computed the mean and standard deviation for the maximum percent errors of each of these group.
            Figure \ref{fig:XDisp_YDisp_cPhi_Angles_Length_BarChart} shows the corresponding errors of x-displacement (Figure \ref{subfig:XDisp_Angles_Length_BarChart}), y-displacement (Figure \ref{subfig:YDisp_Angles_Length_BarChart}), and scalar damage field (Figure \ref{subfig:cPhi_Angles_Length_BarChart}) predictions for each parametric group.  
            
            From Figure \ref{subfig:XDisp_Angles_Length_BarChart}, \textit{XDisp}-GNN shows a clear distinction in error for small versus large crack length.
            For smaller cracks, the crack orientation does not seem to play a significant role in the percent error.
            For instance, for smaller crack lengths with negative and positive angles the errors are $0.56 \pm 0.07 \%$ and $0.55 \pm 0.09 \%$, respectively, which differ only by approximately $0.01 \%$. 
            When the crack length is increased above $0.24$ m, the errors increase.
            Additionally, the crack angle does affect \textit{XDisp}-GNN's accuracy for cases with large crack lengths.
            When considering large cracks, the highest observed error of $1.00 \pm 0.18 \%$ was for the group consisting of negative angles, while for the group consisting of positive angles the error decreased to $0.78 \pm 0.13 \%$.
            A possible explanation for why cases with smaller cracks result in lower errors for \textit{XDisp}-GNN, may be due to the reasoning discussed in Section \ref{subsec:Errors}. 
            The mesh-wise errors of \textit{XDisp}-GNN showed to be highest during the initial time-steps - when the applied displacement load is constantly increased until crack begins to propagate smoothly.
            Once the crack began to propagate smoothly the errors decreased throughout the remaining time-steps.
            Following this observation, a smaller crack will result in a larger number of time-steps in order to fully propagate throughout the domain.
            In essence, a smaller crack will consist of more time-steps where the crack is propagating smoothly, thus, consisting of more time-steps where the errors are low in comparison to larger cracks.
            Therefore, \textit{XDisp}-GNN performed best for cases involving smaller cracks regardless of initial crack orientation, however, for larger cracks it achieved better accuracy for cases with positive angles.
            
            The parametric results of crack angle and crack length for \textit{YDisp}-GNN are shown in Figure \ref{subfig:YDisp_Angles_Length_BarChart}.
            Unlike \textit{XDisp}-GNN, the group with the highest \textit{YDisp}-GNN error was for positive angles + smaller cracks at $1.83 \pm 0.38 \%$, while the group with lowest error was for negative angles + larger cracks at $1.41 \pm 0.18 \%$.
            An interesting observation to make is that both groups with positive angles showed similar results.
            The error obtained for the group involving positive angles + larger cracks was $1.81 \pm 0.30 \%$, approximately $0.02 \%$ lower compared to the group involving positive angles + smaller cracks.
            Therefore, \textit{YDisp}-GNN performed similarly for cases with positive angles regardless of the initial crack length (i.e., small versus large length), however, for cases with negative angles the best performance was seen when considering larger cracks. 
            
            For \textit{cPhi}-GNN results shown in Figure \ref{subfig:cPhi_Angles_Length_BarChart}, {it can be seen that the group with the highest error was for negative angles + smaller cracks at $0.21 \pm 0.11 \%$, while the lowest error was for the group of negative angles + larger cracks at $0.06 \pm 0.04 \%$.}

        \subsubsection{Crack angle and initial edge position}\label{subsubsection:Parametric_angle_position}
\begin{figure} 
                \begin{subfigure}[c]{0.32\textwidth}
                    \centering
                    \begin{subfigure}[t]{1\textwidth}    
                      \centering \includegraphics[width=0.98\linewidth]{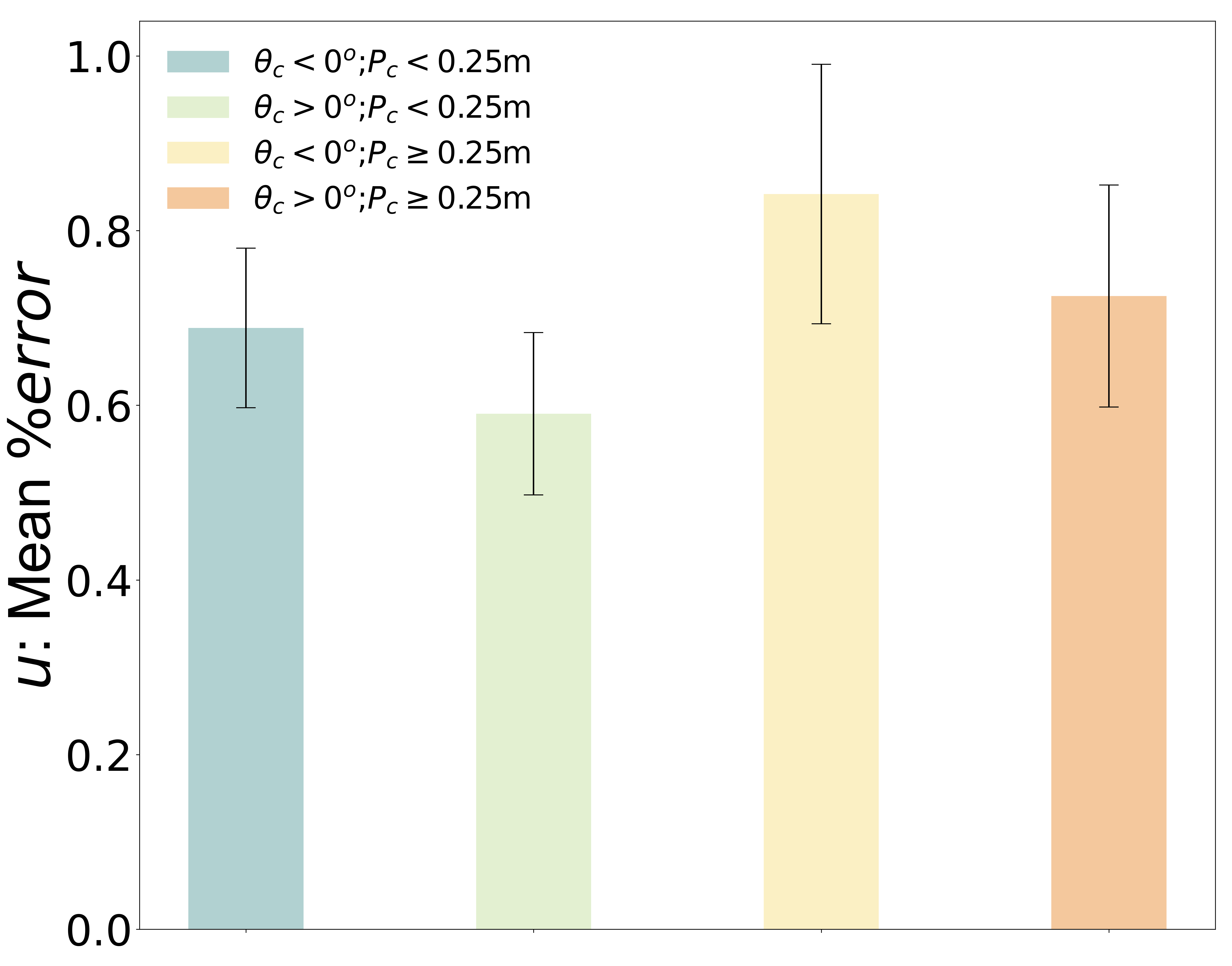}
                        \caption{Percent Errors in $u$}
                        \label{subfig:XDisp_Angles_Position_BarChart}
                    \end{subfigure}
                \end{subfigure}
                \begin{subfigure}[c]{0.32\textwidth}
                    \centering
                    \begin{subfigure}[t]{1\textwidth}    
                      \centering \includegraphics[width=0.98\linewidth]{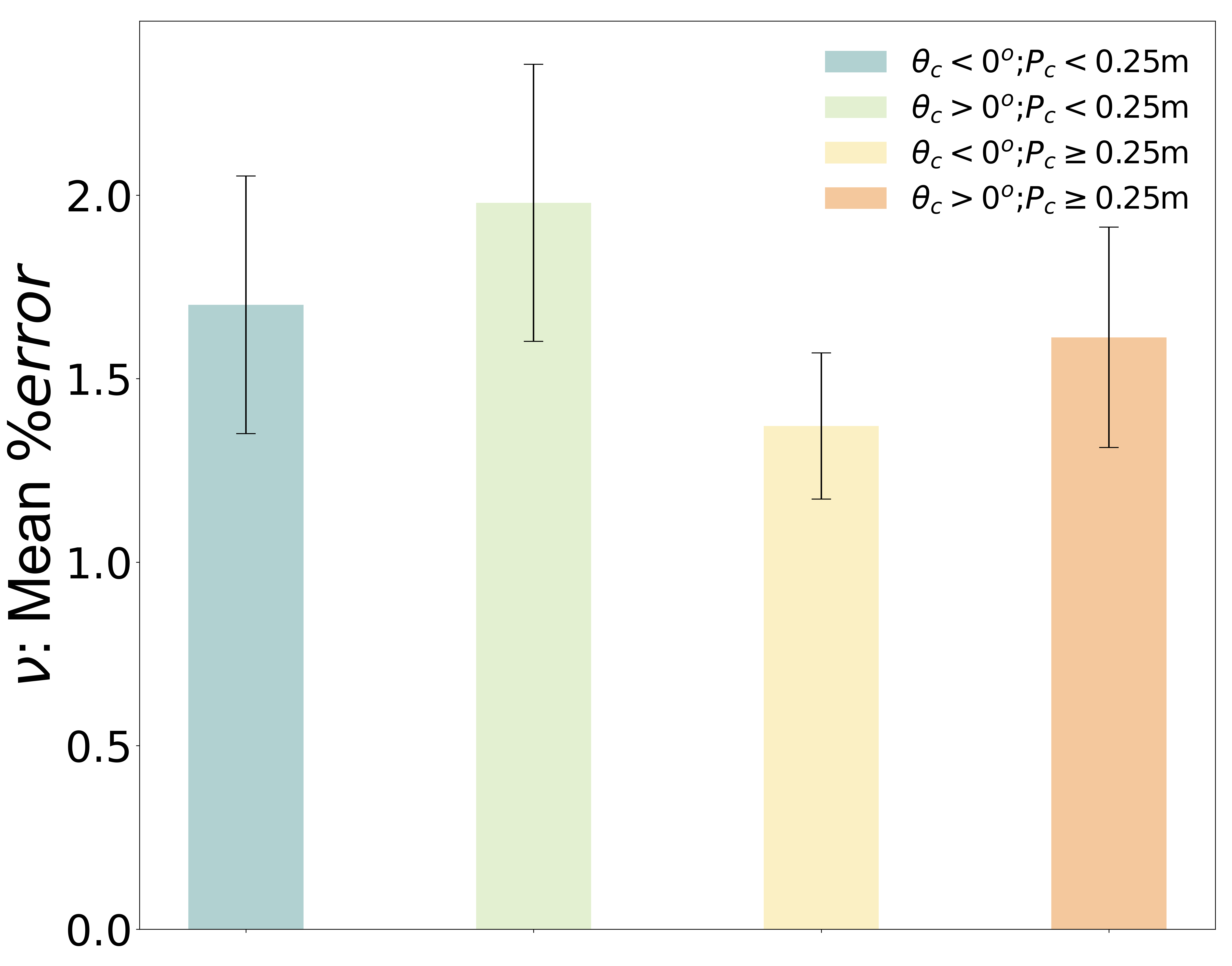}
                        \caption{Percent Errors in $\nu$}
                        \label{subfig:YDisp_Angles_Position_BarChart}
                    \end{subfigure}
                \end{subfigure}
                \begin{subfigure}[c]{0.32\textwidth}
                    \centering
                    \begin{subfigure}[t]{1\textwidth}    
                      \centering
                      \includegraphics[width=0.98\linewidth]{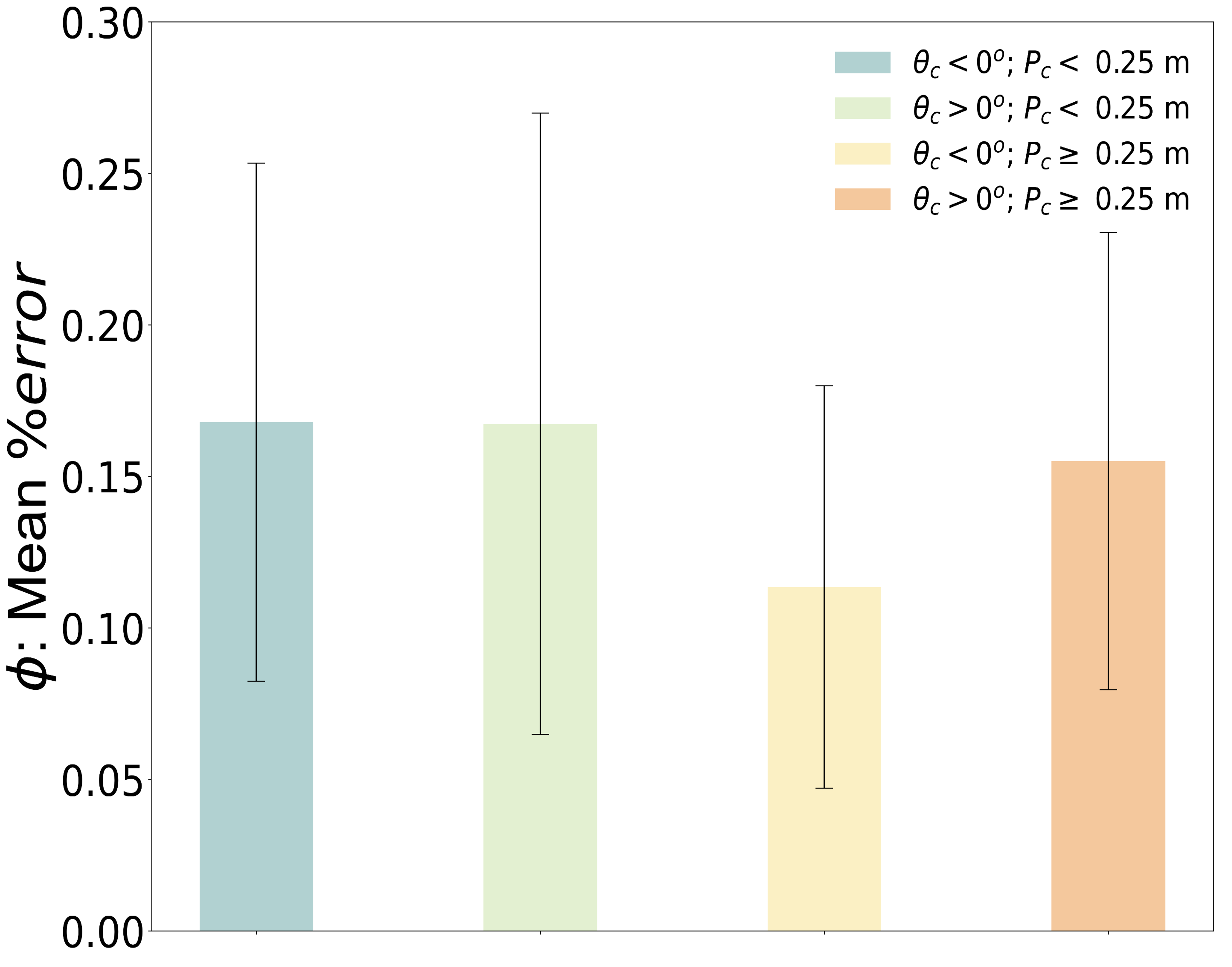}
                        \caption{Percent Errors in $\phi$}
                        \label{subfig:cPhi_Angles_Position_BarChart}
                    \end{subfigure}
                \end{subfigure}
                \caption{Parametric error analysis of the contribution of initial crack angles and initial edge positions on (a) $u$ predictions, (b) $\nu$ predictions, and (c) $\phi$ predictions.}
                \label{fig:XDisp_YDisp_cPhi_Angles_Position_BarChart}
            \end{figure}

            Next, we analyze the effects of varying initial crack orientations along with initial edge position on the prediction errors.
            Using a similar approach as in Section \ref{subsubsection:Parametric_angle_length}, we split the test dataset into four groups: (i) negative crack angle + bottom edge position defined as $\theta_{c} < 0^{o}$; $P_{c} < 0.25$ m, (ii) positive crack angle + bottom edge position defined as $\theta_{c} > 0^{o}$; $P_{c} < 0.25$ m, (iii) negative crack angle + top edge position defined as $\theta_{c} < 0^{o}$; $P_{c} \geq 0.25$ m, and (iv) positive crack angle + top edge position defined as $\theta_{c} > 0^{o}$; $P_{c} \geq 0.25$ m.
            Figure \ref{fig:XDisp_YDisp_cPhi_Angles_Position_BarChart} shows the corresponding errors of x-displacement (Figure \ref{subfig:XDisp_Angles_Position_BarChart}), y-displacement (Figure \ref{subfig:YDisp_Angles_Position_BarChart}), and $\phi$ (Figure \ref{subfig:cPhi_Angles_Position_BarChart}) predictions for each parametric group.  
            
            The \textit{XDisp}-GNN results depicted in Figure \ref{subfig:XDisp_Angles_Length_BarChart} show the lowest errors for configurations involving cracks with bottom edge position (i.e., $P_{c} < 0.25$ m).
            For instance, the group with positive crack angle + bottom edge position resulted in errors of $0.59 \pm 0.09 \%$, and the group with negative crack angle + bottom edge position showed errors of $0.69 \pm 0.09 \%$.
            Moreover, the highest error of \textit{XDisp}-GNN was for negative crack angles + top edge position at $0.84 \pm 0.15 \%$.  
            When considering large cracks the highest error of $1.00 \pm 0.18 \%$ was for the group with negative angles, while for the group with positive angles the error decreased to $0.78 \pm 0.13 \%$.
            From Figure \ref{subfig:XDisp_Angles_Length_BarChart} discussed in Section \ref{subsubsection:Parametric_angle_length}, and Figure \ref{subfig:XDisp_Angles_Position_BarChart}, we conclude that \textit{XDisp}-GNN performed best for cases involving positive crack angle, small crack length, and bottom edge position. 
            
            In contrast to \textit{XDisp}-GNN, the groups resulting in the lowest errors for \textit{YDisp}-GNN involved cases with top edge position.
            For $\{\theta_{c} < 0^{o}$; $P_{c} \geq 0.25\}$ m and $\{\theta_{c} > 0^{o}$; $P_{c} \geq 0.25\}$ m the resulting errors were $1.37 \pm 0.2 \%$ and $1.61 \pm 0.3 \%$, respectively.
            The highest obtained error was for cases with positive crack angles + bottom edge position at $1.98 \pm 0.38 \%$.   
            To understand why cracks located at the top edge position resulted in lower prediction errors using \textit{YDisp}-GNN, we emphasize that for this work the load applied was a uniform tensile displacement load along the top edge of the boundary (i.e., positive y-direction). 
            Because \textit{YDisp}-GNN predicts the evolution of the y-displacement field, $\nu$, an initial crack located closer to the top edge of the domain - where the applied displacement load is located - may help \textit{YDisp}-GNN's prediction accuracy.
            Therefore, from \ref{subfig:YDisp_Angles_Length_BarChart} discussed in Section \ref{subsubsection:Parametric_angle_length}, and Figure \ref{subfig:YDisp_Angles_Position_BarChart} we show that \textit{YDisp}-GNN performed best for cases consisting of negative crack angle, large crack length, and top edge position. 
            
            Lastly, from Figure \ref{subfig:cPhi_Angles_Position_BarChart} the resulting errors for \textit{cPhi}-GNN depict a similar relation compared to Sections \ref{subsubsection:Parametric_angle_length}-\ref{subsubsection:Parametric_angle_position}.
            We found previously that predictions of $\phi$ {showed that crack angles did not significantly affect the model's accuracy, while the crack lengths showed to play a higher role in the model's accuracy.}
            From Figure \ref{subfig:cPhi_Angles_Position_BarChart}, the lowest error was found for the case of {negative crack angle + top edge position with error of $0.11 \pm 0.07 \%$}, while the highest error found was for positive crack angle + bottom edge position with error of {$0.17 \pm 0.10 \%$}.
            Therefore, from \ref{subfig:cPhi_Angles_Length_BarChart} discussed in Section \ref{subsubsection:Parametric_angle_length}, and Figure \ref{subfig:cPhi_Angles_Position_BarChart} we conclude that \textit{cPhi}-GNN performed best for cases involving {negative crack angle, larger crack lengths, and top edge position}. 

   \subsection{Simulation time analysis}
   
\begin{figure} 
        \centering
        \includegraphics[width=0.7\linewidth]{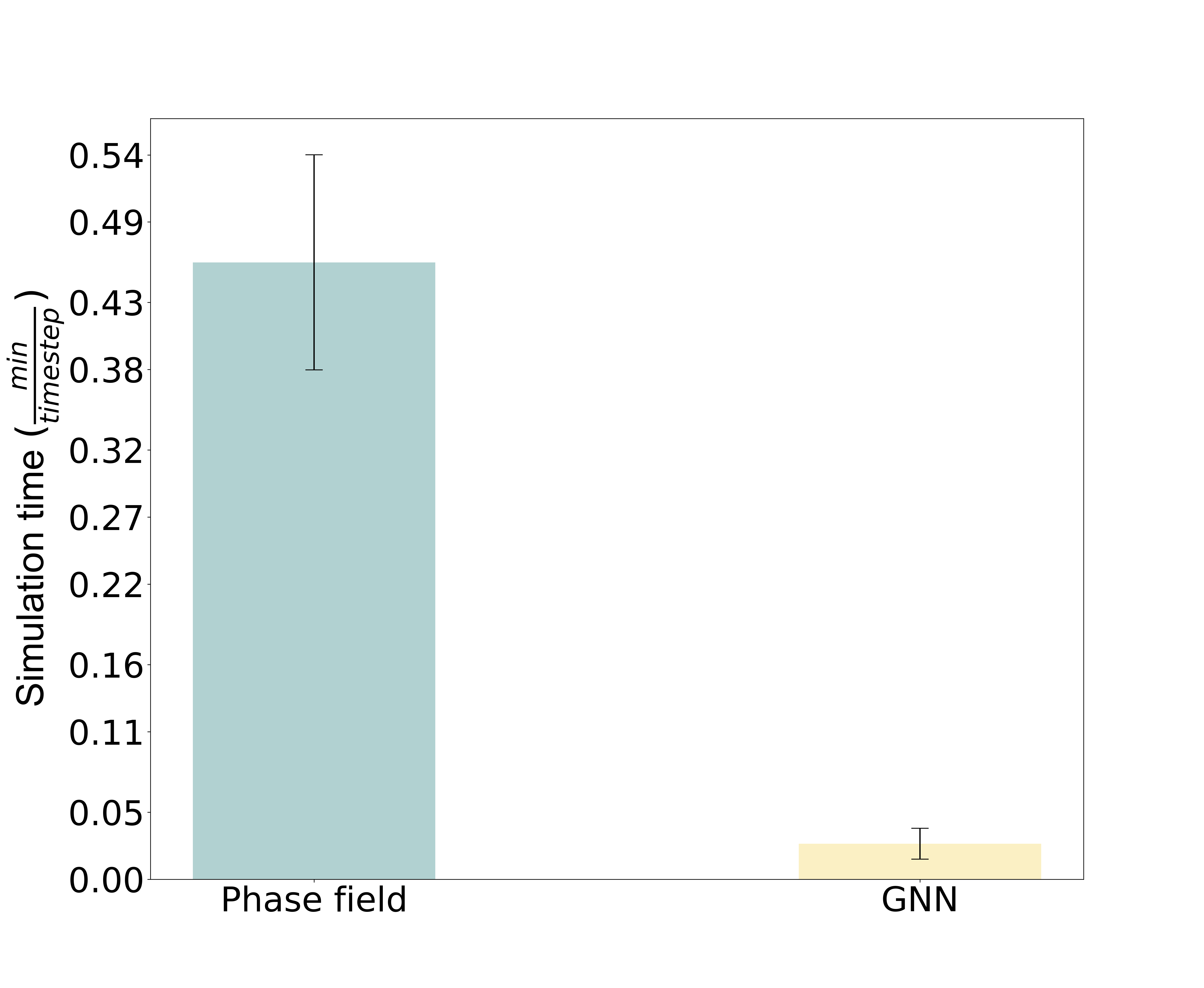}
        \caption{Simulation time analysis for PF fracture model versus GNN framework resulting in a 36x speed-up by \textit{ADAPT}-GNN.}
        \label{fig:Simulaion_Time}
    \end{figure}

        To evaluate the performance of the \textit{ADAPT}-GNN framework, we compared the simulation time to the PF fracture model.
        Towards this goal, we computed the simulation time of \textit{ADAPT}-GNN for the entire test dataset using an Nvidia GeForce RTX 3070 ti GPU on a personal computer system.
        We initialized the simulation time for \textit{ADAPT}-GNN prior to loading each model (\textit{XDisp}-GNN, \textit{YDisp}-GNN, and \textit{cPhi}-GNN), and finalized at the final time-step. 
        We took a similar approach for the PF fracture model.
        The mean and standard deviation of the obtained simulation time per time-step are shown in Figure \ref{fig:Simulaion_Time} for the PF fracture model versus \textit{ADAPT}-GNN.
        It can be seen that \textit{ADAPT}-GNN outperformed the PF fracture model achieving 15x-36x faster simulation time.
        Additionally, we could significantly improve \textit{ADAPT}-GNN's performance by using better GPU units.
        
        We also note that while \textit{ADAPT}-GNN outperformed the PF fracture model in this case, the PF model used in this work is not CPU parallelized.
        A PF fracture model with an ideal parallel scaling may outperform the developed GNN framework when using greater than 16 or 32 processors.
        {Additionally, it is important to consider the long training times required for each model in the \textit{ADAPT}-GNN framework.
        For instance, \textit{XDisp}-GNN and \textit{YDisp}-GNN required 9 hours and 22 minutes each, while \textit{cPhi}-GNN required 10 hours and 57 minutes for a total of 20 epochs.
        This equates to a total of 29 hours and 41 minutes of training time for \textit{ADAPT}-GNN.
        In the case that training was required for simulating cases of 100 time-steps each where the ML framework was 15x faster than PF, \textit{ADAPT}-GNN would begin to outperform the PF model for 34+ simulations.}
        \added[id=R1R2,comment={Q18,Q3}]{Another crucial drawback of data-driven ML methods is the required time for data set collection.
        For instance, it required approximately 30 days to generate 1245 simulations by running 3-4 PF models simultaneously.}
        This shows that conventional fracture models, such as the PF approach, are vital for developing new ML algorithms able to speed up computational times in the future. 
        We emphasize that this work is not intended to substitute conventional PF fracture models but to demonstrate the ability to use ML to speed up computational times.

\section{Conclusion}\label{sec:Conclusion}
    To conclude, the development of mesh-based GNN models for simulating complex fracture problems is a recent area of
    research which has shown significant speed-ups compared to existing high-fidelity computational models.      
    However, integrating this technique into PF simulations with AMR has not been explored in previous works.
    As a result, this work develops an adaptive mesh-based GNN framework (\textit{ADAPT}-GNN), capable of emulating PF fracture models for single-edge notched cracks subjected to tensile loadings.  
    As shown in Figure \ref{fig:GNN_Flowchart}, \textit{ADAPT}-GNN first predicts the x- and y-displacement fields, followed by prediction of the scalar damage field ($\phi$) at the future time-step.
    We note that the predicted displacement fields and $\phi$ can then be utilized to compute the stress evolution in the material.
    Another key feature of \textit{ADAPT}-GNN is its ability to benefit from both computational efficiencies of AMR and ML techniques by representing each instantaneous graph as the refined mesh itself.
    This dynamic graph implementation resulted in simulation speed-ups up to 36x faster than a conventional PF fracture model using an NVIDIA GeForce RTX 3070 Ti GPU on a personal computer.   
    The framework showed good prediction accuracies in the test dataset with maximum percent errors of $1.98 \pm 0.27\%$, $2.74 \pm 0.26\%$ and $0.19 \pm 0.13 \%$ for the x-displacements, y-displacements, and $\phi$, respectively. 
    
    While the \textit{ADAPT}-GNN framework predicts displacements and $\phi$ with overall good accuracy, we point out various limitations.
    {The} conventional PF model used in this work did not have parallel CPU capability.
    A parallelized PF model may outperform the developed GNN framework when using 16 or 32 processors.
    {Also, when re-training each model is required, the framework begins to outperform the PF model only for approximately 34+ simulations (of 100 time-steps each).}
    \textit{ADAPT}-GNN is not able to predict unseen cases involving shear loadings, center cracks, and cracks located at the right edge of the domain. 
    Therefore, the limitations of the developed framework demonstrate how conventional fracture models are essential for developing new ML algorithms.
    
    Ultimately, PF fracture models for simulating crack propagation are one of the most computationally demanding PF models. 
    This work presents the development of a new adaptive mesh GNN capable of predicting PF fracture models of {single-edge notched} crack propagation with good accuracies and computational speed-ups.
    Transfer learning approaches {such as} \cite{Perera2022Genralized} may be employed in future work to extend the current framework's capability of predicting unseen cases with shear loading, center cracks, and cracks located at the right edge.
    The developed framework can also be extended to various PF models other than fracture models by implementing a similar methodology.
    As new GNN techniques are developed, new models and methods, such as subgraphs, can be explored to increase computational speed.  

\section{Data availability}
    The trained models with examples can be found in the following GitHub repository \url{https://github.com/rperera12/Phase-Field-ADAPT-GNN}.
    Supplementary data containing animations have been included along with the manuscript.
\section{Acknowledgements}
    The authors are grateful for the financial support provided by the U.S. Department of Defense in conjunction with the Naval Air Warfare Center Weapons Division (NAWCWD) through the SMART scholarship Program (SMART ID: 2021--17978).

\bibliographystyle{ieeetr}
\bibliography{library}

\begin{thebibliography}{10}

\bibitem{Sedmak2018Review}
A.~Sedmak, ``Computational fracture mechanics: An overview from early efforts
  to recent achievements,'' {\em Fatigue \& Fracture of Engineering Materials
  \& Structures}, vol.~41, no.~12, pp.~2438--2474, 2018.

\bibitem{SUTULA2018205}
D.~Sutula, P.~Kerfriden, T.~{van Dam}, and S.~P. Bordas, ``Minimum energy
  multiple crack propagation. part i: Theory and state of the art review,''
  {\em Engineering Fracture Mechanics}, vol.~191, pp.~205--224, 2018.

\bibitem{SUTULA2018225}
D.~Sutula, P.~Kerfriden, T.~{van Dam}, and S.~P. Bordas, ``Minimum energy
  multiple crack propagation. part-ii: Discrete solution with xfem,'' {\em
  Engineering Fracture Mechanics}, vol.~191, pp.~225--256, 2018.

\bibitem{SUTULA2018257}
D.~Sutula, P.~Kerfriden, T.~{van Dam}, and S.~P. Bordas, ``Minimum energy
  multiple crack propagation. part iii: Xfem computer implementation and
  applications,'' {\em Engineering Fracture Mechanics}, vol.~191, pp.~257--276,
  2018.

\bibitem{Belytschko1999XFEM}
N.~Moës, J.~Dolbow, and T.~Belytschko, ``A finite element method for crack
  growth without remeshing,'' {\em International Journal for Numerical Methods
  in Engineering}, vol.~46, no.~1, pp.~131--150, 1999.

\bibitem{Belytschko2000RockXFEM}
N.~Sukumar, N.~Moës, B.~Moran, and T.~Belytschko, ``Extended finite element
  method for three-dimensional crack modelling,'' {\em International Journal
  for Numerical Methods in Engineering}, vol.~48, no.~11, pp.~1549--1570, 2000.

\bibitem{li2018review}
H.~Li, J.~Li, and H.~Yuan, ``A review of the extended finite element method on
  macrocrack and microcrack growth simulations,'' {\em Theoretical and Applied
  Fracture Mechanics}, vol.~97, pp.~236--249, 2018.

\bibitem{FRANCFORT19981319}
G.~Francfort and J.-J. Marigo, ``Revisiting brittle fracture as an energy
  minimization problem,'' {\em Journal of the Mechanics and Physics of Solids},
  vol.~46, no.~8, pp.~1319--1342, 1998.

\bibitem{app9122436}
A.~Egger, U.~Pillai, K.~Agathos, E.~Kakouris, E.~Chatzi, I.~A. Aschroft, and
  S.~P. Triantafyllou, ``Discrete and phase field methods for linear elastic
  fracture mechanics: A comparative study and state-of-the-art review,'' {\em
  Applied Sciences}, vol.~9, no.~12, 2019.

\bibitem{Ambati2014Review}
M.~Ambati, T.~Gerasimov, and L.~De~Lorenzis, ``A review on phase-field models
  of brittle fracture and a new fast hybrid formulation,'' {\em Computational
  Mechanics}, vol.~55, 12 2014.

\bibitem{Ambati2016Phase}
M.~Ambati, R.~Kruse, and L.~De~Lorenzis, ``A phase-field model for ductile
  fracture at finite strains and its experimental verification,'' {\em
  Computational Mechanics}, vol.~57, 01 2016.

\bibitem{ERNESTI2020112793}
F.~Ernesti, M.~Schneider, and T.~Böhlke, ``Fast implicit solvers for
  phase-field fracture problems on heterogeneous microstructures,'' {\em
  Computer Methods in Applied Mechanics and Engineering}, vol.~363, p.~112793,
  2020.

\bibitem{ZHANG2022114282}
G.~Zhang, T.~F. Guo, K.~I. Elkhodary, S.~Tang, and X.~Guo, ``Mixed graph-fem
  phase field modeling of fracture in plates and shells with nonlinearly
  elastic solids,'' {\em Computer Methods in Applied Mechanics and
  Engineering}, vol.~389, p.~114282, 2022.

\bibitem{clayton2022stress}
T.~Clayton, R.~Duddu, M.~Siegert, and E.~Mart{\'\i}nez-Pa{\~n}eda, ``A
  stress-based poro-damage phase field model for hydrofracturing of creeping
  glaciers and ice shelves,'' {\em Engineering Fracture Mechanics}, vol.~272,
  p.~108693, 2022.

\bibitem{RUNNELS2021110065}
B.~Runnels, V.~Agrawal, W.~Zhang, and A.~Almgren, ``Massively parallel finite
  difference elasticity using block-structured adaptive mesh refinement with a
  geometric multigrid solver,'' {\em Journal of Computational Physics},
  vol.~427, p.~110065, 2021.

\bibitem{AGRAWAL2021114011}
V.~Agrawal and B.~Runnels, ``Block structured adaptive mesh refinement and
  strong form elasticity approach to phase field fracture with applications to
  delamination, crack branching and crack deflection,'' {\em Computer Methods
  in Applied Mechanics and Engineering}, vol.~385, p.~114011, 2021.

\bibitem{Ribot_2019}
J.~G. Ribot, V.~Agrawal, and B.~Runnels, ``A new approach for phase field
  modeling of grain boundaries with strongly nonconvex energy,'' {\em Modelling
  and Simulation in Materials Science and Engineering}, vol.~27, p.~084007, oct
  2019.

\bibitem{Norton2001PYRAMID}
C.~D. Norton, T.~A. Cwik, and J.~Z. Lou, ``Status and directions for the
  pyramid parallel unstructured amr library,'' in {\em Parallel and Distributed
  Processing Symposium, International}, vol.~4, (Los Alamitos, CA, USA),
  p.~30120b, IEEE Computer Society, apr 2001.

\bibitem{MACNEICE2000330}
P.~MacNeice, K.~M. Olson, C.~Mobarry, R.~{de Fainchtein}, and C.~Packer,
  ``Paramesh: A parallel adaptive mesh refinement community toolkit,'' {\em
  Computer Physics Communications}, vol.~126, no.~3, pp.~330--354, 2000.

\bibitem{feng2020stochastic}
Y.~Feng, Q.~Wang, D.~Wu, W.~Gao, and F.~Tin-Loi, ``Stochastic nonlocal damage
  analysis by a machine learning approach,'' {\em Computer Methods in Applied
  Mechanics and Engineering}, vol.~372, p.~113371, 2020.

\bibitem{capuano2019smart}
G.~Capuano and J.~J. Rimoli, ``Smart finite elements: A novel machine learning
  application,'' {\em Computer Methods in Applied Mechanics and Engineering},
  vol.~345, pp.~363--381, 2019.

\bibitem{GU201819}
G.~X. Gu, C.-T. Chen, and M.~J. Buehler, ``De novo composite design based on
  machine learning algorithm,'' {\em Extreme Mechanics Letters}, vol.~18,
  pp.~19--28, 2018.

\bibitem{C8MH00653A}
G.~X. Gu, C.-T. Chen, D.~J. Richmond, and M.~J. Buehler, ``Bioinspired
  hierarchical composite design using machine learning: simulation{,} additive
  manufacturing{,} and experiment,'' {\em Mater. Horiz.}, vol.~5, pp.~939--945,
  2018.

\bibitem{ZHANG2022115233}
Z.~Zhang, Z.~Zhang, F.~{Di Caprio}, and G.~X. Gu, ``Machine learning for
  accelerating the design process of double-double composite structures,'' {\em
  Composite Structures}, vol.~285, p.~115233, 2022.

\bibitem{D1MH01792F}
S.~Lee, Z.~Zhang, and G.~X. Gu, ``Generative machine learning algorithm for
  lattice structures with superior mechanical properties,'' {\em Mater.
  Horiz.}, vol.~9, pp.~952--960, 2022.

\bibitem{ZHANG2020112725}
Y.~Zhang, Z.~Wen, H.~Pei, J.~Wang, Z.~Li, and Z.~Yue, ``Equivalent method of
  evaluating mechanical properties of perforated ni-based single crystal plates
  using artificial neural networks,'' {\em Computer Methods in Applied
  Mechanics and Engineering}, vol.~360, p.~112725, 2020.

\bibitem{hanna2022residual}
J.~M. Hanna, J.~V. Aguado, S.~Comas-Cardona, R.~Askri, and D.~Borzacchiello,
  ``Residual-based adaptivity for two-phase flow simulation in porous media
  using physics-informed neural networks,'' {\em Computer Methods in Applied
  Mechanics and Engineering}, vol.~396, p.~115100, 2022.

\bibitem{ren2022phycrnet}
P.~Ren, C.~Rao, Y.~Liu, J.-X. Wang, and H.~Sun, ``Phycrnet: Physics-informed
  convolutional-recurrent network for solving spatiotemporal pdes,'' {\em
  Computer Methods in Applied Mechanics and Engineering}, vol.~389, p.~114399,
  2022.

\bibitem{wang2022structural}
M.~Wang, S.~Feng, A.~Incecik, G.~Kr{\'o}lczyk, and Z.~Li, ``Structural fatigue
  life prediction considering model uncertainties through a novel digital
  twin-driven approach,'' {\em Computer Methods in Applied Mechanics and
  Engineering}, vol.~391, p.~114512, 2022.

\bibitem{mangal2018applied}
A.~Mangal and E.~A. Holm, ``Applied machine learning to predict stress hotspots
  i: Face centered cubic materials,'' {\em International Journal of
  Plasticity}, vol.~111, pp.~122--134, 2018.

\bibitem{mangal2019applied}
A.~Mangal and E.~A. Holm, ``Applied machine learning to predict stress hotspots
  ii: Hexagonal close packed materials,'' {\em International Journal of
  Plasticity}, vol.~114, pp.~1--14, 2019.

\bibitem{he2021deep}
X.~He, Q.~He, and J.-S. Chen, ``Deep autoencoders for physics-constrained
  data-driven nonlinear materials modeling,'' {\em Computer Methods in Applied
  Mechanics and Engineering}, vol.~385, p.~114034, 2021.

\bibitem{yang2021deep}
Z.~Yang, C.-H. Yu, and M.~J. Buehler, ``Deep learning model to predict complex
  stress and strain fields in hierarchical composites,'' {\em Science
  Advances}, vol.~7, no.~15, p.~eabd7416, 2021.

\bibitem{saha2021hierarchical}
S.~Saha, Z.~Gan, L.~Cheng, J.~Gao, O.~L. Kafka, X.~Xie, H.~Li, M.~Tajdari,
  H.~A. Kim, and W.~K. Liu, ``Hierarchical deep learning neural network
  (hidenn): An artificial intelligence (ai) framework for computational science
  and engineering,'' {\em Computer Methods in Applied Mechanics and
  Engineering}, vol.~373, p.~113452, 2021.

\bibitem{IM2021114030}
S.~Im, J.~Lee, and M.~Cho, ``Surrogate modeling of elasto-plastic problems via
  long short-term memory neural networks and proper orthogonal decomposition,''
  {\em Computer Methods in Applied Mechanics and Engineering}, vol.~385,
  p.~114030, 2021.

\bibitem{HUNTER201987}
A.~Hunter, B.~A. Moore, M.~Mudunuru, V.~Chau, R.~Tchoua, C.~Nyshadham,
  S.~Karra, D.~O’Malley, E.~Rougier, H.~Viswanathan, and G.~Srinivasan,
  ``Reduced-order modeling through machine learning and graph-theoretic
  approaches for brittle fracture applications,'' {\em Computational Materials
  Science}, vol.~157, pp.~87--98, 2019.

\bibitem{HSU2020197}
Y.-C. Hsu, C.-H. Yu, and M.~J. Buehler, ``Using deep learning to predict
  fracture patterns in crystalline solids,'' {\em Matter}, vol.~3, no.~1,
  pp.~197--211, 2020.

\bibitem{lew2021deep}
A.~J. Lew, C.-H. Yu, Y.-C. Hsu, and M.~J. Buehler, ``Deep learning model to
  predict fracture mechanisms of graphene,'' {\em npj 2D Materials and
  Applications}, vol.~5, no.~1, pp.~1--8, 2021.

\bibitem{rovinelli2018using}
A.~Rovinelli, M.~D. Sangid, H.~Proudhon, and W.~Ludwig, ``Using machine
  learning and a data-driven approach to identify the small fatigue crack
  driving force in polycrystalline materials,'' {\em npj Computational
  Materials}, vol.~4, no.~1, pp.~1--10, 2018.

\bibitem{elapolu2022novel}
M.~S. Elapolu, M.~I.~R. Shishir, and A.~Tabarraei, ``A novel approach for
  studying crack propagation in polycrystalline graphene using machine learning
  algorithms,'' {\em Computational Materials Science}, vol.~201, p.~110878,
  2022.

\bibitem{wang2021stressnet}
Y.~Wang, D.~Oyen, W.~G. Guo, A.~Mehta, C.~B. Scott, N.~Panda, M.~G.
  Fern{\'a}ndez-Godino, G.~Srinivasan, and X.~Yue, ``Stressnet-deep learning to
  predict stress with fracture propagation in brittle materials,'' {\em Npj
  Materials Degradation}, vol.~5, no.~1, pp.~1--10, 2021.

\bibitem{Zhang2020High}
K.~Zhang, J.~Wang, Y.~Huang, L.-Q. Chen, and Y.~Cao, ``High-throughput
  phase-field simulations and machine learning of resistive switching in
  resistive random-access memory,'' {\em npj Computational Materials}, vol.~6,
  12 2020.

\bibitem{Zhu2021Linear}
Y.~Zhu, T.~Xu, Q.~Wei, J.~Mai, H.~Yang, H.~Zhang, T.~Shimada, T.~Kitamura, and
  T.-Y. Zhang, ``Linear-superelastic ti-nb nanocomposite alloys with ultralow
  modulus via high-throughput phase-field design and machine learning,'' {\em
  npj Computational Materials}, vol.~7, 12 2021.

\bibitem{SAMANIEGO2020112790}
E.~Samaniego, C.~Anitescu, S.~Goswami, V.~Nguyen-Thanh, H.~Guo, K.~Hamdia,
  X.~Zhuang, and T.~Rabczuk, ``An energy approach to the solution of partial
  differential equations in computational mechanics via machine learning:
  Concepts, implementation and applications,'' {\em Computer Methods in Applied
  Mechanics and Engineering}, vol.~362, p.~112790, 2020.

\bibitem{TEICHERT2019666}
G.~H. Teichert and K.~Garikipati, ``Machine learning materials physics:
  Surrogate optimization and multi-fidelity algorithms predict precipitate
  morphology in an alternative to phase field dynamics,'' {\em Computer Methods
  in Applied Mechanics and Engineering}, vol.~344, pp.~666--693, 2019.

\bibitem{FENG2021113885}
S.~Feng, Y.~Xu, X.~Han, Z.~Li, and A.~Incecik, ``A phase field and
  deep-learning based approach for accurate prediction of structural residual
  useful life,'' {\em Computer Methods in Applied Mechanics and Engineering},
  vol.~383, p.~113885, 2021.

\bibitem{karniadakis2022learning}
V.~Oommen, K.~Shukla, S.~Goswami, R.~Dingreville, and G.~E. Karniadakis,
  ``Learning two-phase microstructure evolution using neural operators and
  autoencoder architectures,'' {\em arXiv preprint arXiv:2204.07230}, 2022.

\bibitem{montes2021accelerating}
D.~Montes~de Oca~Zapiain, J.~A. Stewart, and R.~Dingreville, ``Accelerating
  phase-field-based microstructure evolution predictions via surrogate models
  trained by machine learning methods,'' {\em npj Computational Materials},
  vol.~7, no.~1, pp.~1--11, 2021.

\bibitem{sanchez2020learning}
A.~Sanchez-Gonzalez, J.~Godwin, T.~Pfaff, R.~Ying, J.~Leskovec, and
  P.~Battaglia, ``Learning to simulate complex physics with graph networks,''
  in {\em International Conference on Machine Learning}, pp.~8459--8468, PMLR,
  2020.

\bibitem{frankel2022mesh}
A.~L. Frankel, C.~Safta, C.~Alleman, and R.~Jones, ``Mesh-based graph
  convolutional neural networks for modeling materials with microstructure,''
  {\em Journal of Machine Learning for Modeling and Computing}, vol.~3, no.~1,
  2022.

\bibitem{dai2021graph}
M.~Dai, M.~F. Demirel, Y.~Liang, and J.-M. Hu, ``Graph neural networks for an
  accurate and interpretable prediction of the properties of polycrystalline
  materials,'' {\em npj Computational Materials}, vol.~7, no.~1, pp.~1--9,
  2021.

\bibitem{cryst12020280}
C.~Shu, J.~He, G.~Xue, and C.~Xie, ``Grain knowledge graph representation
  learning: A new paradigm for microstructure-property prediction,'' {\em
  Crystals}, vol.~12, no.~2, 2022.

\bibitem{choudhary2021atomistic}
K.~Choudhary and B.~DeCost, ``Atomistic line graph neural network for improved
  materials property predictions,'' {\em npj Computational Materials}, vol.~7,
  no.~1, pp.~1--8, 2021.

\bibitem{Stylianos2022workflow}
S.~Tsopanidis and S.~Osovski, ``A graph-based workflow for extracting
  grain-scale toughness from meso-scale experiments,'' {\em Materials \&
  Design}, vol.~213, p.~110272, 2022.

\bibitem{fung2021benchmarking}
V.~Fung, J.~Zhang, E.~Juarez, and B.~G. Sumpter, ``Benchmarking graph neural
  networks for materials chemistry,'' {\em npj Computational Materials},
  vol.~7, no.~1, pp.~1--8, 2021.

\bibitem{rosen2022high}
A.~S. Rosen, V.~Fung, P.~Huck, C.~T. O’Donnell, M.~K. Horton, D.~G. Truhlar,
  K.~A. Persson, J.~M. Notestein, and R.~Q. Snurr, ``High-throughput
  predictions of metal--organic framework electronic properties: theoretical
  challenges, graph neural networks, and data exploration,'' {\em npj
  Computational Materials}, vol.~8, no.~1, pp.~1--10, 2022.

\bibitem{HEIDER2020112875}
Y.~Heider, K.~Wang, and W.~Sun, ``So(3)-invariance of informed-graph-based deep
  neural network for anisotropic elastoplastic materials,'' {\em Computer
  Methods in Applied Mechanics and Engineering}, vol.~363, p.~112875, 2020.

\bibitem{Gu2022Perovskite}
G.~Gu, J.~Jang, J.~Noh, A.~Walsh, and Y.~Jung, ``Perovskite synthesizability
  using graph neural networks,'' {\em npj Computational Materials}, vol.~8,
  p.~71, 04 2022.

\bibitem{wang2021inverse}
Q.~Wang and L.~Zhang, ``Inverse design of glass structure with deep graph
  neural networks,'' {\em Nature communications}, vol.~12, no.~1, pp.~1--11,
  2021.

\bibitem{BLACK2022115120}
N.~Black and A.~R. Najafi, ``Learning finite element convergence with the
  multi-fidelity graph neural network,'' {\em Computer Methods in Applied
  Mechanics and Engineering}, vol.~397, p.~115120, 2022.

\bibitem{park2021accurate}
C.~W. Park, M.~Kornbluth, J.~Vandermause, C.~Wolverton, B.~Kozinsky, and J.~P.
  Mailoa, ``Accurate and scalable graph neural network force field and
  molecular dynamics with direct force architecture,'' {\em npj Computational
  Materials}, vol.~7, no.~1, pp.~1--9, 2021.

\bibitem{VLASSIS2020113299}
N.~N. Vlassis, R.~Ma, and W.~Sun, ``Geometric deep learning for computational
  mechanics part i: anisotropic hyperelasticity,'' {\em Computer Methods in
  Applied Mechanics and Engineering}, vol.~371, p.~113299, 2020.

\bibitem{mayr2021boundary}
A.~Mayr, S.~Lehner, A.~Mayrhofer, C.~Kloss, S.~Hochreiter, and J.~Brandstetter,
  ``Boundary graph neural networks for 3d simulations,'' {\em arXiv preprint
  arXiv:2106.11299}, 2021.

\bibitem{perera2022graph}
R.~Perera, D.~Guzzetti, and V.~Agrawal, ``Graph neural networks for simulating
  crack coalescence and propagation in brittle materials,'' {\em Computer
  Methods in Applied Mechanics and Engineering}, vol.~395, p.~115021, 2022.

\bibitem{pfaff2020learning}
T.~Pfaff, M.~Fortunato, A.~Sanchez-Gonzalez, and P.~W. Battaglia, ``Learning
  mesh-based simulation with graph networks,'' {\em arXiv preprint
  arXiv:2010.03409}, 2020.

\bibitem{GOSWAMI2020112808}
S.~Goswami, C.~Anitescu, and T.~Rabczuk, ``Adaptive fourth-order phase field
  analysis for brittle fracture,'' {\em Computer Methods in Applied Mechanics
  and Engineering}, vol.~361, p.~112808, 2020.

\bibitem{Hu2019GINEConv}
W.~Hu, B.~Liu, J.~Gomes, M.~Zitnik, P.~Liang, V.~Pande, and J.~Leskovec,
  ``Strategies for pre-training graph neural networks,'' 2019.

\bibitem{klicpera2020directional}
J.~Klicpera, J.~Gro{\ss}, and S.~G{\"{u}}nnemann, ``Directional message passing
  for molecular graphs,'' {\em CoRR}, vol.~abs/2003.03123, 2020.

\bibitem{zhang2020dynamic}
L.~Zhang, D.~Xu, A.~Arnab, and P.~H. Torr, ``Dynamic graph message passing
  networks,'' in {\em Proceedings of the IEEE/CVF Conference on Computer Vision
  and Pattern Recognition (CVPR)}, June 2020.

\bibitem{gilmer2017neural}
J.~Gilmer, S.~S. Schoenholz, P.~F. Riley, O.~Vinyals, and G.~E. Dahl, ``Neural
  message passing for quantum chemistry,'' in {\em Proceedings of the 34th
  International Conference on Machine Learning} (D.~Precup and Y.~W. Teh,
  eds.), vol.~70 of {\em Proceedings of Machine Learning Research},
  pp.~1263--1272, PMLR, 06--11 Aug 2017.

\bibitem{Zhu2020A3TGCN}
J.~Zhu, Y.~Song, L.~Zhao, and H.~Li, ``A3t-gcn: Attention temporal graph
  convolutional network for traffic forecasting,'' 2020.

\bibitem{Zhao2020TGCN}
L.~Zhao, Y.~Song, C.~Zhang, Y.~Liu, P.~Wang, T.~Lin, M.~Deng, and H.~Li,
  ``T-gcn: A temporal graph convolutional network for traffic prediction,''
  {\em IEEE Transactions on Intelligent Transportation Systems}, vol.~21,
  no.~9, pp.~3848--3858, 2020.

\bibitem{kingma2017adam}
D.~P. Kingma and J.~Ba, ``Adam: A method for stochastic optimization,'' 2017.

\bibitem{Hamilton2020Graph}
W.~L. Hamilton, ``Graph representation learning,'' {\em Synthesis Lectures on
  Artificial Intelligence and Machine Learning}, vol.~14, no.~3, pp.~1--159,
  2020.

\bibitem{Fushiki2011K-Fold}
T.~Fushiki, ``Estimation of prediction error by using k-fold
  cross-validation,'' {\em Statistics and Computing}, vol.~21, p.~137–146,
  apr 2011.

\bibitem{Perera2022Genralized}
R.~Perera and V.~Agrawal, ``A generalized machine learning framework for
  brittle crack problems using transfer learning and graph neural networks,''
  2022.

\end{thebibliography}

\end{document}